\documentclass[11pt]{article}

\usepackage[left=2.7cm,bottom=2.7cm,right=2.7cm,top=2.7cm]{geometry}

\usepackage{amsthm,amsmath,amsfonts,amssymb,xr}
\usepackage[numbers]{natbib}
\usepackage{hyperref}
\usepackage{graphicx,verbatim,paralist,multirow,color}

\newtheorem{cor}{Corollary}
\newtheorem{thm}{Theorem}
\newtheorem{assmp}{Assumption}
\newtheorem{lem}{Lemma}
\newtheorem{prop}{Proposition}

\newtheorem{remark}{Remark}

\newtheorem{definition}{Definition}
\theoremstyle{remark}

\newcommand{\B}{\boldsymbol}
\newcommand{\M}{\mathbf}

\newcommand{\sbt}{\mathrm{s.t. }}

\newcommand{\MU}{{\mathcal M}_{\textsc{u}}}

\DeclareMathOperator*{\argmin}{arg\,min}
\DeclareMathOperator*{\argmax}{arg\,max}
\usepackage{booktabs}
\usepackage{pifont}

\newcommand{\bei}{\begin{itemize}}
\newcommand{\eei}{\end{itemize}}

\newcommand{\btheta}{\boldsymbol \theta}
\newcommand{\bepsilon}{\boldsymbol \epsilon}

\newcommand{\bbeta}{\boldsymbol \beta}

\newcommand{\bY}{{\bf y}}

\newcommand{\bh}{{\bf h}}
\newcommand{\bX}{{\bf X}}

\newcommand{\bx}{{\bf x}}

\newcommand{\bv}{{\bf v}}

\newcommand{\bff}{{\bf f}}
\def\RR{\mathbb{R}}
\newcommand{\grp}{{Group $\ell_0$}}

\begin{document}

\title{Grouped Variable Selection with Discrete Optimization: Computational and Statistical Perspectives}

\author{Hussein Hazimeh\thanks{Google Research;  work done while at the Massachusetts Institute of Technology, \texttt{hazimeh@google.com}}~~~~Rahul Mazumder\thanks{Massachusetts Institute of Technology, \texttt{rahulmaz@mit.edu} }~~~~Peter Radchenko\thanks{University of Sydney,~\texttt{peter.radchenko@sydney.edu.au}}}

\maketitle \sloppy

\begin{abstract}
We present a new algorithmic framework for grouped variable selection that is based on discrete mathematical optimization. While there exist several appealing approaches based on convex relaxations and nonconvex heuristics, we focus on optimal solutions for the $\ell_0$-regularized formulation, a problem that is relatively unexplored due to computational challenges. 
Our methodology covers both high-dimensional linear regression and nonparametric sparse additive modeling with smooth components. Our algorithmic framework consists of approximate and exact algorithms. The approximate algorithms are based on coordinate descent and local search, with runtimes comparable to popular sparse learning algorithms. Our exact algorithm is based on a standalone branch-and-bound (BnB) framework, which can solve the associated mixed integer programming (MIP) problem to certified optimality. By exploiting the problem structure, our custom BnB algorithm can solve to optimality problem instances with $5 \times 10^6$ features \textcolor{black}{and $10^3$ observations} in minutes to hours -- over $1000$ times larger than what is currently possible using state-of-the-art commercial MIP solvers. We also explore statistical properties of the $\ell_0$-based estimators. 
We demonstrate, theoretically and empirically, that our proposed estimators have an edge over popular group-sparse estimators in terms of statistical performance in various regimes. We provide an open source implementation of our proposed framework.
\end{abstract}

\section{Introduction}
Sparsity plays a ubiquitous role in modern statistical regression, especially when the number of predictors is large relative to the number of observations. In this paper, we focus on the case where predictors have a  natural group structure. Typical examples where such a structure appears are models with multilevel categorical predictors and models that represent nonlinear effects of continuous variables using basis functions~\cite{Bbuhl1,yuan4,hastie2015statistical}.  Grouping may also arise from scientifically meaningful prior knowledge about the collection of the predictor variables. More specifically, we consider the usual linear regression framework with response $\M{y}_{n \times 1}$ and model matrix $\M{X}_{n \times p}=[\M{x}_{1}, \ldots, \M{x}_{p}]$.  We suppose that the $p$ predictors are divided into~$q$ pre-specified, non-overlapping groups. For a given $\bbeta\in\RR^p$ and each $g\in\{1,...,q\}$, we denote by $\bbeta_g$ the sub-vector of $\bbeta$ whose coefficients correspond to the predictors in group~$g$. Following the traditional approach in high-dimensional regression, we assume that few of the regression coefficients are nonzero, i.e., the model is sparse.
This leads to a natural generalization of the classical  best subset selection problem in linear regression~\cite{miller2002subset,bertsimas2015best} to the group setting:
\begin{equation}\label{grp-l0-1-const1}
\min_{\B\beta}~~~~\| \M{y} - \M{X} \B\beta \|_{2}^2 + \lambda_0  \sum_{g=1}^{q} \M{1} ( \B\beta_{g}  \neq \M{0}),
\end{equation}
where $\M{1}(\cdot)$ is the indicator function, and $\lambda_0$ is a non-negative regularization parameter that controls the number of nonzero groups selected. We will refer to Problem \eqref{grp-l0-1-const1} as the \grp~problem. 

{Problem~\eqref{grp-l0-1-const1} is NP-Hard~\cite{natarajan1995sparse} and poses computational  challenges. A rich body of prior work explores sparsity-inducing methods to obtain approximate solutions to~\eqref{grp-l0-1-const1}. Popular methods include: convex optimization based procedures, such as Group Lasso~\citep{yuan4}, which is a generalization of the Lasso approach \citep{tibshirani3} to the grouped setting, and local solutions to nonconvex optimization problems arising from group-nonconvex regularizers, such as SCAD, MCP and others~\cite{mcp,huang4}.
Despite the appeal of these approaches,  the statistical and computational aspects of {\emph{optimal}} solutions to~\eqref{grp-l0-1-const1} remain to be understood at a deeper level. 
To this end, we 
 aim to advance the computational frontiers of  Problem~\eqref{grp-l0-1-const1} using novel tools from discrete optimization. Our proposed combinatorial optimization-based algorithms are scalable. In particular, they can deliver optimal solutions
 to~\eqref{grp-l0-1-const1} for instances that are much larger than state-of-the-art approaches. 
 We also develop a better understanding of the statistical properties of Problem~\eqref{grp-l0-1-const1} both theoretically and empirically.}

\smallskip

\noindent \textbf{Computation.} We propose new algorithms based on combinatorial optimization for solving 
Problem~\eqref{grp-l0-1-const1} and its variants. First we present approximate algorithms: they deliver high-quality solutions using a combination of cyclic coordinate descent and local combinatorial optimization~\citep{fastsubset}. These algorithms have runtimes comparable to popular approaches for grouped variable selection (for example, Group Lasso or MCP), but deliver solutions with considerably improved statistical performance (for example, in terms of prediction and variable selection), as we demonstrate in our experiments. Our approximate algorithms deliver good-quality feasible solutions to~\eqref{grp-l0-1-const1} but are unable to certify (global) optimality of solutions via matching lower bounds on the optimal objective value of~\eqref{grp-l0-1-const1}. Certifying optimality is not only important from a methodological perspective but can also be beneficial in practice for mission-critical applications. For example, having certifiably optimal solutions can engender trust and provide transparency in consequential applications such as healthcare. Thus, we propose a new tailored branch-and-bound based optimization framework for solving~\eqref{grp-l0-1-const1} to certifiable optimality.

In our exact (global optimization) framework, we formulate the Group $\ell_0$ problem as a Mixed Integer Program (MIP). However, in a departure from earlier work~\cite{bertsimas2015best,bertsimas2020sparse}, we propose a custom branch-and-bound (BnB) algorithm to solve the MIP. Indeed, MIP-based techniques have gained considerable traction recently to solve to (near) optimality the best subset selection problem, where all groups are of size one~\citep{bertsimas2015best,bertsimas2020sparse,PVR.L0D,mazumder2017subset,fastsubset,xie2020scalable,hazimeh2020sparse}. All these works, with the exception of~\cite{hazimeh2020sparse},  leverage capabilities of powerful commercial MIP solvers such as Gurobi and CPLEX. These solvers have gained wide adoption in the past two decades due to major advances in algorithms and software development~\citep{bixby,junger200950}. However, these general-purpose solvers may take several hours to certify optimality on small instances (for example, with $p = 1000$). In contrast, our custom BnB algorithm exploits problem-specific structure to scale to much larger instances. For example, it can solve to optimality instances with $p = 5 \times 10^6$ -- this is 1000 times larger than what can be handled using Gurobi's MIP-solver. Our BnB algorithm generalizes to the grouped setting the approach of~\citep{hazimeh2020sparse} developed for the best subset selection problem.

\smallskip

\noindent \textbf{Statistical properties.}
Statistical properties of Group Lasso have been extensively studied, and it has been shown, both empirically and theoretically, that it performs well in sparse high-dimensional settings \citep{chesneau1, bach1, nardi1, huang2, wei1, lounici1, obozinski1}, under certain assumptions on the data.
However, Group Lasso also has its shortcomings, similar to those of Lasso in high dimensional linear regression~\cite{bertsimas2015best,Bbuhl1,fastsubset}.
More specifically, depending on the penalty weight, the resulting model may either be very dense or, alternatively, comes with overly shrunk nonzero coefficients. This problem is aggravated when the groups are correlated with each other, as  Group Lasso tends to bring in all of the correlated groups in lieu of searching for a more parsimonious model.
For further discussions of these issues in the special case of Lasso see, for example, \cite{zhang2,mhf-09-jasa,Bbuhl1,bertsimas2015best}, and the references therein.
In this paper, we demonstrate, both empirically and theoretically, that the \grp~methodology has advantages over its Group Lasso counterpart in a variety of regimes. In particular, as a consequence of directly controlling the sparsity level in the optimization problem, our framework leads to substantially sparser models under similar data fidelity.  Moreover, in many scenarios where the predictors are highly correlated, our approach performs better in terms of both estimation and prediction.

\smallskip

\noindent \textbf{Additive models with $\ell_0$-sparsity.} 
In addition to linear models, we also study an important example of regression with group structure that arises in high-dimensional sparse additive modeling~\cite{hastie2015statistical,HT90}.
Here, we estimate a nonparametric multivariate regression function in $q$ covariates, $(x_{1}, \ldots, x_{q})$, which we model as a sparse additive sum of the form $\sum_{j \in S } f_{j} (x_{j})$, where $S \subset \{ 1, \ldots, q \}$.
In this setting, each group generally corresponds to the basis representation of a given additive component, one for each of the~$q$ predictors.
Because the groups are allowed to be large, additional regularization needs to be imposed, typically in the form of a roughness type penalty on the regression functions.
A number of successful Group Lasso-based approaches have been proposed and analyzed in this setting -- see, for example,
\cite{meier1, ravikumar1, huang1, koltchinskii2010sparsity,PVR.vanish,yuan2016minimax} and the references therein. To our knowledge, this is the first paper to explore statistical and computational aspects of \grp-based formulations in the context of sparse additive modeling. 
We show theoretically and empirically that \grp~based methods enjoy certain statistical advantages when compared to the Group Lasso-based counterparts.

\smallskip

\noindent {\textbf{Contributions.}} The focus of this paper is on Problem \eqref{grp-l0-1-const1} and the sparse additive modeling problem (which can be formulated as a variant of Problem \eqref{grp-l0-1-const1}, as we discuss in Section \ref{sec:methodology}). Our main contributions for these two problems can be summarized as follows:
\begin{itemize}
    \item We develop fast approximate algorithms, based on first-order and local combinatorial optimization. We establish convergence guarantees for these algorithms and provide useful characterizations of the corresponding local minima. Our experiments indicate that these algorithms 
    can have an edge in terms of statistical performance over popular alternatives for grouped variable selection.
    \item We present mixed integer second order cone program (MISOCP) formulations for the~\grp-based estimators; and design a novel specialized, nonlinear branch-and-bound (BnB) framework for solving the MISOCP to global optimality. Our custom BnB solver can handle instances with $5 \times 10^6$ features and $10^3$ observations -- more than a  1000  times  larger  than what can be handled by state-of-the-art commercial MISOCP solvers. 
    \item We establish non-asymptotic prediction and estimation error bounds for our proposed estimators, for both the high-dimensional linear regression and sparse additive modeling problems. We show that under the assumption of sparsity, these error bounds compare favorably with the ones for Group Lasso.
    \item We demonstrate empirically that our approach appears to outperform the state of the art (for example, Group Lasso and available algorithms for nonconvex penalized estimators) in a variety of high-dimensional regimes and under different statistical metrics (for example, prediction, estimation, and variable selection). We provide open-source implementations of both our approximate and exact optimization algorithms\footnote{Our open-source code is available on github at \url{https://github.com/hazimehh/L0Group}.}.
\end{itemize}

\smallskip

\noindent {\textbf{Organization.}} In Section~\ref{sec:methodology}, we present formulations for the \grp~and sparse additive modeling problems. Section \ref{section:appx_algorithms} presents approximate algorithms based on first-order and local combinatorial optimization  algorithms. Then, in Section \ref{section:MIP}, we present our exact MIP algorithm. Statistical properties of our approach are investigated in Section~\ref{sec:theory}.  Section~\ref{sec:simulations} presents  computational experiments. Technical proofs and additional computational details are provided in the supplement.

\smallskip

\noindent {\textbf{Notation.}} For any non-negative integer $k$, we denote the set $\{1,...,k\}$ by $[k]$. The complement of a set $A$ is denoted by $A^c$. We denote the index sets corresponding to the~$q$ groups of predictors by
${\mathcal G}_{g}$, for $g \in [q]$ so that
$\cup_{g =1}^{q} {\mathcal G}_{g} = [p]$ and ${\mathcal G}_{g} \cap {\mathcal G}_{\ell} = \emptyset$ for all $g \neq \ell$. For a vector $\B\theta$, we use the notation Supp($\B\theta$) to denote the group support, i.e., Supp$(\B\theta) = \{ g \ | \ \B\theta_g \neq 0, g \in [q] \}$. We also define a measure of $\ell_0$-group sparsity (i.e., number of nonzero groups): $G(\B\theta) := \sum_{g=1}^{q} \M{1} ( \B\theta_{g}  \neq \M{0})$. We denote the gradient of a scalar-valued function, say $J(\B\theta)$, by $\nabla J(\B\theta)$. Moreover, we use the notation $\nabla_{\B\theta_g} J(\B\theta)$ to refer to the subvector of $\nabla J(\B\theta)$ corresponding to the variables in $\B\theta_g$. Vectors and matrices are denoted in boldface.

\section{Optimization problems considered}
\label{sec:methodology}


In this section, we present optimization formulations for the \grp~approach (and its variants), as well as the $\ell_0$-sparse additive function estimation approach.

\subsection{\grp~with ridge regularization}\label{sec:grp-l0-reg}
The algorithms discussed in this paper apply to the \grp~estimator~\eqref{grp-l0-1-const1} with an optional ridge regularization term:
\begin{equation}\label{eq:group_l0l2}
\min_{\B\beta}~~~~\| \M{y} - \M{X} \B\beta \|_{2}^2 + \lambda_0  \sum_{g=1}^{q} \M{1} ( \B\beta_{g}  \neq \M{0}) + \lambda_2 \| \B\beta \|_2^2,
\end{equation}
where $\lambda_0>0$ controls the number of selected groups, and $\lambda_2 \geq 0$ controls the strength of the ridge regularization. 
Our proposed algorithms apply to {\emph{both}} settings: $\lambda_2 = 0$ and $\lambda_2 > 0$ in Problem \eqref{eq:group_l0l2}. 
The choice of the ridge term in~\eqref{eq:group_l0l2} is motivated by earlier work in the context of best-subset selection~\cite{mazumder2017subset,fastsubset}, which suggest  that when the signal-to-noise ratio (SNR) is low, additional ridge regularization can improve the prediction performance of best-subset selection (both theoretically and empirically).   
Additionally, as discussed in Section~\ref{sec:BnB}, the choice $\lambda_2 > 0$, allows for deriving stronger MIP formulations by appealing to perspective formulations~\cite{frangioni2006perspective,gunluk2010perspective}.

\subsection{Nonparametric additive models with $\ell_0$-sparsity}\label{sec:spam-1}
In the multivariate setting, estimating the conditional mean function ${\mathbb E}( y | \M{x}) = f(x_{1}, \ldots, x_{q})$
becomes notoriously difficult, due
to curse of dimensionality. To overcome this problem, additive approximation schemes~\citep{HT90} are commonly used as an effective methodology: $f(\M{x})= \sum_{j=1}^{q} f_{j}(x_{j})$.
A popular approach \citep[see, for example,][]{Wa90} is to choose $f_{j}$ from some smooth functional class ${\mathcal C}_{j}$, such as the class of twice continuously differentiable functions. Given the observations $(y_i, \M{x}_i)$, $i \in [n]$, the additive model $f(\M{x})$ can be estimated by solving the following optimization problem:
\begin{equation}\label{add-pen-1}
\min_{f}~~  \sum_{i=1}^{n} ( y_{i} - \sum_{j=1}^{q} f_{j}(x_{ij}) )^2 + \lambda \sum_{j=1}^{q} \text{Pen}( f_{j} ),
\end{equation}
where $\text{Pen}(f_j)$ is a roughness penalty that controls the amount of smoothness in function~$f_j$.

A key ingredient in the additive function fitting framework is the estimation of a univariate smooth regression function based on
observations $(y_{i}, u_{i}), i \in [n]$. Suppose, for simplicity, that the $u_{i}$s are distinct and $u_{i} \in [0,1]$ for all $i$. For illustration, let us take $\text{Pen}(g) = \int_{0}^{1} (g''(u))^2 du$.  Then, the solution to the corresponding (infinite dimensional) univariate problem is of the form:
$ g(u) = \alpha_{0} + \alpha_{1} u + \sum_{j=1}^{n} \gamma_{j} N_{j}(u),$
where $N_{j}(u)$ are some cubic spline basis functions, such as truncated power series functions, natural cubic splines or the B-spline basis functions,
with knots chosen at the distinct data points $u_{i}, i \in [n]$.
Note that $\int_{0}^{1} (g''(u))^2 du = \B\gamma' \M{\Omega} \B\gamma$, where $\M{\Omega}$ is an $n \times n$ positive definite matrix with the elements  $\omega_{ij}  = \int_{0}^{1} N''_{i} (u) N''_{j}(u) du$.  If we refer to the corresponding functional class as~$\mathcal{C}$, define the elements of~$\M{g}$ as $g_{i} := g(u_i)$, for $i = 1, \ldots, n$,
and let $ \| \M{g} \|_{\mathcal{C}}^2 := \B\gamma' \B{\Omega} \B\gamma$, then the univariate optimization problem is equivalent to
\begin{equation}\label{one-dim-g1}
 (\hat{g}_{1}, \ldots, \hat{g}_{m}) = \hat{\M{g}} \in \argmin ~ \| \M{y} - \M{g} \|_{2}^2 + \lambda \| \M{g} \|_{\mathcal{C}}^2.
 \end{equation}
 Problem~\eqref{one-dim-g1} is a generalized least squares problem in $(\alpha_0, \alpha_1, \B\gamma)$.
A direct extension to the additive model setting is given by the following formulation:
\begin{equation}\label{add-pen-1-1}
\min~~ \| \M{y}  - \sum_{j=1}^{q} \M{f}_{j} \|_{2}^2 + \lambda \sum_{j=1}^{q} \| \M{f}_{j} \|_{\mathcal{C}_j}^2,
\end{equation}
where we minimize over $f_j\in\mathcal{C}_j$ for all $j$, and $\M{f}_{j} = (f_{j}(x_{ij}), \ldots, f_{j}(x_{nj}))$.

We wish to impose sparsity on the additive components $f_j$, $j \in [q]$, which naturally leads to the following optimization problem:
\begin{equation}\label{add-pen-1-sp}
\min~~ \| \M{y}  - \sum_{j=1}^{q} \M{f}_{j} \|_{2}^2 + \lambda_0 \sum_{j=1}^{q} \M{1} ( \M{f}_{j} \neq 0 ) +  \lambda \sum_{j=1}^{q} \| \M{f}_{j} \|_{\mathcal{C}_j}^2. 
\end{equation}
We note that the choice $\text{Pen}(f_{j}) = \sqrt{\int (f_{j}''(u))^2 du}$ leads to the optimization problem
\begin{equation}\label{cubic-spline-add-1-2}
\min~~ \| \M{y}  - \sum_{j=1}^{q} \M{f}_{j} \|_{2}^2 +  \lambda_0 \sum_{j=1}^{q} \M{1} ( \M{f}_{j} \neq 0 ) +  \lambda \sum_{j=1}^{q} \| \M{f}_{j} \|_{\mathcal{C}_j}.  
\end{equation}
Problems~\eqref{add-pen-1-sp} and~\eqref{cubic-spline-add-1-2} are close cousins and result in similar estimators. The terms $\sum_j \| \M{f}_{j} \|_{\mathcal{C}_j}$ and $\sum_j \| \M{f}_{j} \|_{\mathcal{C}_j}^2$ encourage smoothness in each of the additive components, while the sum of indicators directly controls the number of included predictors. In Section~\ref{sec:theory}, we establish theoretical error bounds for the estimator that corresponds to Problem~\eqref{cubic-spline-add-1-2}.

\smallskip

\noindent{\bf{Connections with Group Lasso-type penalization schemes.}} 
For Grouped Lasso-type penalization schemes, the choice of the penalty becomes rather subtle.
Problem~\eqref{add-pen-1} with
$\text{Pen}(f_{j}) = \| \M{f}_{j} \|_{\mathcal{C}_j}^2$ does \emph{not} induce sparsity in $\| \M{f}_{j} \|_{\mathcal{C}_j}$'s for finite $\lambda$.  Alternatively,
the choice $\text{Pen}(f_{j}) = \| \M{f}_{j} \|_{\mathcal{C}_j}$  does result in several components $\| \M{f}_{j} \|_{\mathcal{C}_j}$ being set to zero when~$\lambda$ is large. Note, however,
that $\| \M{f}_{j} \|_{\mathcal{C}_j}=0$ does not imply $f_{j} = 0$.   This is because $\| \M{f}_{j} \|_{\mathcal{C}_j}$ is a seminorm that is not affected by the linear components of $f_j$.  To set $f_{j} =0$ one needs to include the linear components into the penalty.  To overcome these limitations, alternatives have been proposed -- here we mention some penalization schemes that are used to encourage selection and smoothness.
One possible choice \citep{meier1} is $\text{Pen}(f_j) = \sqrt{ \| \M{f}_{j} \|_{2}^2 + \lambda'  \| \M{f}_{j} \|_{\mathcal{C}_j}^2}$, where $\|\M{f}_{j} \|_{2}$ denotes the usual $\ell_{2}$ norm of the vector $\M{f}_{j}$.
The corresponding penalization term is $\lambda \sum_{j} \text{Pen}(f_j)$, and, hence, the parameters $\lambda$ and $\lambda'$ jointly control smoothness and sparsity.
The sum of $\| \M{f}_{j} \|_{2}^2$ and $\lambda'  \| \M{f}_{j} \|_{\mathcal{C}_j}^2$ leads to  double penalization, thereby potentially resulting in unwanted shrinkage that may interfere with variable selection.  Similar issues arise with the choices $\text{Pen}(f_{j}) = \| \M{f}_{j} \|_{2} + \lambda'  \| \M{f}_{j} \|_{\mathcal{C}_j}$, considered in \cite{Bbuhl1}, and $\text{Pen}(f_j) = \sqrt{ \| \M{f}_{j} \|_{2}^2 + \lambda'  \| \M{f}_{j} \|_{\mathcal{C}_j}^2} + \tilde{\lambda} \| \M{f}_{j} \|_{\mathcal{C}_j}^2$, which appears in \cite{meier1}.

Thus, the choice of $\text{Pen}(f_{j})$ plays an important role in obtaining sparsity
for Lasso-type regularization methods.  In contrast, the levels of smoothness and sparsity are controlled separately
in the $\ell_{0}$-formulations: Problems~\eqref{add-pen-1-sp} and~\eqref{cubic-spline-add-1-2}. 
Group Lasso-type penalization schemes may be interpreted as convex relaxations of the $\ell_0$-penalty appearing in Problem~\eqref{cubic-spline-add-1-2}, as discussed in the Supplement~\ref{relax-1}.

\smallskip

\noindent{\bf{Other choices of smooth function classes.}} We note that the above framework, where each additive component is taken to be a cubic spline, can be generalized to more flexible smooth nonparametric models, depending upon the choice of $\text{Pen}(\cdot)$ and the functional classes ${\mathcal C}_{j}$s.
For example, one may consider the class of functions that are~$\tau$ times continuously differentiable, together with the choice
$\text{Pen}(f_{j}) = \int f^{(\tau)}_{j}(u)d u$, where $f^{(\tau)}$ denotes the $\tau$th derivative of $f_{j}$  -- solutions to these problems are given by splines of order $\tau$~\citep{Wa90}.

Another popular paradigm pursued in several works \cite{koltchinskii2010sparsity,lin2,raskutti2012minimax} is the Reproducing Kernel Hilbert Space  (RKHS) framework, wherein every ${\mathcal C}_{j}$ is taken to be a Hilbert space encouraging some
form of smoothness on~$f_{j}$. Here, $\text{Pen}(f_{j}) = \| \M{f}_{j} \|_{{K}_{j}}$ is an appropriate Hilbert space norm.   

\subsection{General problem formulation considered in this paper}
Our focus in this paper is on Problem \eqref{eq:group_l0l2} and the sparse additive modeling problems defined in \eqref{add-pen-1-sp} and \eqref{cubic-spline-add-1-2}. These three problems can all be formulated as follows:
\begin{equation}\label{conv-1-smooth}
\min_{\B\beta} ~~~ \B\beta' \M{P} \B\beta + \langle \M{a}, \B\beta \rangle  + \lambda_0 G(\B\beta) + \lambda_1 \sum_{g=1}^{q} \| \M{P}_{g} \B\beta_{g}\|_{2},
\end{equation}
for suitable choices of $\B{a}$, $\M{P} \succeq \M{0}$, $\M{P}_{g} \succ \M{0}$, $g \in [q]$, where we recall that  $G(\B\beta) := \sum_{g=1}^{q} \M{1} ( \B\beta_{g}  \neq \M{0})$. The term $\sum_{g=1}^{q} \| \M{P}_{g} \B\beta_{g}\|_{2}$ is only used for the sparse additive modeling problem in \eqref{cubic-spline-add-1-2}. Problems~\eqref{grp-l0-1-const1} and~\eqref{add-pen-1-sp} can be obtained by setting $\lambda_1 = 0$ and choosing $\M{P}$ and $\B a$ appropriately.

To simplify the presentation, we apply a change of variable in Problem \eqref{conv-1-smooth}:  $\B\theta_{g} = \M{P}^{\frac12}_{g} \B\beta_{g}$ for $g \in [q]$. This leads to the following equivalent problem:
\begin{equation}\label{conv-1-smooth1}
\min_{\B\theta} ~~~  h(\B\theta) := \underbrace{\B\theta' \M{W} \B\theta + \langle \M{b}, \B\theta \rangle}_{:=\ell(\B\theta)}~ +~\underbrace{\lambda_0 G(\B\theta) + \lambda_1 \sum_{g=1}^{q} \| \B\theta_{g}\|_{2}}_{:=\Omega(\B\theta)},
\end{equation}
for appropriately defined\footnote{Let $\M{D}_1=\text{diag}(\M{P}_{1}^{\frac12},\ldots, \M{P}_{q}^{\frac12})$ be a block diagonal matrix. Then
$\M{W} = \M{D}^{-1}_{1} \M{P} \M{D}^{-1}_{1}$, and $\M{b} =  \M{D}^{-1}_{1} \M{a}$.} $\M{W}$ and $\M{b}$. \textcolor{black}{For notational convenience, we define:
$$\ell(\B\theta):=\B\theta' \M{W} \B\theta + \langle \M{b}, \B\theta \rangle~~~\text{and}~~~ \Omega(\B\theta):=\lambda_0 G(\B\theta) + \lambda_1 \sum_{g=1}^{q} \| \B\theta_{g}\|_{2}.$$}
Our algorithmic development will focus on Problem~\eqref{conv-1-smooth1}. 


\smallskip

\noindent \textbf{Overview of our algorithms: } Problem \eqref{conv-1-smooth1} is nonconvex due to the discontinuity in $G(\B\theta)$. In Section \ref{section:appx_algorithms}, we design fast algorithms that can obtain high-quality approximate solutions for this problem. In Section \ref{section:MIP}, we develop an exact algorithmic framework, based on a custom MIP solver, which obtains certifiably optimal solutions to~\eqref{conv-1-smooth1}. 
Our algorithm constructs: (i) a sequence of feasible solutions, whose objective values are valid upper bounds, and (ii) a sequence of lower bounds (a.k.a. dual bounds). As our BnB  algorithm progresses, these upper and lower bounds converge towards the optimal objective of Problem \eqref{conv-1-smooth1}. The solver terminates and certifies optimality when the upper and lower bounds match\footnote{In practice, MIP solvers terminate when the difference between the upper and lower bounds are below a small, user-defined threshold.}. 
Our experiments indicate that high-quality initial solutions, as available from the algorithms presented in Section~\ref{section:appx_algorithms}, can significantly speed up convergence and reduce memory requirements in our BnB algorithm. 

\section{Approximate Algorithms} \label{section:appx_algorithms}
In this section, we develop fast approximate algorithms to obtain high quality local minimizers for Problem \eqref{conv-1-smooth1}. While these algorithms do not deliver certificates of optimality (via dual bounds), they attain nearly-optimal (and at times optimal) solutions to many statistically challenging instances,  in running times comparable to group Lasso-based algorithms.

A main workhorse of our approximate algorithms is a nonstandard application of 
cyclic block coordinate descent (BCD) to the  discontinuous objective function~\eqref{conv-1-smooth1}. We draw inspiration from the appealing scalability properties of coordinate descent in sparse learning problems \citep[see, for example,][]{glmnet,BeckSparsityConstrained,fastsubset}. Our second algorithm is based on local combinatorial search and is used to improve the quality of solutions obtained by BCD. We establish convergence guarantees for these two algorithms. 

Our algorithms arise from studying necessary optimality conditions for Problem~\eqref{conv-1-smooth1}.
To this end, we show that the quality of solutions obtained by BCD are of higher quality than local solutions corresponding to the popular proximal gradient descent (PGD)~\cite{parikh2014proximal} algorithm\footnote{Though PGD is popularly used in the context of convex optimization problems, it also leads to useful algorithms for nonconvex sparse learning problems. In particular, 
PGD for our problem can be viewed as a generalization of the iterative hard thresholding (IHT) algorithm \cite{blumensath2008iterative} to the group setting.}.
The local minimizers corresponding to local combinatorial search form a smaller subset of those available from BCD.
In this section, we establish the following hierarchy among the classes of local minima:
\begin{equation}\label{eq:hierarchy}
\text{Global Minima} ~\subseteq~ \text{Local Search Minima}  ~\subseteq~  \text{BCD Minima}  ~\subseteq~ \text{PGD Minima}.
\end{equation}
Above, PGD minima correspond to the fixed points of the PGD algorithm; they include all the fixed points of our proposed BCD algorithm. As we move from right to left in the above hierarchy, the classes become smaller, i.e., they impose stricter necessary optimality conditions. At the top of the hierarchy we have the global minimizers of the problem, which can be obtained using our exact MIP-based framework (we discuss this in Section \ref{section:MIP}). Our approximate algorithms are inspired by recent work~\cite{fastsubset} on the sparse regression problem, but the approach presented here has notable differences.
In particular, the coordinate descent algorithm in \cite{fastsubset} performs exact minimization per coordinate, which can be computationally expensive when extended to the group setting. Thus, our proposed BCD algorithm performs inexact minimization per group. In addition, the presence of 
$\ell_2$ norms in our objective function makes the analysis for the rate of convergence for our algorithm different.

\subsection{Block Coordinate Descent}
We present a cyclic BCD algorithm to obtain good feasible solutions to Problem~\eqref{conv-1-smooth1} 
and establish convergence guarantees.
We first introduce a useful upper bound for $\ell(\B\theta)$.
For every $g \in [q]$, we define $S_g = \{ (\B\theta, \tilde{\B\theta}) \ | \ \B\theta_i = \tilde{\B\theta}_i, \  \forall i \in [q] \text{ s.t. } i \neq g \} $. By the Block Descent Lemma \citep{bertsekas2016nonlinear}, the following upper bound holds for every $g \in [q]$:
\begin{align} \label{eq:groupdescent}
    \ell(\B\theta) \leq  \ell(\tilde{\B\theta}) + \langle \nabla_{\B\theta_g} \ell(\tilde{\B\theta}), \B\theta_g - \tilde{\B\theta}_g \rangle + \frac{{L}_g}{2} \|\B\theta_g - \tilde{\B\theta}_g \|_2^2, \qquad \forall (\B\theta, \tilde{\B\theta}) \in S_g,
\end{align}
where $L_g$ is the ``group-wise'' Lipschitz constant of $\nabla \ell({\B\theta})$, i.e., $L_g$ is a constant which satisfies: $\|\nabla_{\B\theta_g} \ell({\B\theta}) - \nabla_{\B\theta_g} \ell(\tilde{\B\theta}) \|_2 \leq L_g \| \B\theta_g - \tilde{\B\theta}_g \|_2$, for all $(\B\theta, \tilde{\B\theta}) \in S_g$. Since $\ell(\B\theta)$ is a quadratic function, 
$L_g = 2\sigma_{\max}(\M{W}_{g})$, where $\M{W}_{g}$ is the submatrix of $\M{W}$ with columns and rows restricted to group $g$, and $\sigma_{\max}(\cdot)$ denotes the largest eigenvalue.

Cyclic BCD sequentially minimizes the objective of~\eqref{conv-1-smooth1} with respect to one group of variables while the other groups are held fixed. Let $\B\theta^{l}$ be the iterate obtained by the algorithm after the $l$-th iteration. Then, in iteration $l+1$, the variables in a group $g$ (say), are updated while the other groups are held fixed. Specifically, we have $(\B\theta^{l}, \B\theta^{l+1}) \in S_g$. 
Using~\eqref{eq:groupdescent} with $\tilde{\B\theta} = \B\theta^{l}$, $\hat{L}_g > {L}_g$ and adding $\Omega(\B\theta)$ to both sides we get:
\begin{align}\label{majori-zation-expression-1}
    h(\B\theta) \leq \tilde{g}(\B\theta; \B\theta^l) :=  \ell({\B\theta}^{l}) + \langle \nabla_{\B\theta_g} \ell({\B\theta}^{l}), \B\theta_g - {\B\theta}_g^{l} \rangle + \frac{\hat{{L}}_g}{2} \|\B\theta_g - {\B\theta}^{l}_g \|_2^2 + \Omega(\B\theta).
\end{align}
Note that the left hand side of~\eqref{majori-zation-expression-1} is the objective function of Problem \eqref{conv-1-smooth1}. 
We obtain $\B\theta^{l+1}_g$ by minimizing the upper bound on our objective, $\tilde{g}(\B\theta; \B\theta^l)$, with respect to $\B\theta_{g}$:
\begin{align} \label{eq:nextiterate}
    \B\theta^{l+1}_g \in \argmin_{\B\theta_g} \tilde{g}(\B\theta; \B\theta^l) = \argmin_{\B\theta_g} \frac{\hat{{L}}_g}{2} \Bigg \|\B\theta_g - \Big({\B\theta}^{l}_g - \frac{1}{\hat{L}_g} \nabla_{\B\theta_g} \ell({\B\theta}^{l}) \Big) \Bigg\|_2^2 + \Omega(\B\theta_g).
\end{align}
{Although nonconvex, the minimization problem in \eqref{eq:nextiterate} admits a closed-form solution, which can be obtained via the operator $H: \mathbb{R}^{u} \to \mathbb{R}^{u}$ defined as follows:}
\begin{align} \label{eq:thresholdingop}
    H(\M z; \B\lambda; \hat{L}_g) = 
    \begin{cases}
    \frac{\M z}{\|\M z\|_2} \Big[ \| \M z\|_2 - \frac{\lambda_1}{\hat{L}_g} \Big] & \text{ if } \|\M z\|_2 > \sqrt{\frac{2 \lambda_0}{\hat{L}_g}} + \frac{\lambda_1}{\hat{L}_g} \\
    0 & \text{ otherwise }
    \end{cases}
\end{align}
where $\B\lambda = (\lambda_0, \lambda_1)$. It can be readily seen that an optimal solution of \eqref{eq:nextiterate} is given by $H(\M z; \B\lambda; \hat{L}_g)$, where $\M z = {\B\theta}^{l}_g - \frac{1}{\hat{L}_g} \nabla_{\B\theta_g} \ell({\B\theta}^{l})$. 
Below we summarize our proposed cyclic BCD algorithm.
\begin{itemize}
\item[] \emph{ \underline{Algorithm 1: Cyclic Block Coordinate Descent (BCD)}}
\item \textbf{Input:} Initialization $\B\theta^0$ and $\hat{L}_g$ for every $g \in [q].$
\item \textbf{Repeat} Steps 1, 2 for $l=0, 1, 2, \dots$ until convergence:
    \begin{enumerate}
        \item $g \gets 1 + (l\mod q)$ and $\B\theta^{l+1}_j \gets \B\theta^{l}_j$ for all $j \neq g$
        \item $\B\theta^{l+1}_g \gets H(\M z; \B\lambda; \hat{L}_g)$, where $\M z = {\B\theta}^{l}_g - ({1}/{\hat{L}_g}) \nabla_{\B\theta_g} \ell({\B\theta}^{l})$.
    \end{enumerate}
\end{itemize}

\noindent \textbf{Convergence Analysis.} To establish convergence of the sequence $\{\B \theta^{l}\}$ in Algorithm 1, we make use of the following assumption. 
\begin{assmp} \label{CDAssumptions}
At least one of the following conditions holds:
\begin{itemize}
    \item[(a)] Strong Convexity: $\M W \succ 0$.
    \item[(b)] Restricted Strong Convexity: Let $\hat{\B\theta}$ be a (Group Lasso) solution defined as $\hat{\B\theta} \in \argmin_{\B\theta} \ell({\B\theta}) + \lambda_1 \sum_{g=1}^{q} \| \B\theta_{g}\|_{2}$. Let $k = \max_{\B\theta} \{ \| \B\theta \|_0~|~G(\B\theta) \leq G(\hat{\B\theta})\}$. Every collection of $k$ columns in $\M W$ are linearly independent, and the initial solution $\B \theta^{0}$ (in Algorithm 1) satisfies $h(\B {\theta}^{0}) \leq h(\hat{\B\theta})$.
\end{itemize}{}
\end{assmp}
Assumption \ref{CDAssumptions}(a) holds if a ridge regularization term is used, i.e., it holds for Problem \eqref{eq:group_l0l2} with $\lambda_2 > 0$. Assumption \ref{CDAssumptions}(b) is less restrictive because we can have $\M W \succeq \M{0}$. 
Suppose that for some non-negative integer $u$, every set of $u$ columns in $\M W$ are linearly independent. Then, in the Group Lasso problem (defined in Assumption \ref{CDAssumptions}(b)), $\lambda_1$ can be chosen sufficiently large so that some Group Lasso solution $\hat{\B\theta}$ satisfies $k \leq u$. If $\hat{\B\theta}$ is used to initialize Algorithm 1, then Assumption  \ref{CDAssumptions}(b) is satisfied.


The following theorem establishes a linear convergence guarantee for the sequence generated by Algorithm~1. 
\begin{thm} \label{theorem:CDConvergence}
Let $\{ \B\theta^{l}\}$ be the sequence generated by Algorithm 1 and suppose that Assumption \ref{CDAssumptions} holds. Then,
\begin{enumerate}
    \item The group support stabilizes after a finite number of iterations, i.e., there exists an integer $K$ and a support $S \subseteq [q]$ such that $\text{Supp}(\B\theta^{l}) = S$ for all $l \geq K$.
    \item  The sequence $\{\B\theta^{l}\}$ converges to a solution $\B\theta^{*}$, with $\text{Supp}(\B\theta^{*}) = S$ (as defined in Part 1), satisfying:
    \begin{align}
        & \B\theta^{*}_S \in \argmin_{\B\theta_S
        } ~~ \ell(\B\theta_S) + \lambda_1 \sum_{g \in S} \| \B\theta_{g}\|_{2} \label{eq:betastar} \\
        & \|\B\theta^{*}_g\|_2 \geq \sqrt{\frac{2\lambda_0}{\hat{L}_g}}, \quad \forall g \in S \label{eq:insuppineq} \\
        & \| \nabla_{\B\theta_g} \ell(\B\theta^{*}) \|_2 \leq \sqrt{2 \lambda_0 \hat{L}_g} + \lambda_1, \quad \forall g \in S^c. \label{eq:outsuppineq}
    \end{align}
        \item The function $\B\theta_S \mapsto \ell(\B\theta_S)$ is strongly convex with a   strong convexity parameter $\sigma_S > 0$. Let $L_S$ be the Lipschitz constant of $\nabla_{\B\theta_S} \ell(\B\theta_S)$. Define $\hat{L}_{\text{max}} = \max_{g \in S} \hat{L}_g + 2\lambda_1$ and $\hat{L}_{\text{min}} = \min_{g \in S} \hat{L}_g + 2 \lambda_1$. Then, for $l \geq K$, the following holds:
        \begin{align}
            h(\B\theta^{(l+1)q}) - h(\B\theta^{*}) \leq \Bigg( 1 - \frac{\sigma_S}{\eta} \Bigg)  \Big( h(\B\theta^{lq}) - h(\B\theta^{*}) \Big),
        \end{align}
        where $\eta = 2 \hat{L}_{\text{max}} (1 + |S| {(L_S + 2\lambda_1 |S|)^2} \hat{L}^{-2}_{\text{min}}).$
\end{enumerate}
\end{thm}
The proof of Theorem \ref{theorem:CDConvergence} is in the supplement. We present here a high-level sketch of the proof. We establish part 1 by proving a sufficient decrease condition. For part 2, we show that the objective function restricted to the group support $S$ is strongly convex, and thus convergence follows from standard results on cyclic BCD, e.g., \cite{bertsekas2016nonlinear}. To establish the linear rate of convergence in part 3 of the theorem, we extend the result of \cite{BeckConvergence} who show that cyclic BCD can achieve a linear rate of convergence on smooth and strongly convex functions: note that our objective function after support stabilization is {\emph{not}} smooth due to the presence of the term $\sum_{g \in S} \| \B\theta_{g}\|_{2}$.

\smallskip

\noindent \textbf{Optimality conditions of BCD and PGD.} The conditions in Theorem \ref{theorem:CDConvergence} (part 2) characterize a fixed point of Algorithm 1. These are necessary optimality conditions for Problem \eqref{conv-1-smooth1} since any global minimizer must be a fixed point for Algorithm 1. In what follows, we will show that the necessary optimality conditions imposed by PGD (which is a generalization of \cite{blumensath2008iterative} to the group setting) are generally less restrictive compared to those imposed by Algorithm 1. Note that PGD is an iterative algorithm whose updates for Problem \eqref{conv-1-smooth1} are given by:
\begin{align} \label{eq:ihtpenupdate}
    \B{\theta}^{l+1} \in \argmin_{\B{\theta}}~ \left\{ \frac{1}{2\tau} \| \B{\theta} - (\B{\theta}^{l} - \tau \nabla \ell(\B{\theta}^{l}))  \|_2^2 + \Omega(\B{\theta}) \right\},
\end{align}
where $\tau>0$ is a step size. 
Let $L$ be the Lipschitz constant of $\nabla \ell(\B\theta)$. For a constant step size, the update in \eqref{eq:ihtpenupdate} converges if $\tau = {1}/{\hat{L}}$ where $\hat{L}$ is a constant chosen such that $\hat{L} > L$ \citep[see, for example,][]{fastsubset, PenalizedIHT}. For the choice $\tau = {1}/{\hat{L}}$, it can be readily checked that any fixed point of PGD satisfies the three optimality conditions in Theorem \ref{theorem:CDConvergence} (part 2), but with $\hat{L}_g$ replaced by~$\hat{L}$. The group-wise Lipschitz constant $L_g$ satisfies $L_g \leq L$ (for any $g$). In many high-dimensional problems, we can have $L_g \ll L$ \citep[see][]{BeckSparsityConstrained,fastsubset}. Thus, Algorithm 1 generally imposes more restrictive necessary optimality conditions compared to PGD, which can lead to higher quality local minima in practice. This establishes a part of the hierarchy in \eqref{eq:hierarchy}.

\subsection{Local Combinatorial Search}
In this section, we introduce a local combinatorial search algorithm to improve the quality of solutions obtained by cyclic BCD (Algorithm 1). The algorithm performs the following two steps in the $t$-th iteration:
\begin{compactenum}
    \item \textbf{Block Coordinate Descent:} We run Algorithm 1 initialized at the current solution $\B\theta^{t-1}$ to obtain a solution $\B\theta^{t}$. We denote the indices of the nonzero groups in $\B\theta^{t}$ by $\text{Supp}(\B\theta^{t})=S$.
    \item \textbf{Group Combinatorial Search:} We attempt to improve the solution $\B\theta^{t}$ by \textsl{swapping} groups of variables from inside and outside the support $S$. In particular, we search for two subsets $S_1 \subseteq S$ and $S_2 \subseteq{S}^c$ such that removing $S_1$ from the support, adding $S_2$ to the support, and then optimizing over the groups in $S_2$, improves the current objective. To ensure that the local search problem is computationally feasible, we restrict our search to subsets satisfying $|S_1| \leq m$ and $|S_2| \leq m$, where $m$ is a pre-specified integer that takes relatively small values (for example, in the range $1$ to $10$).
\end{compactenum}

We present a formal description of the optimization problem in step 2 (above). We denote the standard basis of ${\mathbb{R}}^{p}$ by $\{\M e_1, \dots, \M e_p\}$. Given a set $J \subseteq [q]$, we define the $p \times p$ matrix $\M U^J$ as follows: the $i$-th column of $\M U^J$ is $\M{e}_i$ if $i \in \cup_{g \in J} {\mathcal G}_{g}$ and $\M 0$ otherwise. In other words, for any $\B\theta \in \mathbb{R}^p$, we have $(\M U^J \B\theta)_i = \theta_i$ if $i \in \cup_{g \in J} {\mathcal G}_{g}$ and $0$ otherwise. The optimization problem in Step 2 is given by: 
\begin{equation} \label{eq:swaps}
 \min_{S_{1}, S_{2}, \B{\theta}} ~~ h(\B{\theta}^{t} - \M U^{S_1} \B{\theta}^{t} + \M U^{S_2} \B{\theta}) ~~~~~ \text{s.t. } ~~~~ S_1 \subseteq S, S_2 \subseteq S^c, |S_1| \leq m, |S_2| \leq m,
\end{equation}
where we recall that $S = \text{Supp}(\B\theta^{t})$. If there is a feasible solution $\hat{\B\theta}$ to \eqref{eq:swaps} satisfying $h(\hat{\B\theta}) < h(\B{\theta}^{t})$, then we move to the improved solution $\hat{\B\theta}$; otherwise, we terminate the algorithm. We summarize the algorithm below:
\begin{itemize}
\item[] \emph{ \underline{Algorithm 2: Local Combinatorial Search}}
\item \textbf{Input:} Initial solution $\B\theta^0$ and swap subset size $m$.
\item \textbf{Repeat} Steps 1--3 for $t=1, 2, \dots$ until convergence:
    \begin{enumerate}
        \item Run Algorithm 1 initialized from $\B\theta^{t-1}$ to obtain a solution $\B\theta^t$.
        \item Search for a feasible solution $\hat{\B\theta}$ to \eqref{eq:swaps} satisfying $h(\hat{\B\theta}) < h(\B{\theta}^{t})$.
        \item If step 2 succeeds, $\B\theta^t \gets \hat{\B\theta}$. Otherwise, terminate.
    \end{enumerate}
\end{itemize}

Theorem~\ref{theorem:localsearch} establishes that Algorithm 2 converges in a finite number of iterations and characterizes the corresponding solution.
\begin{thm} \label{theorem:localsearch}
Let $\{ \B{\theta}^{t} \}$ be the sequence of iterates generated by 
Algorithm~2 and suppose Assumption~\ref{CDAssumptions} holds. Then, $\B{\theta}^{t}$ converges in a finite number of iterations to a solution that we denote by~$\B\theta^{\dagger}$. Let $S = \text{Supp}(\B\theta^{\dagger})$. Then, $\B\theta^{\dagger}$ satisfies the necessary optimality conditions in part 2 of Theorem \ref{theorem:CDConvergence}. In addition, $\B\theta^{\dagger}$ satisfies:
\begin{equation} \label{eq:inescapable}
 h(\B\theta^{\dagger}) \leq \min_{S_{1}, S_{2}, \B{\theta}} ~~ h(\B\theta^{\dagger} - \M U^{S_1} \B\theta^{\dagger} + \M U^{S_2} \B{\theta}) ~~~~~ \text{s.t. } ~~~~ S_1 \subseteq S, S_2 \subseteq S^c, |S_1| \leq m, |S_2| \leq m.
\end{equation}
\end{thm}

Theorem \ref{theorem:localsearch} shows that the solutions obtained by Algorithm 2 impose more restrictive necessary optimality conditions (in particular, condition \eqref{eq:inescapable}) compared to Algorithm 1, which justifies part of the hierarchy in \eqref{eq:hierarchy}. This is expected, as every iteration of Algorithm 2 improves over a solution obtained by Algorithm 1.  The quality of solutions returned by Algorithm 2 depends on the swap subset size $m$. For a sufficiently large choice of $m$, the algorithm will return a global minimizer. Intuitively, the computational cost of the local search in step 2 of Algorithm 2 increases with $m$. In our experiments, we observe that small choices such as $m=1$ can lead to significant improvements in solution quality compared to algorithms that do not incorporate combinatorial optimization. These improvements are most pronounced in settings where $n \ll p$ or the predictors across groups are highly correlated. In Section \ref{sec:swap_search_MIP}, we present a MIP formulation for the local search problem in Algorithm 2 for $ m > 1$. For the special case of $m=1$, we use our own custom implementation that is more efficient than using a MIP-based approach. 

\subsection{Algorithms for the cardinality constrained formulation} \label{sec:cardinality}
Algorithms 1 and 2 provide solutions for the (penalized) formulation in \eqref{conv-1-smooth1}.  While this leads to a family of high-quality estimators across a range of model sizes, it does not allow for explicit control over the number of nonzero groups $G(\B\theta)$. To this end, 
we consider the cardinality constrained variant of problem~\eqref{conv-1-smooth1}:
\begin{align} \label{eq:cons}
    \min_{\B \theta}~~ E(\B \theta):= \ell(\B\theta) + \lambda_1 \sum_{g \in [q]} \| \B\theta_{g}\|_{2} ~~~ \text{s.t. }~~~ G(\B\theta) \leq k.
\end{align}

In order to obtain a solution to~\eqref{eq:cons} with a desired support size, we propose the following procedure. First, we run Algorithm 2 (say) over a grid of $\lambda_0$-values to obtain a sequence of solutions. Then, if a desired support size, say $k$, is missing, we obtain it by applying proximal gradient descent (PGD) to Problem \eqref{eq:cons}:
\begin{align} \label{eq:ihtconsupdate}
    \B{\theta}^{l+1} \in \argmin_{\B\theta:~ G(\B\theta) \leq k}~ \left\{ \frac{1}{2\tau} \| \B{\theta} - (\B{\theta}^{l} - \tau \nabla \ell(\B{\theta}^{l}))  \|_2^2 + \lambda_1 \sum_{g \in [q]} \| \B\theta_{g}\|_{2} \right\},
\end{align}
where $\tau > 0$ is a step size and the initial solution $\B\theta^{0}$ can be obtained from Algorithm 2 (for example, we take a solution with group support size closest to $k$).

The next proposition establishes the convergence of update \eqref{eq:ihtconsupdate} and describes its fixed points.

\begin{prop} \label{prop:ihtconvergence}
Let $\{ \B{\theta}^{l} \}$ be the sequence of iterates generated the PGD updates~\eqref{eq:ihtconsupdate}. Let $L$ be the Lipschitz constant of $\nabla \ell(\B{\theta})$ and a scalar $\hat{L}$ such that $\hat{L}> L$. Then, $\{ \B{\theta}^{l} \}$ converges for a step size $\tau ={1}/{\hat{L}}$. Moreover, a solution $\B{\theta}^{*}$ with group support $S$ is a fixed point of~\eqref{eq:ihtconsupdate} iff $G(\B{\theta}^{*}) \leq k$, and
\begin{align*}
\B{\theta}^{*}_S \in \argmin_{\B{\theta}_S}~E(\B{\theta}_S)  \quad \text{ and }
\quad \|\nabla_{\B\theta_g} \ell(\B{\theta}^{*})\|_2 \leq  \gamma_{(k)} \quad \text{ for } g \in S^c,
\end{align*}
where $\gamma_{g} = \|\hat{L} \B\theta^{*}_{g} - {\nabla_{\B\theta_g} \ell(\B{\theta}^{*})}\|_2 $, and $\gamma_{(k)}$ denotes the $k$th largest value in the sequence $\{ \gamma_g \}_{g=1}^{q}$.
\end{prop}
We omit the proof of Proposition~\ref{prop:ihtconvergence} as it can be established by a simple extension to the standard results on the convergence of IHT \citep[for example, those in][]{blumensath2008iterative, BeckSparsityConstrained}.

\section{Mixed Integer Programming} \label{section:MIP}
In this section, we propose MIP formulations and algorithms to solve~\eqref{conv-1-smooth1} and the combinatorial search problem in Algorithm 2. Section \ref{sec:formulations} introduces MIP formulations, and Section~\ref{sec:BnB} presents a new BnB algorithm for solving the corresponding problems to optimality.

\subsection{MIP Formulations} \label{sec:formulations}

\subsubsection{Formulations for Problem \eqref{conv-1-smooth1}} Below we present two MIP-formulations for~\eqref{conv-1-smooth1}. 

\noindent \textbf{Big-M Formulation: } We first present a Big-M based MIP formulation for Problem \eqref{conv-1-smooth1}:
\begin{subequations} \label{eq:MIP_bigM}
\begin{align} 
    \min_{\B\theta, \M{z}} ~~~  & \ell(\B\theta) + \lambda_0 \sum_{g=1}^q z_g + \lambda_1 \sum_{g=1}^{q} \| \B\theta_{g}\|_{2}  \\
    \text{s.t.}~~& \|\B\theta_{g}\|_2 \leq \MU z_{g}, ~~ g \in [q] \label{eq:MIP_bigM_constraint} \\
    & z_g \in \left\{0,1\right\}, ~~ g \in [q] \label{eq:MIP_bigM_feasible}
\end{align}
\end{subequations}
where, the optimization variables are $\B\theta$ (continuous) and $\M{z}$ (binary). 
Above, $\MU$ is an a-priori specified constant (leading to the name ``Big-M'') such that some optimal solution, say $\B{\theta}^{*}$, to \eqref{conv-1-smooth1} satisfies $\max_{g \in [q]} \| \B{\theta}^{*}_g \|_{2} \leq \MU$. 
In \eqref{eq:MIP_bigM}, the binary variable $z_{g}$ controls whether all the regression coefficients in group $g$ are zero or not: $z_{g} = 0$ implies that $\B\theta_{g} = \M{0}$, and $z_{g}=1$ implies that $\| \B{\theta}_g \|_{2} \leq \MU$. Such Big-M formulations are commonly used in mixed integer programming to model relations between discrete and continuous variables, and have been recently used in $\ell_0$-regularized regression~\cite{bertsimas2015best,xie2020scalable} (for example).
Various techniques have been proposed to estimate the constant $\MU$ in practice; see \cite{bertsimas2015best} for a discussion on estimating the Big-M in the context of linear regression. 
The constraints in \eqref{eq:MIP_bigM_constraint} are second order cones~\cite{BV2004}. Moreover, the objective function in \eqref{eq:MIP_bigM} can be written as a linear function, with additional second order cone constraints to express the quadratic function   $\ell({\B\theta})$ and the terms $\| \B\theta_g \|_2, ~ g \in [q]$. Thus, Problem \eqref{eq:MIP_bigM} can be reformulated as a Mixed Integer Second Order Cone Program (MISOCP), which can be modeled and solved (for small/moderate problem instances) with commercial MIP solvers such as  Gurobi, CPLEX, and MOSEK. We present an efficient, standalone BnB algorithm for~\eqref{eq:MIP_bigM} in Section~\ref{sec:BnB}.  

\smallskip

\noindent \textbf{Perspective reformulation: } Recall that 
Problem~\eqref{conv-1-smooth1} contains a ridge term in its objective. The ridge term can be used to derive stronger MIP formulations for~\eqref{conv-1-smooth1} based on the perspective formulation~\cite{frangioni2006perspective,gunluk2010perspective}. As we discuss below, the perspective-based formulation differs from the Big-M formulation~\eqref{eq:MIP_bigM}---when $\lambda_{2}>0$, it usually leads to tighter convex relaxations and consequently, reduced MIP runtimes.
First, we rewrite~\eqref{conv-1-smooth1} as 
\begin{align} \label{eq:MIP_BnB}
    \min_{\B\theta, \M{z}} ~~~  & \tilde{\ell}(\B\theta) + \lambda_0 \sum_{g=1}^q z_g + \lambda_1 \sum_{g=1}^{q} \| \B\theta_{g}\|_{2} + \lambda_2 \sum_{g =1}^{q} \| \B\theta_{g} \|_2^2 ~~~~~  \text{s.t.} ~~~~ \eqref{eq:MIP_bigM_constraint}, \eqref{eq:MIP_bigM_feasible}
\end{align}
where $\ell(\B\theta) = \tilde{\ell}(\B\theta) + \lambda_{2} \| \B\theta\|_{2}^2$.
Using the perspective 
reformulation~\citep{frangioni2006perspective,gunluk2010perspective,hongbo} for the ridge term
$\sum_{g \in [q]} \| \B\theta_{g} \|_2^2$
in the objective, we can reformulate~\eqref{eq:MIP_BnB} as
\begin{subequations}  \label{eq:MIP_hybrid}
\begin{align}
    \min_{\B\theta, \M{z}, \B{s}} ~~~  & \tilde{\ell}(\B\theta) + \lambda_0 \sum_{g=1}^q z_g + \lambda_1 \sum_{g=1}^{q} \| \B\theta_{g}\|_{2} + \lambda_2 \sum_{g=1}^q s_g, \\
    \text{s.t.}~~& \|\B\theta_{g}\|_2 \leq \MU z_{g},~ g \in [q] \label{eq:bigm_constraint}\\
    &  \|\B\theta_{g}\|_2^2 \leq s_g z_g, ~~ g \in [q] \label{eq:cone_constraint}\\ 
    & z_g \in \left\{0,1\right\}, s_g \geq 0, ~~ g \in [q]. 
\end{align}
\end{subequations}
Compared to \eqref{eq:MIP_BnB}, 
formulation~\eqref{eq:MIP_hybrid} uses additional auxiliary variables $s_g \in \mathbb{R}_{\geq 0}, ~ g \in [q]$ and rotated second order cone constraints: $\|\B\theta_{g}\|_2^2 \leq s_g z_g$ for $g \in [q]$. Each $s_g$ takes the place of the term $\| \B\theta_g\|_2^2$ in the objective function in~\eqref{eq:MIP_bigM}. Specifically, any optimal solution $(\B\theta^{*}, \M{z}^{*}, \B{s}^{*})$ to \eqref{eq:MIP_hybrid} must satisfy $s_g^{*} = \|\B\theta_{g}^{*}\|_2^2$. 

Although the MIP formulations \eqref{eq:MIP_hybrid} and \eqref{eq:MIP_BnB} are equivalent, their continuous relaxations are generally different. The following proposition states that the relaxation of \eqref{eq:MIP_hybrid} is generally tighter (i.e., has a higher objective) than the relaxation of \eqref{eq:MIP_BnB}.

\begin{prop} \label{prop:relaxation_quality}
Let $v_1$ and $v_2$ be the objective values of \eqref{eq:MIP_BnB} and \eqref{eq:MIP_hybrid} upon relaxing the binary variable $z_g$ to $[0,1]$ for all $g \in [q]$. Let $(\B\theta^{*}, \M{z}^{*}, \M{s}^{*})$ be an optimal solution to the relaxation corresponding to $v_2$. Then, the following holds:
$$
v_2 - v_1 \geq \lambda_2 \sum_{g \in [q] | z_g^{*} > 0} \| \B\theta^{*}_g \|_2^2 \Big ((z_g^{*})^{-1} - 1 \Big).
$$
\end{prop}
Proposition \ref{prop:relaxation_quality} implies that using formulation \eqref{eq:MIP_hybrid}  (over formulation~\eqref{eq:MIP_bigM}) can lead to tighter lower bounds for the root node relaxation; and hence tighter dual bounds for the node relaxations in the BnB tree. This can result in improved runtimes in the overall BnB solver (as we demonstrate in our experiments). 
Thus, in our algorithmic framework in Section \ref{sec:BnB}, we focus on formulation \eqref{eq:MIP_hybrid}. To be clear, our BnB procedure applies even without the presence of a ridge term (i.e., $\lambda_{2}=0$). Specifically, if $\lambda_2 = 0$ in \eqref{eq:MIP_hybrid}, the conic constraints~\eqref{eq:cone_constraint} can be removed and  formulation~\eqref{eq:MIP_hybrid} reduces to the Big-M formulation in \eqref{eq:MIP_bigM}.

\subsubsection{MIP formulation for local combinatorial search} \label{sec:swap_search_MIP}
We present a MIP formulation for the local search problem\footnote{We recommend the use of the MIP formulations when $m\geq 2$. When $m=1$ a solution to the local search procedure can be computed efficiently from first principles.} that arises in Algorithm~2. Problem~\eqref{eq:swaps} can be formulated using the following Big-M based MIP:
\begin{subequations} \label{eq:localsearcMIP}
\begin{align}
\min\limits_{  \M{u}, \M{z}, \B{\theta}} \quad  & \ell(\M{u}) + \lambda_0 \sum\limits_{g=1}^{q} z_g + \lambda_1 \sum_{g=1}^{q} \| \M{u}_g \|_2 \nonumber \\
\text{s.t.}~~ & \M{u} = \B{\theta}^{t} - \sum_{g \in S} \M U^g \B\theta^{t} (1 - z_g) + \sum_{g \in S^c} \M U^g \B\theta \label{eq:dummyvar}\\
& \| \M{u}_g \|_2 \leq \MU z_g, ~ g \in S^c \label{eq:swaps-big-M}\\
& \sum_{g \in S} z_g \geq |S| - m,~ \sum_{g \in S^c} z_g \leq m \label{eq:cut12} \\
& z_g \in \left\{0,1\right\}, ~~ g \in [q]. 
\end{align}
\end{subequations}
In the formulation above, we assume that $\MU$ is chosen sufficiently large so that some optimal solution to \eqref{eq:swaps}, say $\B \theta^{*}$, satisfies $\| \B\theta^{*}_g \|_2 \leq \MU, ~ g \in S^c$. As we discuss below, the objective in~\eqref{eq:localsearcMIP} represents $h(\M{u})$ with $\M{u} = \B{\theta}^{t} - \M U^{S_1} \B{\theta}^{t} + \M U^{S_2} \B{\theta}$, where $h(\M{u})$, $S_{1}$ and $S_{2}$ are as defined in~\eqref{eq:swaps}.
Note that the variable $\M{u}$ is an auxiliary variable introduced to simplify the presentation. The binary variables $z_g, g \in [q]$ are used to select the subsets $S_1 \subseteq S$ and $S_2 \subseteq S^c$. In particular, for $g \in S$, $z_g = 0$ iff $g \in S_1$, and this is encoded by constraint \eqref{eq:dummyvar}. On the other hand, for $g \in S^c$, $z_g = 1$ iff $g \in S_2$, and this is encoded by constraints \eqref{eq:dummyvar} and \eqref{eq:swaps-big-M}. Therefore, $\sum_{g=1}^{q} z_g$ is equal to $G(\M{u})$. The constraints \eqref{eq:cut12} enforce $|S_1| \leq m$ and $|S_2| \leq m$. 

The local search MIP-formulation~\eqref{eq:localsearcMIP} has a {\emph{smaller}} search space compared 
to the full problem~\eqref{eq:MIP_bigM}. This is due to the additional constraints appearing in~\eqref{eq:cut12}. 
Furthermore, Problem~\eqref{eq:localsearcMIP} effectively uses 
$|S^c|$-many `free' continuous group-variables---this is in contrast to $|S| + |S^c|$ continuous group-variables appearing in the full problem. Thus, for small values of $m$, Problem \eqref{eq:localsearcMIP} can be typically solved faster than the MIP formulation of~\eqref{conv-1-smooth}. While~\eqref{eq:localsearcMIP} is based on a Big-M formulation,  
in the presence of an additional ridge regularizer, one can also derive a perspective reformulation using ideas similar to~\eqref{eq:MIP_hybrid}. 

\subsection{Exact optimization via a custom nonlinear Branch-and-Bound algorithm} \label{sec:BnB}

High-performance commercial MIP solvers, such as Gurobi and CPLEX, often deliver state-of-the-art performance for a variety of MIP problems.
These solvers are based on a BnB framework, which can solve MIP problems to global optimality, typically without having to explicitly enumerate all (exponentially many) solutions in the search space. These solvers are general-purpose and do not take into account the specific structure of the problems we consider here. Therefore, their performance can suffer:
we have empirically observed that they may require several hours to solve (to certifiable optimality) instances of~\eqref{eq:MIP_hybrid} with $p \sim 10^3$, and larger problems can take  much longer. 

To address this lack of scalability in general-purpose MIP solvers, we propose a specialized, nonlinear BnB framework for solving~\eqref{eq:MIP_hybrid} to certifiable optimality. Our framework takes into account problem structure to achieve scalability. As we demonstrate in the experiments section, our BnB can solve instances with $p \sim 5 \times 10^6$ to certifiable optimality in minutes to hours, whereas Gurobi takes prohibitively long (at least a day) for $p \sim 10^3$. An important feature of our proposal is an open-source, standalone implementation of the BnB solver, which does not rely on sophisticated and proprietary BnB-capabilities of commercial MIP solvers (e.g., Gurobi).
We first  give a high-level overview of our novel nonlinear BnB framework and then dive into specific technical details.

\smallskip

\noindent {\bf{Overview of nonlinear BnB:}}~  Nonlinear BnB is a general framework for solving mixed integer nonlinear programs \cite{bkl}. This framework constructs a search tree to partition the set of feasible solutions of the given MIP (Problem \eqref{eq:MIP_hybrid} in our case). Instead of explicitly enumerating all the (exponentially many) feasible solutions, BnB uses intelligent enumeration and methods to prune parts of the tree by using lower bounds (dual bounds) on the optimal objective value.
In what follows, we briefly describe how the tree is constructed and pruned. Starting at the root node, the algorithm solves a nonlinear convex relaxation of Problem \eqref{eq:MIP_hybrid}, where all binary variables are relaxed to $[0,1]$ -- this is usually referred to as the \textsl{root relaxation}. Then, the algorithm chooses a \textsl{branching variable}, say $z_g$, and creates two child nodes (optimization subproblems): one with $z_g = 0$ and another with $z_g = 1$, where all other binary variables are relaxed to $[0,1]$. The algorithm then proceeds recursively: for every unvisited node, it solves the corresponding optimization problem and checks if there is any fractional (i.e., non-binary) variable $z_{g}$. If there is any fractional  $z_{g}$, the branching process must continue --- to this end, the algorithm branches on one fractional $z_g$, generating two new child nodes. Thus, every node in the search tree corresponds to an optimization subproblem and every edge represents a branching decision.  

While growing the search tree, BnB maintains an upper bound on the objective function (which can be obtained from any feasible solution to the problem). If the optimization subproblem at the current node leads to an objective value that exceeds the upper bound, then the node is pruned (i.e., no children are generated for this node), because none of its descendants can have a better objective value than the upper bound. Another case where BnB can safely prune a node is when the corresponding subproblem leads to an integral solution, i.e., a binary $\M{z}$ (since there will be no variables to branch on). For further discussion on nonlinear BnB, see~\cite{bkl}.

\smallskip

\noindent \textbf{Specific details: } 
There are many delicate details in BnB that can critically affect its  scalability: for example, the choice of the algorithm for solving the continuous node subproblems, obtaining upper bounds, branching, and tree-search strategies. We discuss our choices below: 
\begin{itemize}
    \item \textbf{Subproblem solver:} The optimal solutions of the continuous optimization subproblems encountered in the course of BnB are typically sparse (see Section \ref{sec:relaxation_reformulation} for further discussions). To solve these subproblems, we propose an active-set algorithm, which  exploits sparsity by considering a reduced problem restricted to a small subset of groups.
    Moreover, we share information on the active sets across the BnB tree to speed up convergence (see Section~\ref{sec:active_set_solver}).
    \item \textbf{Upper bounds:} Better upper bounds can lead to aggressive pruning in the search tree, which can reduce the overall runtime. We obtain the initial upper bound using the approximate algorithms of Section \ref{section:appx_algorithms}. As we demonstrate in the  experiments, our approximate algorithms typically obtain optimal or near-optimal solutions, making them a good choice to initialize BnB. Moreover, at every node of BnB, we attempt to improve the upper bound by using the sparsity pattern of the solution to the current node's subproblem.
    More concretely, let $S \subseteq q$ denote the group support of the latter subproblem's solution. 
    Then, we obtain a new upper bound, by restricting optimization to $S$, i.e., we solve:
     $$\min_{\B\theta} ~~~  \tilde{\ell}(\B\theta) + \lambda_1 \sum_{g=1}^{q} \| \B\theta \|_{2} + \lambda_2 \| \B\theta \|_2^2 ~~~ \text{ s.t. }~~~ \B\theta_{S^c} = \M{0}, ~ \|\B \theta_g \|_2 \leq \MU, ~g \in [q].$$
    \item \textbf{Branching and search strategies: } The branching strategy selects the next variable to branch on, while the search strategy decides which unexplored node in the search tree to visit next. Many elaborate strategies for branching and search have been proposed in the literature -- see \cite{morrison2016branch} for a survey. When the initial upper bound is of high quality, more aggressive pruning is possible, and simple strategies tend to work relatively well in practice \citep[for example, see the discussion in][]{clausen1999best}. Since our approximate algorithms typically return good upper bounds, we rely on simple strategies. For branching, we use maximum fractional branching \cite{bkl,morrison2016branch}, which branches on the factional variable $z_g$ whose value is closest to $0.5$. For search, we use breadth-first search and switch to depth-first search if memory issues are encountered.
\end{itemize}
Our approach extends our recent work~\cite{hazimeh2020sparse} for the best subset selection problem (with a group size of one). We note that there are important differences as the \grp~problem involves a
different and more challenging optimization formulation. Specifically, the Big-M constraints in \eqref{eq:bigm_constraint} translate to second order cones, instead of box-constraints that appear when the group sizes are one.
Furthermore, in the group setup, we have a non-smooth term $\sum_{g \in [q]} \| \B\theta_{g}\|_{2}$ in the objective of~\eqref{eq:MIP_hybrid}. 
The conic constraints and $\ell_2$ norms in our problem require special care when developing the subproblem solver (for example, when reformulating the subproblems in Section \ref{sec:relaxation_reformulation} and designing the active set algorithm in Section \ref{sec:active_set_solver}). It is also worth mentioning that in the simplest case where  $\lambda_1 = \lambda_2 = 0$, our solver solves a MISOCP, whereas \cite{hazimeh2020sparse} solves a mixed integer quadratic program.


\subsubsection{Relaxation reformulation} \label{sec:relaxation_reformulation}
In this section, we study the convex relaxation arising at a node of the BnB search tree. 
We present a particular reformulation of this problem that leads to (i) useful insights about the sparsity in the solutions of the convex relaxation; and (ii) computational benefits.
To simplify the presentation, we will first focus on the root relaxation of \eqref{eq:MIP_hybrid}, which is obtained by relaxing all the binary variables in~\eqref{eq:MIP_hybrid} to $[0,1]$. 

Note that the root relaxation involves the variables $(\B\beta, \M z, \M s)$. In Proposition  \ref{prop:reformulation}, we show that the root relaxation can be reformulated in the $\B \beta$ space, leading to a regularized least squares problem. The associated regularizer can be characterized in terms of the reverse Huber penalty \cite{owen2007robust} (see also~\cite{hongbo}), which is a function $\mathcal{H}: \mathbb{R} \to \mathbb{R}$ defined as follows:
\begin{align} \label{eq:reverse_huber}
    \mathcal{H}(t) = 
        \begin{cases}
            |t| & \text{if~} |t| \leq 1 \\
            (t^2+1)/{2} & \text{otherwise}.
        \end{cases}
\end{align}
\begin{prop} \label{prop:reformulation}
The root relaxation obtained by relaxing the binary variables in \eqref{eq:MIP_hybrid} to $[0,1]$ is equivalent to:
\begin{align} \label{eq:root_relaxation_simplified}
    \min_{\B\theta} ~~~ F(\B\theta) :=  \tilde{\ell}(\B\theta) + \sum_{g=1}^{q} \Psi(\B\theta_g; \B\lambda, \MU) ~~ \text{s.t.} ~~ \| \B\theta_g \|_{2} \leq \MU, ~g \in [q].
\end{align}
where $\B\lambda=(\lambda_0,\lambda_1,\lambda_2)$ and 
\begin{align*}
\Psi(\B\theta_g; \B\lambda, \MU) := 
\begin{cases}
2 \lambda_0 \mathcal{H} (\sqrt{\lambda_2/\lambda_0} \| \B\theta_g \|_2) + \lambda_1 \| \B\theta_g \|_2 &  \text{if}~\sqrt{{\lambda_0}/{\lambda_2}} \leq  \MU \\
(\lambda_0/\MU + \lambda_1 + \lambda_2 \MU) \| \B\theta_g \|_2 &      \text{if}~\sqrt{{\lambda_0}/{\lambda_2}} >  \MU.
\end{cases}
\end{align*}
\end{prop}

The reformulation in \eqref{eq:root_relaxation_simplified}  eliminates the the conic and Big-M constraints from the root relaxation, at the expense of introducing the non-smooth penalty $\sum_{g=1}^{q} \Psi(\B\theta_g; \B\lambda, \MU)$ which is separable across the blocks $\{\B\theta_{g}\}_{1}^{q}$.
Depending on the choices of $\B \lambda$ and $\MU$, the penalty $\Psi$ is either the $\ell_2$ norm or a combination of the reverse Huber penalty and the $\ell_2$ norm. In either case, the penalty is sparsity-inducing. In essence, Problem \eqref{eq:root_relaxation_simplified} is similar to the Group Lasso problem \cite{yuan4}, with two exceptions: (i) Problem \eqref{eq:root_relaxation_simplified} has the additional constraints: $\| \B\theta_g \|_{2} \leq \MU, ~g \in [q]$, and (ii) when $\sqrt{{\lambda_0}/{\lambda_2}} \leq  \MU$, the penalty involves the reverse Huber penalty. 

\smallskip

\noindent \textbf{Node relaxations within the BnB tree: } The convex relaxation subproblem encountered at a node of the BnB search tree is similar to the root relaxation, except that some of the $z_g$s are fixed to $0$ or $1$.  The fixed $z_g$s are determined by the branching decisions made starting from the root until reaching the node. The convex relaxation at a particular node can be reformulated in the $\B \beta$-space similar to the reformulation of the root relaxation in  \eqref{eq:root_relaxation_simplified}, except that: (i) if $z_g = 0$ then the corresponding group should be removed from the objective function; and (ii) if $z_g = 1$, then the penalty $\Psi(\B\theta_g; \B\lambda, \MU)$ should be replaced with  $\tilde{\Psi}(\B\theta_g; \B\lambda) := \lambda_1 \|\B\theta_g\|_2 + \lambda_2 \|\B\theta_g\|_2^2$. More precisely, let $\mathcal{Z}$ and $\mathcal{N}$ be the sets of indices of the $z_g$s that are fixed to $0$ and $1$, respectively. Then, the following subproblem is solved at the corresponding node:
\begin{equation}\label{eq:root_relaxation_simplified-1}
\min_{\B\theta} ~~  \tilde{\ell}(\B\theta) + \sum_{g \in \mathcal{N}^c} \Psi(\B\theta_g; \B\lambda, \MU) + \sum_{g \in \mathcal{N}} \tilde{\Psi}(\B\theta_g; \B\lambda)  ~~ \text{s.t.} ~~ \B\theta_{\mathcal{Z}} = \M{0},  \| \B\theta_g \|_{2} \leq \MU, ~g \in [q].
\end{equation}
In the next section, we develop a scalable algorithm for solving Problem~\eqref{eq:root_relaxation_simplified}. The BnB subproblem~\eqref{eq:root_relaxation_simplified-1} can be solved similarly after  accounting for the fixed $z_g$s.

\subsubsection{Active-Set subproblem solver} \label{sec:active_set_solver}
As discussed earlier, a solution to Problem~\eqref{eq:root_relaxation_simplified} is expected to be sparse in $\B\theta$ (this will be also true for the node sub-problems in the BnB tree). To exploit this sparsity, we use an active-set algorithm: We start by solving Problem \eqref{eq:root_relaxation_simplified} restricted to a small subset of groups (i.e., the \textsl{active set}). After convergence on the active set, we augment the active set with a collection of groups that violate the optimality conditions for the full problem (if any) and then resolve the problem restricted to the augmented active set. The algorithm keeps iterating between solving a reduced optimization problem and augmenting the active set, until the optimality conditions for the full problem are satisfied. Such active-set algorithms have proven to be effective in scaling up the solvers for group Lasso-type problems \citep[for example, see][]{hazimeh2020hierarchy}---our usage differs in that we use this active-set strategy within every node of the BnB tree.  

Next, we describe our active-set algorithm more formally. Let $\mathcal{A} \subseteq [q]$ be the active set. The algorithm starts by solving \eqref{eq:root_relaxation_simplified} restricted to the active set, i.e., 
\begin{align} \label{eq:restricted_subproblem}
    \hat{\B\theta} \in \argmin_{\B\theta} ~~~ F(\B\theta)  ~~ \text{s.t.} ~~ \| \B\theta_g \|_{2} \leq \MU, ~g \in [q],  ~~\B\theta_{\mathcal{A}^c} = \M{0}.
\end{align}
After solving \eqref{eq:restricted_subproblem}, we check if $\hat{\B\theta}$ satisfies the optimality condition for the full problem. Equivalently, 
for every group $g \in \mathcal{A}^c$, we check if the following holds 
\begin{align} \label{eq:group_optimality_condition}
    \B{0} \in \argmin_{\B\theta_g} ~~~ F(\hat{\B\theta}_1, \dots, \B\theta_g, \dots, \hat{\B\theta}_q ) ~~ \text{s.t.} ~~ \| \B\theta_g \|_{2} \leq \MU.
\end{align}
Since $\B\theta_g = \B 0$ is in the interior of the feasible set, condition \eqref{eq:group_optimality_condition} is equivalent to the zero-subgradient condition:  $\B{0} \in \partial_{\B\theta_g} F(\hat{\B\theta}_1, \dots, \hat{\B\theta}_{g-1}, \B{0},  \hat{\B\theta}_{g+1}, \dots, \hat{\B\theta}_q )$, and can be checked in closed form. 


We repeat the procedure of solving the restricted subproblem in \eqref{eq:restricted_subproblem} and augmenting $\mathcal{A}$ with  groups that violate \eqref{eq:group_optimality_condition}, until there are no more violations. The algorithm is summarized below.
\begin{itemize}
    
\item[] \emph{ \underline{Algorithm 3: An Active-set Algorithm for~\eqref{eq:root_relaxation_simplified} }}
\item \textbf{Input:} Initial solution $\hat{\B\theta}$ and initial active set $\mathcal{A}$.
\item \textbf{Repeat} Steps 1--3 till convergence:
\begin{enumerate}
    \item Solve the restricted problem \eqref{eq:restricted_subproblem} to get a solution $\hat{\B\theta}$.
    \item $\mathcal{V} \gets \{g \in \mathcal{A}^c \ | ~  \eqref{eq:group_optimality_condition} \text{ is violated}  \}$.
    \item If $\mathcal{V}$ is empty \textbf{terminate}, otherwise\footnote{{In some cases, $| \mathcal{V} |$ can be large, which can slow down the solver in Step 1. Thus, if $\mathcal{V}$ has more than $K$ groups, we augment $\mathcal{A}$ with the $K$ groups in $\mathcal{V}$ that have the largest violation (instead of $\mathcal{A} \gets \mathcal{A} \cup \mathcal{V}$). In our experiments we set $K = 10$. We found this helpful to keep the size of the active set manageable during the course of the algorithm.}}, $\mathcal{A} \gets \mathcal{A} \cup \mathcal{V}$.
\end{enumerate}
\end{itemize}
Algorithm 3 is guaranteed to converge to an optimal solution for Problem~\eqref{eq:root_relaxation_simplified} in a finite number of steps, as there are finitely many groups.  

\smallskip

\noindent \textbf{Choice of the active set:} The quality of the initial active set $\mathcal{A}$ can have a important effect on the number of iterations in Algorithm 3. 
Due to the choice of our branching rule, the parent and its two child nodes solve similar subproblems; the only difference between these subproblems is that a single $z_g$ is fixed to 0 or 1 in the children. Thus, the solutions and supports of the parent and its children are unlikely to differ by much. We therefore initialize the active set of every node in the BnB tree (except the root) with the support of its parent. For the root node, we initialize the active set with the support of the warm start, obtained from the approximate algorithms that are discussed in Section~\ref{section:appx_algorithms}.

\smallskip

\noindent \textbf{Solving the restricted subproblem:} The convex sub-problem~\eqref{eq:restricted_subproblem} in Step~1 has a small active set and can be solved with a variety of optimization algorithms: for example,   
BCD,  proximal gradient methods~\citep{bertsekas2016nonlinear} or an interior point solver (as available in Gurobi). In our experiments, we use the latter 
due to its good performance in practice.

\section{Statistical Theory}
\label{sec:theory}

In this section we derive non-asymptotic prediction and estimation error bounds for the Group~$\ell_0$ estimators, and compare them to the bounds that have been established for the corresponding Group Lasso-based approaches.  We focus on linear regression models in Section~\ref{sec:lin.reg} and on nonparametric additive models in Section~\ref{sec:add.mod}.

{\color{black}In our analysis, we focus on constrained specifications of the proposed estimators, leaving the penalized case for future research.} To simplify the presentation, we consider the setting where the model is correctly specified, so that the true regression function is a feasible solution to the corresponding optimization problem. However, our results can be generalized to allow for model misspecification.

We say that a constant is \textit{universal} if it does not depend on other parameters, such as~$n$, $q$ or $k$. We use the notation~$\gtrsim$ and~$\lesssim$ to indicate that inequalities~$\ge$ and~$\le$, respectively, hold up to positive universal multiplicative factors, and write $\asymp$ when the two inequalities hold simultaneously. We use $a\vee b$ to denote $\max(a,b)$.

\subsection{Linear Model}
\label{sec:lin.reg}





We assume that the observed data follows the model $\M{y}=\M{X}\bbeta^*+\bepsilon$, where $\M{X}$ is deterministic and the elements of~$\bepsilon$ are independent $N(0,\sigma^2)$ with $\sigma>0$.  We define $k_*=G(\bbeta^*)$ and refer to $n^{-1}\|\M{X}\widehat\bbeta-\M{X}\bbeta^*\|^2_2$ as the prediction error for estimator~$\widehat{\bbeta}$.  
Given $\bbeta\in\RR^p$ and $J\subseteq [q]$, we write~$\bbeta_J$ for the sub-vector of~$\bbeta$ indexed by $\cup_{g\in J}\mathcal{G}_g$.  Consider the following definition, in which we use the notation $\|\bbeta\|_{2,1} =\sum_{g=1}^{q} \|\B\beta_{g}\|_2$.


\begin{definition}
\label{def.sparse.eigen}
Given a positive integer~$k$ and a constant~$c\ge1$, let
\begin{equation*}
\gamma_k=\min_{\bbeta\ne\M{0},\,G(\bbeta)\le k}\frac{\sqrt{k}\|\M{X}\bbeta\|_2}{\sqrt{n}\|\bbeta\|_{2,1}}~~~\text{and}~~~
\kappa_{k,c}=\min_{J\subseteq[q],|J|\le k}\; \left\{\min_{\bbeta\ne\M{0},\,\|\bbeta_{J^c}\|_{2,1}\le c\|\bbeta_{J}\|_{2,1}}\frac{\sqrt{k}\|\M{X}\bbeta\|_2}{\sqrt{n}\|\bbeta_{J}\|_{2,1}} \right\}.
\end{equation*}
\end{definition}
The above definition is most meaningful under the scaling of the features where $\|\M{x}_j\|_2\asymp \sqrt{n}$ for all~$j$. As we discuss below,  constants $\kappa_{k_*,c}^{-1}$, with $c>1$,  appear in the prediction and estimation error bounds for the Group Lasso estimator, while $\gamma_{2k_*}^{-1}$ appears in the estimation error bound for the Group~$\ell_0$ estimator.  The following result establishes a useful relationship for these quantities.


\begin{prop}
\label{prop.ineqs1}
$\gamma_{2k} \ge \kappa_{k,c}/\sqrt{2}$, for all positive integers~$k$ and all $c\ge1$.
\end{prop}

We study estimator~$\widehat \bbeta$ that solves the following optimization problem:
\begin{equation}\label{grp-l0-1-const1-card} 
\min_{\B\beta}~~~~\| \M{y} - \M{X} \B\beta \|_{2}^2  ~~~~\sbt ~~~  \sum_{g=1}^{q} \M{1} ( \B\beta_{g}  \neq \M{0})  \leq k,
\end{equation}
where $k$ is a fixed parameter that controls the sparsity level. We note that \eqref{grp-l0-1-const1-card} is a special case of 
the cardinality constrained problem considered in Section \ref{sec:cardinality}. {\color{black} We write~$T_g$ for the number of features in group~$g\in[q]$ and define $\bar{T}_k=\max_{|J|\le k}\sum_{g\in J} T_g/k$, noting the following relationships in the special case where every group has the same number of~$T$ features: $\bar{T}_k=T$ 
and $p=qT$.} Our first result provides the prediction error bound for~$\widehat \bbeta$, which holds without any assumptions on the design.


\begin{thm}
\label{gen.thm}
Let $\delta_0\in(0,1)$ and suppose that $\widehat \bbeta$ {\color{black}is a global solution} to optimization problem~\eqref{grp-l0-1-const1-card} for~$k\ge k_*$.
Then,
\begin{equation*}
\frac1n\|\M{X}\widehat\bbeta-\M{X}\bbeta^*\|^2_2
\lesssim \sigma^2k\Big[\frac{\bar{T}_k+\log( q /k)}{n}\Big]+\sigma^2\Big[\frac{\log(1/\delta_0)}{n}\Big]
\end{equation*}
with probability at least $1-\delta_0$.
\end{thm}
Letting $\delta_0=(k/q)^{k}$ and using Definition~\ref{def.sparse.eigen}, we derive the following result.


\begin{cor}
\label{lin.cor}
{\color{black} If~$\widehat \bbeta$ is a global solution to optimization problem~\eqref{grp-l0-1-const1-card}} for~$k=k_*$, then
\begin{eqnarray*}
\frac1n\|\M{X}\widehat\bbeta-\M{X}\bbeta^*\|^2_2
&\lesssim& \sigma^2k_*\Big[\frac{\bar{T}_{k_*}+\log( q /k_*)}{n}\Big]\\
\|\widehat\bbeta-\bbeta^*\|_{2,1} &\lesssim& \sigma k_*\Big[\frac{\bar{T}_{k_*}+\log( q /k_*)}{n}\Big]^{1/2}\big[\gamma_{2k_*}\big]^{-1}
\end{eqnarray*}
with probability at least $1-(k_*/q)^{k_*}$.
\end{cor}

We make several observations regarding the established error bounds, comparing them to the bounds for the Group Lasso estimator, denoted by $\widehat\bbeta_{\text{GL}}$, which replaces the $\ell_0$ constraint in Problem~\eqref{grp-l0-1-const1-card} with a penalty on~$\|\bbeta\|_{2,1}$.  To simplify the comparison of the corresponding rates, we focus on the setting where $T_g=T$ for all~$g\in[q]$ and $k=k_*$.


\begin{remark}
\label{opt.rates}
The Group~$\ell_0$ prediction error rate provided in Corollary~\ref{lin.cor} matches the corresponding optimal prediction error rate established in \cite{lounici1}.  The estimation error rate in Corollary~\ref{lin.cor} is also optimal provided that $\gamma_{2k_*}^{-1}$ is bounded by a universal constant under the aforementioned feature scaling $\|\M{x}_j\|_2\asymp\sqrt{n}$.

\end{remark}


\begin{remark}
Let $\|\M{x}_j\|_2\asymp\sqrt{n}$ for all~$j$ and assume that $\kappa_{k_*,c}^{-1}$ is bounded by a universal constant for some $c>1$.  Then, the error bounds for the Group Lasso estimator \citep[see, for example, Section 8.3 of][]{Bbuhl1} are
\begin{equation}
\label{gr.lasso.bnd1}
n^{-1}\|\M{X}\widehat\bbeta_{\text{GL}}-\M{X}\bbeta^*\|^2_2
\lesssim \sigma^2k_*\Big[\frac{T+\log(q)}{n}\Big]\quad\text{and}\quad
\|\widehat\bbeta-\bbeta^*\|_{2,1} \lesssim \sigma k_*\Big[\frac{T+\log(q)}{n}\Big]^{1/2}. 
\end{equation}
The Group~$\ell_0$ rates discussed in Remark~\ref{opt.rates} are better than those in display~(\ref{gr.lasso.bnd1}), because they replace the $\log(q)$ term with $\log(q/k_*)$.  Moreover, in view of Proposition~\ref{prop.ineqs1}, the assumption on~$\gamma_{2k_*}$ in Remark~\ref{opt.rates} is weaker than the Group Lasso assumption on~$\kappa_{k_*,c}$.  Finally,  the Group~$\ell_0$ prediction error bound holds without any assumptions on the design. 

The last observation represents an important non-trivial advantage of $\ell_0$-based approaches over Lasso-type methods.  \cite{zhang2017optimal} provide examples of design matrices in the usual linear regression context for which the Lasso prediction error is lower-bounded by a constant multiple of $1/\sqrt{n}$, generally leading to a much larger prediction error than the one for the $\ell_0$-based method.\footnote{The lower-bound applies to a wide class of coordinate-separable M-estimators, including local optima of nonconvex regularizers such as SCAD and MCP.}
\end{remark}


\begin{remark}
One advantage of estimator~(\ref{grp-l0-1-const1-card}) is that tuning parameter~$k$ directly controls the sparsity of the proposed estimator.  In particular, the~$\widehat \bbeta$ that achieves the bounds in Corollary~\ref{lin.cor} satisfies $G(\widehat\bbeta)\le k_*$.
On the other hand, the~$\widehat \bbeta_{GL}$ that achieves bounds~(\ref{gr.lasso.bnd1}) is typically much more dense. The following inequality, which holds with high probability, is provided in \cite{lounici1}:
\begin{equation*}
G(\widehat\bbeta_{GL})\le\Big[\frac{64\phi_{\max}}{\kappa_{k_*,3}}\Big]k_*.
\end{equation*}
Here, $\phi_{\max}$ is the maximum eigenvalue of $\M{X}^{\top}\M{X}/n$. Thus, the right-hand side is at least~$64k_*$.
\end{remark}


{\color{black}
The error rates presented above can also apply to approximate solutions obtained after an early termination of the MIO solver. Upon termination, the solver provides the upper and lower bounds on the value of the objective function. We denote these bounds by $UB$ and $LB$, respectively, and write $\tau=(UB-LB)/UB$ for the corresponding optimality gap. The next result considers an approximate solution~$\widetilde{\bbeta}$ and demonstrates that the bounds in Corollary~\ref{lin.cor} hold for~$\widetilde{\bbeta}$ when $\tau$ is bounded away from one and $\tau\lesssim k_*[\bar{T}_{k_*}+\log( q /k_*)]/n$.

\begin{cor}
\label{cor.opt.gap}
Let~$k=k_*$ and suppose that $\tau\le 1-c$ for some positive universal constant~$c$. Then,
\begin{eqnarray*}
\frac1n\|\M{X}\widetilde{\bbeta}-\M{X}\bbeta^*\|^2_2
&\lesssim& \sigma^2k_*\Big[\frac{\bar{T}_{k_*}+\log( q /k_*)}{n}\Big]+\sigma^2\tau
\end{eqnarray*}
with probability at least $1-(k_*/q)^{k_*}$.
\end{cor}
}
 
An attractive feature of Theorem~\ref{gen.thm} is that the uncertainty parameter~$\delta_0$ is independent of the tuning parameter~$k$. This allows us to control the expected prediction error\footnote{{\color{black}An application of Definition~\ref{def.sparse.eigen} yields a corresponding bound on the expected estimation error.}}, as we demonstrate in the following result.

\begin{cor}
\label{lin.cor2}
Under the conditions of Theorem~\ref{gen.thm},
\begin{equation*}
\mathbb{E}\|\M{X}\widehat\bbeta-\M{X}\bbeta^*\|^2_2
\lesssim \sigma^2k\big[\bar{T}_k+\log( q /k)\big].
\end{equation*}
\end{cor}

{\color{black}
\subsubsection{Selecting the model size tuning parameter~$k$}
The results presented above rely on the fact that $k\ge k^*$. We now analyze a BIC-type approach for selecting~$k$ that does not require the knowledge of the true sparsity level. Approaches of this type have been shown to be successful in the setting of high-dimensional linear regression for the purposes of model selection \citep{kim2012consistent,wang2013calibrating}. 

We denote the global solution to Problem~\eqref{grp-l0-1-const1-card} by~$\widehat\bbeta_k$ and let~$\check{T}=\max_{g\in[q]}T_g$.  The next result focuses on the estimator~$\widehat\bbeta^B=\widehat\bbeta_{\hat k}$, where
\begin{equation*}
\hat k=\argmin_{k\le q}~\| \M{y} - \M{X} \widehat\bbeta_k \|_{2}^2 + ak[\check{T}+\log(q/k)]
\end{equation*}
and~$a$ is a non-negative tuning parameter. 

\begin{thm}
\label{BIC.thm}
There exists a universal constant~$a_0$, such that if $ a_0\sigma^2\le a\lesssim\sigma^2$, then
\begin{equation*}
\mathbb{E}\|\M{X}\widehat\bbeta^B-\M{X}\bbeta^*\|^2_2
\lesssim \sigma^2(k_*\vee1)\big[\check{T}+\log( q /k^*)\big]\quad\text{and}\quad 
\mathbb{E} G(\widehat\bbeta^B)\lesssim k_*\vee 1.
\end{equation*}
\end{thm}

Theorem~\ref{BIC.thm} demonstrates that, for an appropriate choice of~$a$, estimator~$\widehat\bbeta^B$ achieves the optimal prediction error rate in Remark~\ref{opt.rates} and has the same order of group sparsity as~$\bbeta^*$.

\subsubsection{The case of large group sizes} The results presented so far illustrate that, at least on average, the group sizes need to be of a smaller order than~$n$ to achieve prediction consistency. We now consider the challenging setting where some groups may have more than~$n$ features. In this setting, additional regularization within each group is required for good predictive performance. One way to impose additional regularization is to encourage sparsity within groups and, thus, perform bi-level variable selection \citep{huang4}. This can be achieved by including an additional~$\ell_0$ penalty on the number of features within each group or an additional~$\ell_1$ penalty on the coefficients within each group. Here, we take the approach of regularizing using the $\ell_2$-penalty, without encouraging additional sparsity. As discussed in Section~\ref{sec:spam-1}, we take a similar approach in nonparametric additive modeling, where groups sizes of order~$n$ arise naturally via the cubic spline representation of the functional components.

We assume, for concreteness, that the features have been normalized to achieve $\|\M{x}_j\|_2=\sqrt{n}$ for all $j\in[p]$, and focus on the following optimization problem:
\begin{equation}\label{grp-l0-1-const1-card-l2} 
\min_{\B\beta}~~~~\| \M{y} - \M{X} \B\beta \|_{2}^2  + \lambda\sum_{g \in [q]} \sqrt{T_g}\| \B\beta_{g}\|_{2}  ~~~~\sbt ~~~  \sum_{g=1}^{q} \M{1} ( \B\beta_{g}  \neq \M{0})  \leq k.
\end{equation}
Because the $\ell_2$-penalty parameter is allowed to vary across groups, \eqref{grp-l0-1-const1-card-l2} is a slight generalization of 
the cardinality constrained problem considered in 
Section~\ref{sec:cardinality}, but the algorithms presented here will apply to~\eqref{grp-l0-1-const1-card-l2} with minor adjustments. We define $\tilde{q}=p/\min_{g\in[q]}T_g$ and note that $\tilde{q}=q$ when all the groups are of the same size. The next result establishes a prediction error bound for the large group-size setting.

\begin{thm}
\label{slow.rt.thm}
Let $\delta_0\in(0,1)$ and suppose that $\widehat\bbeta$ is a global solution to problem~\eqref{grp-l0-1-const1-card-l2} for~$k\ge k_*$.
There exist universal positive constants $c_0$, such that if $\lambda\ge c_0\sigma\sqrt{n\log(2e\tilde{q})}$, then
\begin{equation*}
\frac1n\|\M{X}\widehat\bbeta-\M{X}\bbeta^*\|^2_2
\lesssim \frac{\lambda}{n}\sum_{g \in [q]} \sqrt{T_g}\| \B\beta^*_{g}\|_{2}
\end{equation*}
with probability at least $1-1/[4e\tilde{q}]$.
\end{thm}
Focusing on the case $T_g=T$ for all~$g\in[q]$ and $k=k_*$, for concreteness, we derive the corresponding prediction error rate\footnote{{\color{black}When $T=1$, this rate matches the slow error rate for the Lasso estimator \citep[for example,][]{Bbuhl1}.}} of $\sigma\sqrt{T\log(q)/n} \| \B\beta^*\|_{2,1}$. This error rate demonstrates that prediction consistency can be achieved even when~$T\gg n$, provided that the coefficients in $\bbeta^*$ are sufficiently small. For example, when each coefficient equals $1/T$, the resulting rate is $\sigma k_*\sqrt{\log(q)/n}$. 
}


\subsection{Nonparametric Additive Model}
\label{sec:add.mod}

We study the performance of the proposed approach in the deterministic design setting. We write $\|\cdot\|_{L_2}$ for the $L_2$ norm of a real-valued function on $[0,1]$. Using the notation in Section~2.2,
we let $\mathcal{C}_j=\mathcal{C}$ for all~$j$ and focus on the case where $\mathcal{C}$ is an $L_2$-Sobolev space:
\begin{equation*}
\mathcal{C}=\left\{g:[0,1]\mapsto \RR, \;  \|g\|_{L_2}+\|g^{(m)}\|_{L_2}<\infty\right\}~~~\text{and}~~~\text{Pen}(g)=\|g^{(m)}\|_{L_2}.
\end{equation*}
We define~$\mathcal{C}_\text{gr}=\{f:[0,1]^q\mapsto \RR, \; f(\M{x})=\sum_{j=1}^q f_j(x_j),\; f_j\in\mathcal{C}\}$ as the corresponding space of additive functions.  We associate each $f\in \mathcal{C}_\text{gr}$ with the vector $\bff=\sum_{j=1}^q \bff_j$, where $\bff_j=\big(f_j(x_{1j}),...,f_j(x_{nj})\big)$, and let
\begin{equation*}
G(f)=\sum_{j=1}^{q} \M{1} ( \bff_j  \neq \M{0}), \qquad  \text{Pen}_{\text{gr}}(f)=\sum_{j=1}^q \text{Pen}(f_j).
\end{equation*}

We focus on the estimator that globally solves the following optimization problem:
\begin{equation}
\label{crit.add1}
\min_{f\in\mathcal{C}_\text{gr}}\;\,\|\bY- \bff\|_n^2+\lambda_n \text{Pen}_{\text{gr}}(f) \quad\text{s.t.}\quad G(f)\le k,
\end{equation}
where $\|\cdot\|_n$ denotes the Euclidean norm divided by~$\sqrt{n}$.\footnote{We acknowledge the notational inconsistency when $n\le2$.} To ensure identifiability of the representation $f(\M{x})=\sum_{j=1}^q f_j(x_j)$, additional restrictions are typically imposed.  For example, a popular method is to separate out the constant term and require that $\sum_{i=1}^n f_j(x_{ij})=0$ for each~$j$.  Here we follow the approach of \cite{tan2019doubly} and avoid specifying a particular set of restrictions.  We treat every representation of~$f$ as equivalent, with the understanding that one particular representation is used when evaluating properties of the components, such as $\|\bff_j\|_n$.

We are interested in comparing estimator~(\ref{crit.add1}), denoted by~$\widehat f$, with the widely popular Group Lasso-based approach, which replaces the~$\ell_0$ constraint in Problem~(\ref{crit.add1}) with a penalty on $\sum_{j=1}^q\|\bff_j\|_n$.  Theoretical properties of the latter approach have been investigated extensively \citep[see, for example,][and the references therein]{meier1,koltchinskii2010sparsity,raskutti2012minimax,suzuki2013,yuan2016minimax,tan2019doubly}. To compare the error bounds for the two estimators, we need the following definition.


\begin{definition}
\label{nonparam.def}
Given a positive integer~$k$, a constant $\xi \in (1,\infty]$ and an index set $J\subseteq[q]$, let
\begin{equation*}
\begin{aligned}
A_{k,\xi}=&\{f\in\mathcal{C}_\text{gr}:\; \sum_{j=1}^q\|\bff_j\|_n\ne{0}, \, G(f)\le k, \; 2 n^{-m/(2m+1)} \text{\rm Pen}_{\text{\rm gr}}(f)\le (\xi-1)\sum\nolimits_{j=1}^q\|\bff_{j}\|_n\} &\\
B_{J,\xi}=&\{f\in\mathcal{C}_\text{gr}: \; \sum_{j=1}^q\|\bff_j\|_n\ne{0},\; \sum_{j\notin J}\|\bff_{j}\|_n+n^{-m/(2m+1)}\text{\rm Pen}_{\text{\rm gr}}(f)\le \xi\sum_{j\in J}\|\bff_{j}\|_n\}& \\
\psi(k,\xi)=&\min_{f\in A_{k,\xi}}\frac{\sqrt{k}\|\bff\|_n}{\sum_{j=1}^q\|\bff_j\|_n} ~~~\text{and}~~~
\phi(k,\xi) = \min_{J\subseteq [q],|J|\le k}\left\{\min_{f\in B_{J,\xi}}\frac{\sqrt{k}\|\bff\|_n}{\sum_{j\in J}\|\bff_{j}\|_n} \right\}. &
\end{aligned}
\end{equation*}
\end{definition}
As we discuss below,  constants $\phi(2k,\xi)^{-1}$  appear in the error bounds for the Group Lasso-based approach, while constants $\psi(k,\xi)^{-1}$ appear in some of the bounds that we establish for~$\widehat{f}$. The following result establishes a useful relationship for these quantities.


\begin{prop}
\label{prop.ineqs2}
For all positive integers~$k$ and all $\xi\in(1,\infty]$, $\psi(2k,\xi) \ge \phi(k,\xi) / \sqrt{2}$.
\end{prop}

We assume that the observed data follows the model $\M{y}=\bff^*+\bepsilon$, where $f^*\in\mathcal{C}_{\text{gr}}$, and the elements of~$\bepsilon$ are independent $N(0,\sigma^2)$ with $\sigma>0$.  We refer to $\|\widehat\bff-\bff^*\|^2_n$ as the prediction error for estimator~$\widehat{f}$.
We write $r_n=n^{-m/(2m+1)}$, suppressing the dependence on~$m$ for notational simplicity, noting that $r_n^2$ is the optimal prediction error rate in the univariate regression setting where $f^*\in\mathcal{C}$. For example, in the case where $\mathcal{C}$ is the second order Sobolev space, which corresponds to $m=2$, the above rate is $r_n^2=n^{-4/5}$. We define $\alpha=1/(4m+2)$ and note that $\alpha=1/10$ when $m=2$.  The next result, in which we treat~$m\ge 1$ as a fixed integer, establishes prediction error bounds for the proposed approach.


\begin{thm}
\label{add.thm}
Let $k_*=G(f^*)$ and consider optimization Problem~(\ref{crit.add1}) with~$k\ge k_*$. There exists a universal constant~$c_1$, such that if $\lambda_n\ge c_1\sigma\big[k^{2\alpha}r_n^2+k^{\alpha}r_n\sqrt{\log (eq/k)/n}\big]$, then
\begin{equation}
\label{np.or.bnd.gen}
\|\widehat\bff-\bff^*\|^2_n \lesssim \sigma^2 k\Big[k^{2\alpha}r_n^2+\frac{\log (eq/k)}{n}\Big] + \lambda_n\text{\rm Pen}_{\text{\rm gr}}(f^*)
\end{equation}
with probability at least $1-(k/q)^k$. Furthermore, for every $\xi\in(1,\infty]$,
there exists a finite constant~$c_2$, which depends only on~$\xi$, such that if $\lambda_n\ge c_2\sigma\big[r_n^2+r_n\sqrt{\log (q)/n}\big]$, then
\begin{equation}
\label{np.or.bnd.cc}
\|\widehat\bff-\bff^*\|^2_n\lesssim  \sigma^2 k\Big[r_n^2+\frac{\log (q)}{n}\Big]\big[\psi(2k,\xi)\big]^{-2} + \lambda_n \text{\rm Pen}_{\text{\rm gr}}(f^*)
\end{equation}
with probability at least $1-1/q$.
\end{thm}
%
We make the following observations regarding the established error bounds.  To simplify the comparison of the error rates, we focus on the setting where $k=k_*$ and $\text{\rm Pen}_{\text{\rm gr}}(f^*)\asymp \sigma k_*$. The last relationship holds, for example, when the scaled roughness of each nonzero component, $\text{\rm Pen}(f^*_j)/\sigma$, is bounded above and below by positive universal constants.


\begin{remark}
The expression in error bound~(\ref{np.or.bnd.gen}) is optimized for the setting where $\text{\rm Pen}_{\text{\rm gr}}(f^*)\asymp \sigma k$.  However, as we show in the proof, the bound can be improved when~$\sigma k$ and $\text{\rm Pen}_{\text{\rm gr}}(f^*)$ have different orders of magnitude.
\end{remark}


\begin{remark}
The prediction error rate provided in~(\ref{np.or.bnd.cc}) is analogous to the rate established in~\cite{tan2019doubly} for the Group Lasso-based approach\footnote{
To the best of our knowledge, the bounds in \citep{tan2019doubly} are overall the strongest in the literature for the Group Lasso-based approach, due to the relative weakness of the imposed conditions: see the discussion in Remark~12 of \cite{tan2019doubly}.}, however, the latter rate replaces $\psi(2k_*,\xi)^{-2}$ with $\phi(k_*,\xi)^{-2}$. By Proposition~\ref{prop.ineqs2}, the former rate is at least as good as the latter, with a potential improvement due to the additional $\ell_0$ group sparsity requirement in the definition of~$\psi$.
If for some fixed $\xi>1$ quantity $\psi(2k_*,\xi)^{-1}$ is bounded by a universal constant, then inequality~(\ref{np.or.bnd.cc}) yields the following prediction error rate:
\begin{equation*}
\|\widehat\bff-\bff^*\|^2_n \lesssim \sigma^2 k_* \Big[r_{n}^2+\frac{\log(q)}{n}\Big].
\end{equation*}
This rate matches the one established in~\cite{tan2019doubly} for the Group Lasso-based approach under an analogous (but somewhat stronger) assumption\footnote{{\color{black}For a comprehensive discussion of this assumption, we refer the reader to \cite{meier1,tan2019doubly}, and the references therein.}} on $\phi(k_*,\xi)^{-1}$.
\end{remark}


\begin{remark}
Bound~\eqref{np.or.bnd.gen} yields the following error rate without imposing assumptions on the design:
\begin{equation*}
\|\widehat\bff-\bff^*\|^2_n \lesssim \sigma^2 k_* \Big[k_*^{2\alpha}r_{n}^2+\frac{\log(eq/k_*)}{n}\Big].
\end{equation*}
If $k_*\lesssim 1$ or $k_*^{2\alpha}r_{n}^2\lesssim \log(eq/k_*)/{n}$, then the above expression can be upper-bounded by
\begin{equation*}
\sigma^2 k_* \big[ r_{n}^2+\frac{\log(eq/k_*)}{n}\big].
\end{equation*}
Thus, $\widehat f$ achieves the corresponding minimax lower bound on the prediction error \citep{raskutti2012minimax,suzuki2013,tan2019doubly}.
\end{remark}


\begin{remark}
When $q=k_*$, the prediction error rate given by bound~(\ref{np.or.bnd.gen}) is $k_*^{1+1/(2m+1)}r_n^2$, which improves over the corresponding $k_*^{1+3/(2m+1)}r_n^2$ rate\footnote{
Theorem~1 in~\cite{lin2} treats the number of predictors ($q=k_*$) as fixed and omits it from the expression for the error rate.  However, an examination of the proof of their Theorem~1 and the entropy bound in their Lemma~A.1, which explicitly accounts for the number of predictors, reveals the effect of the dimension $k_*$.} derived in~\cite{lin2}.
In particular, when $m=2$, the former rate is $k_*^{6/5}n^{-4/5}$, while the latter is $k_*^{8/5}n^{-4/5}$.  The improvement in the rate is a consequence of the more refined entropy bounds derived in our proofs.

\end{remark}


\begin{remark}
In the special case of~$m=2$ and $k_*\lesssim 1$, bound~(\ref{np.or.bnd.gen}) yields the prediction error rate of $n^{-4/5}+\log(q)/n$, which matches the optimal univariate rate of~$n^{-4/5}$ when~$\log(q)\lesssim n^{1/5}$.
\end{remark}


\begin{remark}
If for some fixed $\xi>1$ quantity $\psi(2k_*,\xi)^{-1}$ is bounded by a universal constant, then a direct consequence of Theorem~\ref{add.thm} is the following estimation error rate:
\begin{equation*}
\sum_{j=1}^q\|\widehat\bff_j-\bff^*_j\|_n\lesssim \sigma k_*\Big[r_{n}+\sqrt{\frac{\log(q)}{n}}\Big].
\end{equation*}
\end{remark}


\section{Experiments}
\label{sec:simulations}
We present experiments that shed light on the practical performance of our proposals compared to the state of the art. Our algorithms are implemented in Python and are available at
\url{https://github.com/hazimehh/L0Group}.
In Section~\ref{sec:stat-perf1}, we investigate the statistical properties of our algorithms for the \grp~problem. In Section \ref{sec:experiment_timings}, we present computation times of our MIP algorithm. Section \ref{sec:comp-spam} investigates nonparametric sparse additive models. Additional numerical experiments can be found in the supplementary material. 

\subsection{Grouped variable selection}\label{sec:stat-perf1}

We consider both synthetic and real datasets in our experiments, as discussed below.

\smallskip

\noindent{\bf{Synthetic data generation.}} The underlying model is $\M{y} = \M{X}\B\beta^{*} + \B\epsilon$, where $\B\beta^{*} \in \RR^p$ has~$q$ groups, all with the same size. Once we generate $\M{X}$ (see below), every column is standardized to have unit $\ell_{2}$-norm.
The errors $\epsilon_{i} \stackrel{\text{iid}}{\sim} N(0, \sigma^2),i = 1, \ldots, n$, are independent of $\M{X}$, and $\sigma^2$ is chosen to achieve a desired signal-to-noise ratio (SNR)\footnote{For a generative model of the form $y_{i} = \mu_{i} + \epsilon_{i}$, we define $\text{SNR} = \text{Var}{(\mu)}/{\text{Var}(\epsilon)}$.}. We note that the SNR values in our experiments are sufficiently high to make the true model support recovery possible.

Two different types of $\M{X}$ are considered:
(a) {\texttt {example=1}}: We first generate group representatives $\B\gamma_{1}, \ldots, \B\gamma_{q} \sim \text{MVN}_{q}(0, \B\Sigma)$, where, $\B\Sigma_{q \times q} = ((\sigma_{ij}))$, with
$\sigma_{ij} = \rho^{|i-j|}$.  Given a $\B\gamma_{g}$, the covariates $\M{x}_{j}, j \in {\mathcal G}_g$ are generated by adding independent Gaussian noise to a scalar multiple of~$\B\gamma_{g}$, to achieve pairwise correlation of~$0.9$ within the group.
(b) {\texttt{example=2}}: Here we take $\M{X}\sim \text{MVN}_{p}(0, \B\Sigma)$, where $\sigma_{ij} = \rho$, for all $i \neq j$, with $\sigma_{jj} =1$ for all $j$.

To generate the true population regression coefficients, the $k_{*}$ nonzero groups are taken to be equally spaced
in $\{1, \ldots, q\}$. All the nonzero entries of $\B\beta^{*}$ are drawn independently from a standard Gaussian distribution.

\noindent{\bf{Competing algorithms and tuning}.} In the  experiments of this section, we focus on the \grp~problem defined in \eqref{grp-l0-1-const1}, and study the performance of our algorithms. We compare against the following state-of-the-art grouped variable selection methods: Group Lasso (based on $\ell_{2,1}$ regularization), Group MCP, and Group SCAD -- these estimators are computed by using the R package \texttt{grpreg} \cite{breheny2015group}. For synthetic data (Sections~\ref{sec:vary_n} and~\ref{sec:large_instances}), we construct a separate validation set with a fixed design.
We tune the parameters of the different problems to minimize the prediction error on the validation set. Specifically, for each of \grp~and Group Lasso, we tune the regularization parameter over a (one-dimensional) grid with 100 values. For MCP and SCAD, we tune the first parameter $\lambda$ over a grid with 100 values, and leave the second parameter $\gamma$ to its default value in \texttt{grpreg}. 
\textcolor{black}{For MCP and SCAD, high-dimensional BIC (HBIC) is another method  for doing tuning parameter selection~\cite{wang2013calibrating} -- we present results for MCP and SCAD based on HBIC tuning following~\cite{wang2013calibrating} in the Supplementary material (Section~\ref{sec:additional_experiments}).}

\noindent{\bf{Performance measures}.}
Given an estimator $\hat{\B\beta}$, we consider the following performance measures:
\begin{itemize}
\item \textbf{True Positives (TP)}: The number of nonzero groups that are in both $\hat{\B\beta}$ and $\B\beta^{*}$.
\item \textbf{False Positives (FP)}: The number of nonzero groups in $\hat{\B\beta}$ but not in $\B\beta^{*}$
\item \textbf{Recovery F1 Score}: The harmonic mean of precision and recall, i.e., F1 Score = ${2P R}/{(P + R)}$, where $P = \text{TP}/(\text{TP} + \text{FP})$ is precision and $R = \text{TP}/k_{*}$ is recall. We note that an F1 Score of $1$ implies perfect support recovery. 
\item \textbf{Test MSE}: This is defined as $\frac{1}{n} \| \M X \hat{\B\beta} - \M X \B\beta^{*} \|_2^2$.
\end{itemize}

\subsubsection{Statistical performance for varying number of observations} \label{sec:vary_n}
\begin{figure}[h!]
\centering
\includegraphics[width=\textwidth]{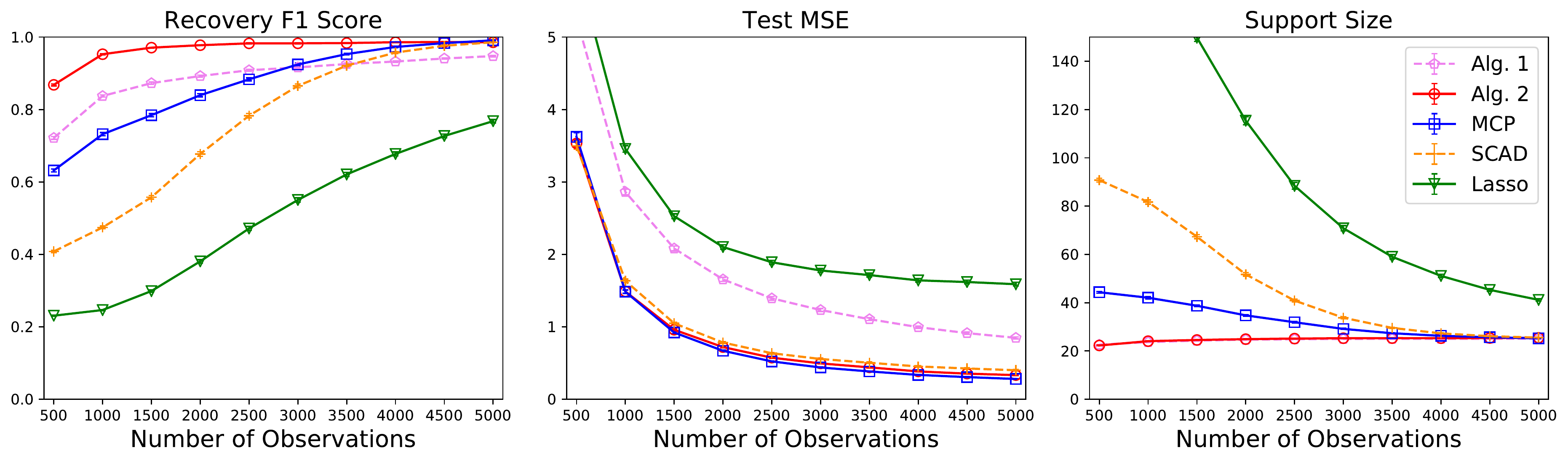}
\caption{Performance measures for varying number of observations on a synthetic dataset with highly correlated features. \textcolor{black}{The measures are averaged over 500 repetitions, and the standard error is represented using error bars}. Alg. 1 and Alg. 2 are our proposed algorithms. Here, ``Lasso" is a shorthand for Group Lasso, we use the same convention for SCAD, MCP. All methods are tuned by minimizing the prediction error on a validation set. 
\textcolor{black}{The average running time (in seconds) across $n$ is 11 for Algorithm 1, 42 for Algorithm 2, and 4 for Group Lasso, MCP, and SCAD. This difference in running time may be in part due to the efficient C implementation used in \texttt{grpreg}, as opposed to our prototype in Python.}}
\label{graph:varyn}
\end{figure}

In this experiment, we study the effect of varying the number of observations $n$ on the performance of \grp~and other state-of-the-art group regularizers (Group Lasso, MCP, and SCAD). We obtain approximate estimators to the \grp~problem using Algorithms 1 and 2 (with $m=1$). \textcolor{black}{We generate 500 datasets} having exponentially decaying correlation (i.e., under \texttt{example=1}) with a correlation parameter $\rho = 0.9$, $p = 5000$, a group size of $4$, number of nonzero groups $k_{*} = 25$, and $\text{SNR}=10$. This setting is relatively difficult for recovery as each group is highly correlated with a few others. \textcolor{black}{We report the mean and the standard error for each performance measure in Figure~\ref{graph:varyn}.} 

Figure~\ref{graph:varyn} shows that Algorithm 2 notably outperforms the other methods in terms of variable selection; it perfectly recovers the support for $n \approx 2000$. Group MCP and SCAD require roughly $4500$ observations to recover the true support,  whereas Group Lasso does not recover the support even when $n=p$. Moreover, Algorithms 1 and 2 attain the smallest support sizes for any $n$, whereas the other methods require much larger supports, especially for small $n$. Algorithm 2 has the lowest test MSE for all $n$. The test MSE of MCP matches that of Algorithm 2 in most of the cases, while the other methods lag behind. We also note that there is a gap between the test MSE of Algorithms 1 and 2. This difference is likely due to Algorithm~2 doing a better job in optimization. 

\textcolor{black}{In Supplementary Material Section~\ref{sec:additional_experiments}, we report the results of the same experiment but with lower feature correlation ($\rho=0.5$ and $\rho=0$). The performance of all methods improve as correlation decreases, but the results are qualitatively similar to what we see in Figure~\ref{graph:varyn}. In addition, we also report results based on HBIC tuning for Group MCP and SCAD regularizers -- with HBIC tuning, the sparsity of Group MCP and SCAD estimators generally improves, but their prediction performance suffers.}

\subsubsection{Statistical performance on high-dimensional instances} \label{sec:large_instances}
We compare the performance of the different methods under two high-dimensional settings. In both settings, we generate data with constant correlation (i.e., under \texttt{example=2}) and $\text{SNR} =10$. Below is a description of the settings:
\begin{itemize}
\item Setting 1: $\rho = 0.9, n=1000, p=100,000, k=10$, and a group size of $10$.
\item Setting 2: $\rho = 0.3, n=1000, p=100,000, k=20$, and a group size of $4$.
\end{itemize}
\textcolor{black}{For each setting, we generate 500 replications}, on which we train and tune the algorithms. The tuning here is based on validation MSE -- \textcolor{black}{results based on HBIC tuning for (group) SCAD and MCP estimators are presented in the supplementary material (Section~\ref{sec:additional_experiments})}.
To ensure a fair comparison in terms of running time, we solve the \grp~problem approximately using Algorithm 2 (with $m=1$), which typically has the same order of running time (seconds in this case) as the other group selection methods considered here. We report the averaged results for Settings $1$ and $2$ in 
Table~\ref{table:setting1}.

\begin{table}[h!]
\centering
\caption{Performance measures for Setting 1 (top panel) and Setting 2 (bottom panel). Means are reported along with their standard errors---we consider 500 replications.} 
\vspace{0.5cm}
\label{table:setting1}
\begin{tabular}{cc}
\rotatebox[origin=c]{90}{Setting 1~~~~~}&
\begin{tabular}{lccccc} 
\toprule
Algorithm &  $\| \hat{\B\beta} \|_0$ &  TP &  FP &  MSE &  $\| \hat{\B\beta} - \B\beta^{*} \|_{\infty}$ \\
\midrule
Group $\ell_0$ &         $ 100.3 ~ (1.2) $ &           $ 8.9 ~ (0.09) $ &              $ 1.1 ~ (0.1) $ &          $ 19.2  ~ (1.1) $ &      $ 1.17 ~ (0.03) $ \\
Group Lasso      &        $ 2086.1 ~ (27.8) $ &  $ 9.6 ~ (0.05)$ &            $ 199.0 ~ (2.8) $ &          $ 26.9  ~ (1.1) $ &      $ 1.44 ~ (0.02) $ \\
Group MCP        &         $ 294.8 ~ (6.7) $ &           $ 9.4 ~ (0.06) $ &            $ 20.1 ~ (0.7) $ &          $ 19.8  ~ (1.2) $ &      $ 1.10 ~ (0.04) $ \\
Group SCAD       &         $ 686.3 ~ (18.6) $ &           $ 9.5 ~ (0.05) $ &            $ 59.1 ~ (1.9) $ &          $ 24.5  ~ (1.2) $ &      $ 1.21 ~ (0.03) $ \\
\bottomrule
\end{tabular} \\
\rotatebox[origin=c]{90}{Setting 2}&
\begin{tabular}{lccccc}
Group $\ell_0$ &           $ 79.4 ~ ( 0.1 ) $ &            $ 19.7 ~ ( 0.03 ) $ &              $ 0.2 ~ ( 0.02 ) $ &           $ 1.12 ~ ( 0.03 ) $ &      $ 0.356 ~ ( 0.007 ) $ \\
Group Lasso      &        $ 1126.5 ~ ( 11.7 ) $ &           $ 19.8 ~ ( 0.02 ) $ &           $ 261.8 ~ ( 2.9 ) $ &           $ 5.06 ~ ( 0.10 ) $ &      $ 0.703 ~ ( 0.007 ) $ \\
Group MCP        &         $ 139.6 ~ ( 2.1 ) $ &            $ 19.8 ~ ( 0.02 ) $ &             $ 15.1 ~ ( 0.5 ) $ &           $ 1.20 ~ ( 0.03 ) $ &      $ 0.352 ~ ( 0.006 ) $ \\
Group SCAD       &         $ 310.4 ~ ( 4.9 ) $ &            $ 19.9 ~ ( 0.02 ) $ &            $ 57.8 ~ ( 1.2 ) $ &           $ 1.56 ~ ( 0.06 ) $ &      $ 0.415 ~ ( 0.008 ) $ \\
\bottomrule
\end{tabular}
\end{tabular}
\end{table}

Under both settings, \grp~selects significantly smaller support sizes and false positives than other methods, and is more consistent across the replications (as evidenced by the small standard error). For example, in Table \ref{table:setting1} (top), \grp~has a support size which is roughly $20$ times smaller than the one for the Lasso and 3 times smaller than one for MCP. For few of the instances, one true positive is missed in \grp, but the difference with the other methods is marginal. In terms of MSE and the estimation error (i.e., $\| \B\beta - \B\beta^{*} \|_{\infty}$), \grp~appears to outperform the other methods, with the differences being most pronounced in the high correlation setting of Table \ref{table:setting1} (top). This aligns with the results in Figure \ref{graph:varyn}, where we saw that \grp~leads to important improvements when features are highly correlated and $n$ is small.


\subsubsection{Real data}
\begin{figure}
    \centering
    \includegraphics[scale=0.4]{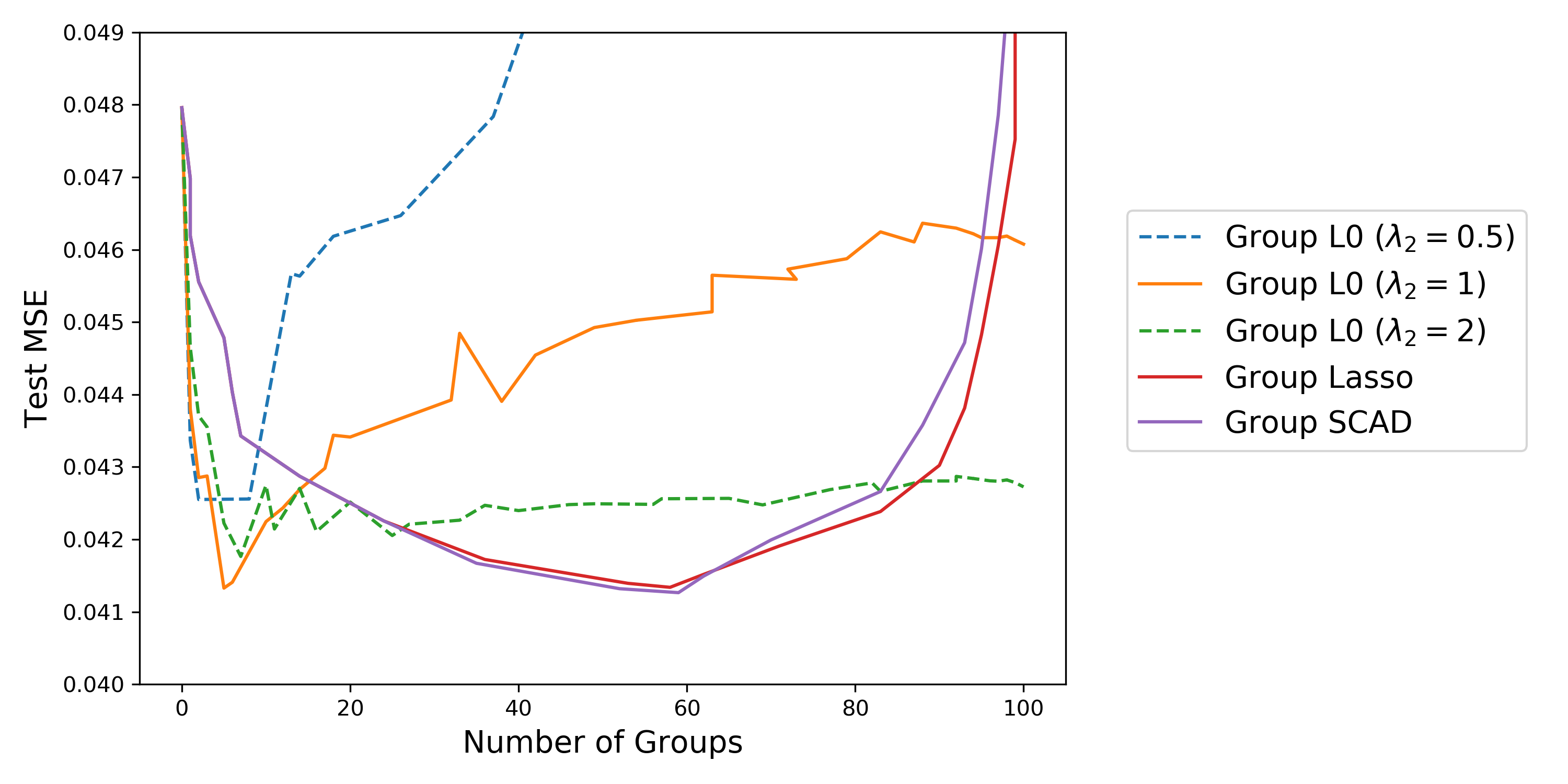}
    \caption{Test MSE on the Amazon Reviews dataset ($n=3500$, $p=3368$, and $q = 100$). For \grp, we consider additional ridge regularization and vary the corresponding regularization parameter $\lambda_2 \in \{0.5, 1, 2\}$. }
    \label{fig:amazon_reviews}
\end{figure}

We study the performance of the different methods on the Amazon Reviews dataset~\cite{fastsubset}. \textcolor{black}{As the focus of this paper is on group variable selection in the case where the non-overlapping groups are pre-specified, we perform pre-processing of the features to obtain a grouping of the features (based on simple exploratory analysis). All the different group-sparse estimators make use of the same group structure -- this allows us to fairly compare their performance given the chosen group-structure.
We note that the downstream results depend upon the input group structure.}
After preprocessing, the dataset consists of $3482$ predictors divided into $100$ groups. We use $3500$ and $2368$ observations for training and testing, respectively. Additional details on the dataset and preprocessing are discussed in the Supplement~\ref{data-examples}. 
On this dataset, we fit regularization paths for Group $\ell_0$, Lasso, and SCAD\footnote{We also tried group MCP, but the solver faced numerical problems -- hence, their results are not reported.}. For Group $\ell_0$, we use an additional ridge regularization term\footnote{This is found to be useful here due to high feature correlations within a group.} and consider $\lambda_2 \in \{0.5, 1, 2 \}$. In Figure \ref{fig:amazon_reviews}, we plot the test MSE at different sparsity levels. The results indicate that the lowest MSE is roughly the same for \grp~($\lambda_2 = 1$), Lasso, and SCAD; with \grp~having a clear advantage in terms of the support size. Specifically, \grp~with $\lambda_2 = 1$ attains the lowest MSE at $5$ groups whereas Group Lasso and SCAD require around $60$ groups to achieve a similar MSE performance.

In the Supplement \ref{sec:birthweight}, we report results on another real dataset; and our conclusions are qualitatively similar to the example in Figure~\ref{fig:amazon_reviews}.

\subsection{MIP-based global optimality certificates: Timing comparisons} \label{sec:experiment_timings}
Here, we compare the running time of our BnB solver with Gurobi for obtaining globally optimal solutions (we note that Algorithms~1, 2 presented earlier are approximate algorithms.)
We generate synthetic data under \texttt{example=2}, and we study the effect of the number of predictors $p$ on the running time. Specifically, we vary $p \in \{ 10^3, 10^4, 10^5, 10^6, 5 \times 10^6 \}$ and fix the other data generation parameters as follows: group size of $10$, $n=10^3$, $\rho = 0.1$, $k_{*} = 5$, SNR $=10$, and set all nonzero coefficients in $\B\beta^{*}$ to 1. \textcolor{black}{We limit our largest problem instance to $n=10^3$ and $p=5 \times 10^6$ due to memory limitations -- as our data-matrix $\M{X}$ is dense, generating and storing a copy of it is memory intensive.} We solve the MIP in \eqref{eq:MIP_hybrid} to optimality, for two cases: \textbf{(i)} with ridge regularization ($\lambda_2 > 0$) and \textbf{(ii)} without ridge regularization ($\lambda_2 = 0$). In both cases, we fix $\lambda_1 = 0$. For case (i), we choose $(\lambda_0, \lambda_2)$ so that the solution obtained has $k_{*}$ nonzero groups and minimizes the $\ell_2$ estimation error. More formally, for a fixed choice of $(\lambda_0, \lambda_2)$, let $\B\theta(\lambda_0, \lambda_2)$ denote a solution of \eqref{eq:MIP_hybrid}. Then, we choose the parameters of case (ii) as follows:
$$
(\lambda_0^{*}, \lambda_2^{*}) \in \argmin_{(\lambda_0, \lambda_2)} \| \B\theta(\lambda_0, \lambda_2) - \B\beta^{*} \|_2 ~~ \text{s.t.} ~~ G(\B\theta(\lambda_0, \lambda_2))= k_{*}.
$$
We estimate $(\lambda_0^{*}, \lambda_2^{*})$ by running Algorithm 2 on a two-dimensional grid with $\lambda_{0} \in \{10^3, 2 \times 10^3, \dots, 10^4\}$ and $\lambda_2 \in \{10^{-5}, 10^{-4}, \dots, 10^{5}\}$. 
For case (ii), we choose $\lambda_0$ so that the corresponding solution has~$k_{*}$ nonzero groups. 
Let $S^{*}$ be the support of the true solution $\B\beta^{*}$, and let $\hat{\B\beta}$ be the solution obtained by solving $\min_{\B\beta} \ell(\B\beta) ~~ \text{s.t. } \B\beta_{(S^{*})^c} = 0$. Then, in both cases,  we set $\MU$ to $\max_{g \in [q]} \|\hat{\B\beta}_g\|_2$.
For the two solvers, we set the optimality gap\footnote{Given an upper bound UB and a lower bound LB, the optimality gap is defined as (UB-LB)/UB.} to $1\%$ and use a warm start obtained from Algorithm 2. The running times were measured on a cluster with CentOS 7. Each job (i.e., a single run of a solver over one dataset) was allocated 4 cores of an Intel Xeon Gold 6130 CPU @ 2.10GHz processor and up to $120$ GB of RAM. For each job, we set a time limit of 24 hours. 

\begin{table}[htbp]
\caption{Running time in seconds for solving Problem \eqref{eq:MIP_hybrid} to optimality. A dash (-) indicates that Gurobi cannot solve the problem in 24 hours and has an optimality gap of~$100\%$ upon termination.}
\label{table:BnB_times}
\centering
\setlength{\tabcolsep}{8pt}
\renewcommand{\arraystretch}{1.2}
\begin{tabular}{|c|cc|cc|}
\hline
\multirow{2}{*}{$p$} & \multicolumn{2}{c|}{Case (i): $\lambda_2 = \lambda_2^{*}$} & \multicolumn{2}{c|}{Case (ii): $\lambda_2 = 0$} \\
                & Ours & Gurobi & Ours & Gurobi \\ \hline
$10^3$          & 96   & 24223  & 373  & 8737   \\
$10^4$          & 199  & -      & 466  & -      \\
$10^5$          & 231  & -      & 1136 & -      \\
$10^6$          & 386  & -      & 1628 & -      \\
$5 \times 10^6$ & 1922 & -      & 11627     & -      \\ \hline
\end{tabular}
\end{table}

In Table \ref{table:BnB_times}, we report the running time (in seconds) for cases (i) and (ii). In both cases, the results indicate that our BnB can solve instances with $p = 5 \times 10^6$ in the order of minutes to hours, whereas Gurobi cannot solve the problem beyond $p=10^3$ within the 24-hour  time limit. Specifically, for $p \geq 10^4$, Gurobi's optimality gap is $100\%$. The reason behind this large gap is that Gurobi cannot solve the root relaxation in the 24-hour time limit, so the best lower bound upon termination is 0. The running times for our BnB solver in case (i) are lower than case (ii), and this can be attributed the perspective reformulation which exploits the presence of the ridge regularizer to speed up computation. It is also worth mentioning that our implementation of BnB is a prototype that does not exploit parallelism (commercial solvers like Gurobi can exploit parallelism). Parallelizing our BnB implementation is expected to make it faster, especially on difficult instances where the search tree is large. In the Supplement Section~\ref{sec:additional_timings}, we report the running times of our BnB algorithm and Gurobi for different choices of $\MU$.

\textcolor{black}{In Supplement Section~\ref{sec:additional_timings}, we present an example showing how the runtime of our BnB algorithm changes as~$n$ is increased.}

\subsection{Nonparametric Additive Models}\label{sec:comp-spam}

We study an expanded version of the popular Boston Housing dataset\footnote{The dataset was downloaded from~\url{https://archive.ics.uci.edu/ml/datasets/Housing}.} as an application of our MIP framework to $\ell_0$-sparse additive modeling. The dataset consists of $13$ covariates. To get a better idea about the performance in the presence of irrelevant covariates, we augmented the data with $50$ irrelevant covariates. Specifically, we selected $5$ covariates uniformly at random. For each selected covariate, we randomly permuted the entries of the covariate vector and augmented the data with the permuted vector---we repeated this step $10$ times.
This led to $63$ covariates in total. We randomly sampled $406$ observations for training and $50$ observations for validation, and we standardized the response and the covariates. We predict house price using the 63 covariates. 

\begin{figure}[htbp]
    \centering
    \includegraphics[scale=0.5]{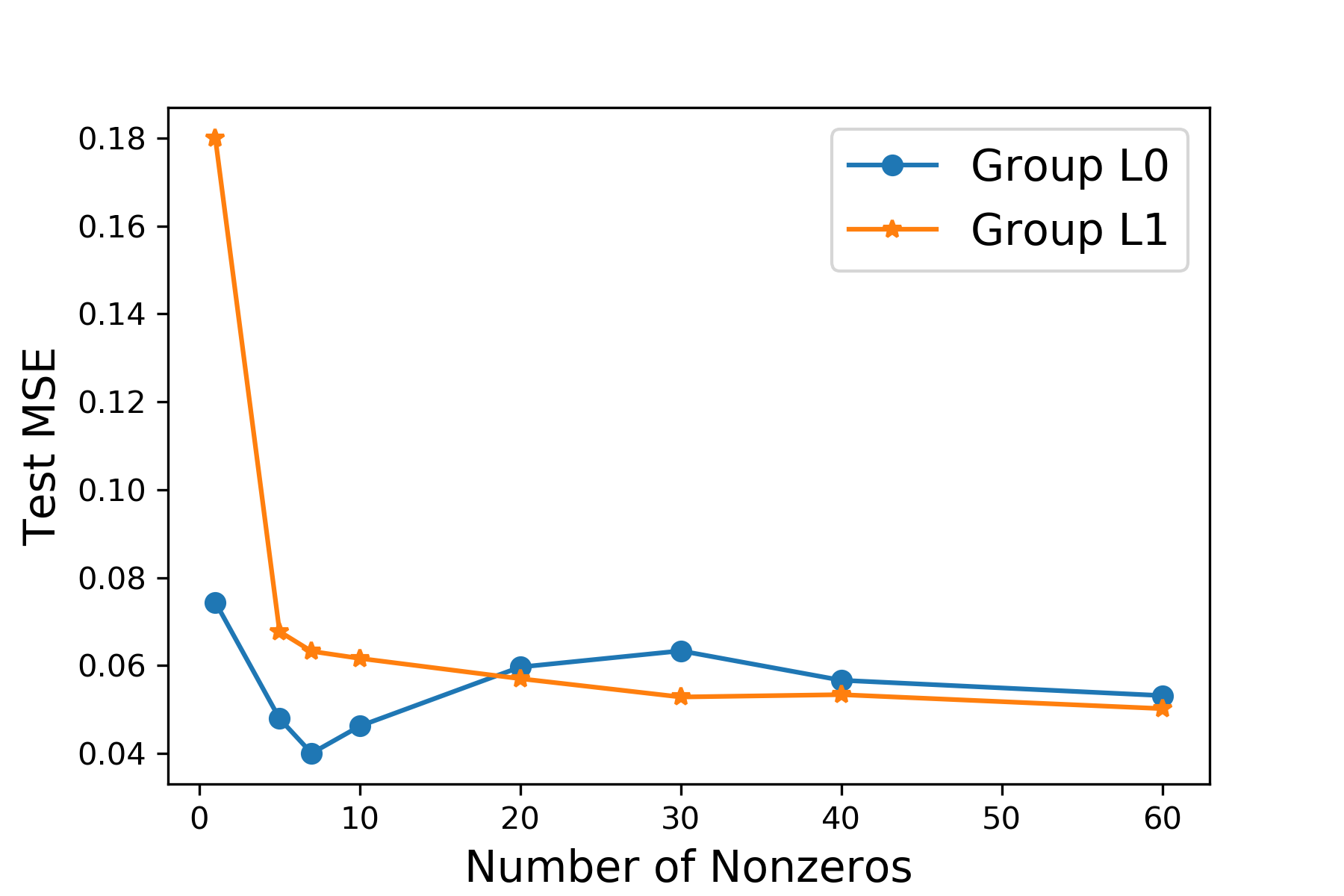}
    \caption{Test MSE versus the number of nonzeros on the Boston Housing dataset (with additional noisy covariates).}
    \label{fig:Boston_Housing}
\end{figure}

We compare the performance of sparse additive models based on \grp~and Group Lasso. In both approaches, we used B-splines of degree 3 for the basis functions, with $10$ knots equi-spaced in the covariates. For the \grp-based approach, we used formulation \eqref{add-pen-1-sp} and tuned $\lambda$ over a grid of $100$ values between $10^{-5}$ and $10^{-2}$ (equi-spaced on a logarithmic scale). We obtained the Group Lasso-based approach by relaxing all the binary variables in the MIP formulation of \eqref{add-pen-1-sp} to the interval $[0,1]$, and we tuned $\lambda$ over a grid of $100$ values ranging from $10^{-4}$ to $1$  (equi-spaced on a logarithmic scale). In Figure \ref{fig:Boston_Housing}, we plot the test MSE versus the number of nonzeros, for each of the two models. The results indicate that the \grp-based approach achieves the minimum test MSE at $7$ nonzeros, whereas the Group Lasso-based method achieves its  minimum MSE at around $60$ nonzeros (without matching the performance of Group L0).

\section{Concluding Remarks}
{\color{black} We revisit a well-known family of problems in sparse learning where the variables are naturally organized into a collection of pre-specified non-overlapping groups. We study both the associated linear regression problem and the problem of nonparametric additive modelling with smooth components. 
In contrast to the earlier work, we pursue MIP-based methods to solve the underlying discrete optimization problem to optimality, and at scale. Our algorithmic contributions include (i) fast algorithms based on coordinate descent and combinatorial local search to obtain good feasible solutions; and (ii) an exact algorithm based on a custom branch-and-bound procedure, which exhibits significant speedups compared to 
off-the-shelf commercial MIP solvers. We present statistical theory for our estimators encompassing both the linear model and the nonparametric additive model settings. 

Our paper contributes to the growing body of work exploring MIP-based tools to address a broad range of computational problems that arise in statistics. 
An incomplete list of recent references includes classification~\cite{dedieu2021learning},
factor driven regression~\cite{lee2021factor},
learning acyclic graphs~\cite{manzour2021integer},
sparse PCA~\cite{dey2021using}, 
signal estimation~\cite{atamturk2021sparse,lee2021sparse}, among others -- we also refer the reader to a recent book~\cite{bertsimas2019machine} and survey~\cite{tillmann2021cardinality}. 

There are several directions for future work. One important application of group-sparse models arises in sparse representation of multiple measurement vectors, where the response is multivariate and the regression coefficients form a 
matrix~\cite{tropp2006algorithms,malioutov2005sparse,cotter2005sparse}. In a series of works, \cite{chen2006theoretical,stojnic2009reconstruction,van2010theoretical} study different convex relaxations of group-$\ell_0$-type formulations and their theoretical properties.
Other examples of group-sparse problems can be found in~\cite{bach2012structured,zhao2009composite}. It will be interesting to extend the approaches presented here to these more general settings.}


\section{Acknowledgements} 
We would like to thank the Associate Editor and the referees for their thoughtful and constructive comments that helped us improve the paper. 
We thank Shibal Ibrahim for his help with the Boston Housing dataset experiment. The research was partially supported by the Office of 
Naval Research ({ONR-N000141512342, ONR-N000141812298}), National Science Foundation ({NSF-IIS-1718258}). 

\section{Convex Relaxation of Problem~\eqref{cubic-spline-add-1-2}}\label{relax-1}
Consider Problem~\eqref{cubic-spline-add-1-2} and suppose that the solution to this problem is bounded. Moreover, we assume that the
$\ell_{2}$-norms of every group $\M{f}_{j}$ satisfies: $\| \M{f}_{j}\|_{2} \leq \MU$.
Then it follows that the problem is equivalent to:
 \begin{equation}\label{cubic-spline-add-1-3}
\min~~ \| \M{y}  - \sum_{j=1}^{q} \M{f}_{j} \|_{2}^2 + \lambda_0  \sum_{j \in [q]} z_{j} + \lambda \sum_{j=1}^{q} \| \M{f}_{j} \|_{C_j}  ~~~ \sbt ~~~ \| \M{f}_{j} \|_{2} \leq \MU z_{j}, z_{j} \in \{0, 1\}, j \in [q].
\end{equation}
Relaxing the $z_j$'s in the above to $[0,1]$, leads to the following formulation:
\begin{equation}\label{cubic-spline-add-1-5}
\min~~ \| \M{y}  - \sum_{j=1}^{q} \M{f}_{j} \|_{2}^2 + \lambda \sum_{j=1}^{q} \| \M{f}_{j} \|_{C_j}  + \lambda_1 \sum_{j=1}^{q}  \| \M{f}_{j} \|_{2} ~~~ \sbt ~~~ \| \M{f}_{j} \|_{2} \leq \MU
\end{equation}
where $\lambda_1 := \frac{\lambda_0}{\MU}$. Next, we (i) drop the constraints in the above,  and (ii) rewrite the resulting problem as follows:
 \begin{equation}\label{cubic-spline-add-1-51}
\Gamma_{1}:=\min~~ \| \M{y}  - \sum_{j=1}^{q} \M{f}_{j} \|_{2}^2 + \lambda (\sum_{j=1}^{q} \| \M{f}_{j} \|_{C_j}  + \frac{\lambda_{1}}{\lambda} \sum_{j=1}^{q}  \| \M{f}_{j} \|_{2}).
\end{equation}
Note that \eqref{cubic-spline-add-1-51} is a relaxation of \eqref{cubic-spline-add-1-3} (and consequently of \eqref{cubic-spline-add-1-2}). Now, using the fact that
$$  (\| \M{f}_{j} \|_{C_j} +  \frac{\lambda_{1}}{\lambda} \| \M{f}_{j} \|_{2})/ \sqrt{2}  \leq  \sqrt{\| \M{f}_{j} \|^2_{C_j} + \left(\frac{\lambda_{1}}{\lambda}\right)^2 \| \M{f}_{j} \|^2_{2}} ,$$
it follows that the following
 \begin{equation}\label{cubic-spline-add-1-6}
\Gamma_{2} := \min~~ \| \M{y}  - \sum_{j=1}^{q} \M{f}_{j} \|_{2}^2 + \sqrt{2}\lambda  \sum_{j=1}^{q} \left( \sqrt{\| \M{f}_{j} \|^2_{C_j} +   \left(\frac{\lambda_{1}}{\lambda}\right)^2 \| \M{f}_{j} \|^2_{2}}\right),
\end{equation}
is an upper bound to Problem~\eqref{cubic-spline-add-1-51} (with the tuning parameters kept fixed).
Note that Problem~\eqref{cubic-spline-add-1-6} is indeed the penalty considered in~\cite{meier1}, with the choice of
$\ \text{Pen}(f_{j}) = \sqrt{\| \M{f}_{j} \|^2_{C_j} +   {\lambda'} \| \M{f}_{j} \|^2_{2}}$, where $\lambda'$ is appropriately chosen to match~\eqref{cubic-spline-add-1-6}.

We note that the penalty chosen in formulation~\eqref{cubic-spline-add-1-51} is similar to the penalty considered in~\cite{raskutti2012minimax}, wherein the authors consider an RKHS framework with penalization:
$$  \lambda \sum_{j=1}^{q} \sqrt{\B\beta_{j}' \M{K}^{j} \B\beta_{j}} + \lambda' \sum_{j=1}^{q} \| \M{K}^{j} \B\beta_{j}\|_{2},$$
where $\M{K}^{j}$ indicates the kernel basis matrix for the $j$th coordinate.

\begin{appendix}
\renewcommand\thesection{\Alph{section}}
\renewcommand{\theequation}{\thesection.\arabic{equation}}
\setcounter{equation}{0}
\renewcommand{\thefigure}{\thesection.\arabic{figure}}
\setcounter{figure}{0}
\renewcommand{\thetable}{\thesection.\arabic{table}}
\setcounter{table}{0}

\section{Proofs}

\subsection{Proof of Theorem \ref{theorem:CDConvergence}}

The following lemma shows that there is a sufficient decrease in the objective after every group update in Algorithm 1. The result of this lemma will be used in the  proof of Theorem \ref{theorem:CDConvergence}.
\begin{lem}{(Sufficient Decrease)} \label{lemma:suffdec}
The sequence of iterates $\{\B\theta^{l}\}$ in Algorithm 1 satisfies the following for every $l$ and $g = 1 + (l\mod q)$:
\begin{align}
    h(\B\theta^{l}) - h(\B\theta^{l+1}) \geq \frac{\hat{L}_g - L_g}{2} \|\B\theta^{l}_g - \B\theta^{l+1}_g  \|^2_2.
\end{align}
\end{lem}

\textbf{Proof of Lemma \ref{lemma:suffdec}}. 
Fix some $l \geq 0$ and let $g = 1 + (l\mod q)$. Applying \eqref{eq:groupdescent} to $(\B\theta^{l+1}, \B\theta^{l})$ and adding $\Omega(\B\theta^{l+1})$ to both sides, we get:
\begin{align}
        h(\B\theta^{l+1}) \leq  \ell({\B\theta}^{l}) + \langle \nabla_{\B\theta_g} \ell({\B\theta}^{l}), \B\theta^{l+1}_g - {\B\theta}_g^{l} \rangle + \frac{{{L}}_g}{2} \|\B\theta^{l+1}_g - {\B\theta}^{l}_g \|_2^2 + \Omega(\B\theta^{l+1}).
\end{align}
By rewriting the term $\frac{{{L}}_g}{2} \|\B\theta^{l+1}_g - {\B\theta}^{l}_g \|_2^2$ in the above as $\frac{{{L}}_g - \hat{L}_g }{2} \|\B\theta^{l+1}_g - {\B\theta}^{l}_g \|_2^2 + \frac{{\hat{L}}_g}{2} \|\B\theta^{l+1}_g - {\B\theta}^{l}_g \|_2^2$ and regrouping terms, we get:
\begin{align} \label{eq:suffdecbd}
     h(\B\theta^{l+1}) \leq \tilde{g}(\B\theta^{l+1}; \B\theta^{l}) + \frac{{{L}}_g - \hat{L}_g }{2} \|\B\theta^{l+1}_g - {\B\theta}^{l}_g \|_2^2.
\end{align}
But $\tilde{g}(\B\theta^{l+1}; \B\theta^{l}) \leq \tilde{g}(\B\theta^{l}; \B\theta^{l})$ (by the definition of $\B\theta^{l+1}$ in \eqref{eq:nextiterate}). Moreover, $\tilde{g}(\B\theta^{l}; \B\theta^{l}) = h(\B\theta^{l})$, which implies $\tilde{g}(\B\theta^{l+1}; \B\theta^{l}) \leq h(\B\theta^{l})$. Using the latter bound in \eqref{eq:suffdecbd}, we arrive to the result of the lemma.

\smallskip

\textbf{Proof of the theorem.}
In the rest of this proof, we utilize the following definition: $E(\B\theta_S) := \ell(\B\theta_S) + \lambda_1 \sum_{g \in S} \| \B\theta_{g}\|_{2}$. 
\begin{itemize}
    \item \textbf{Part 1.} We will show that the event $\text{Supp}(\B\theta^{l}) \neq \text{Supp}(\B\theta^{l+1})$ cannot happen infinitely often. Suppose that $\text{Supp}(\B\theta^{l}) \neq \text{Supp}(\B\theta^{l+1})$ holds for some $l$. Then, either one of the following cases must hold for $g = 1 + (l\mod q)$: (I) $\B\theta^{l}_g = 0 \neq \B\theta^{l+1}_g$ or (II) $\B\theta^{l}_g \neq 0 = \B\theta^{l+1}_g$. Next, we will consider Case (I). Since $\B\theta^{l+1}_g \neq 0$, then from the definition of the thresholding operator in \eqref{eq:thresholdingop}, we have $\| \B\theta^{l+1}\|_2 > \sqrt{\frac{2 \lambda_0}{\hat{L}_g}}$. Plugging the latter inequality into Lemma \ref{lemma:suffdec}, we get:
    \begin{align} \label{eq:changesuppobjdec}
    h(\B\theta^{l}) - h(\B\theta^{l+1}) \geq \frac{\hat{L}_g - L_g}{\hat{L}_g} \lambda_0.
    \end{align}
    The same result in \eqref{eq:changesuppobjdec} applies for Case (II) as well. Thus, whenever the support changes, the objective improves by a positive constant (defined in the r.h.s of \eqref{eq:changesuppobjdec}), which combined with the fact that $h(\B\theta) \geq 0$, implies that the support cannot change infinitely often.
    
    \item \textbf{Part 2.} First, we will show that the function $E(\B\theta_S)$ is strongly convex. This trivially holds under Assumption \ref{CDAssumptions}(a). Next, we will assume that only Assumption \ref{CDAssumptions}(b) is satisfied. In this case, we have $h(\B\theta^{0}) \leq h(\hat{\B\theta})$ (where $\hat{\B\theta}$ is defined in Assumption \ref{CDAssumptions}(b)). Since Algorithm 1 is a descent algorithm, we have $h(\B\theta^{l}) \leq h(\hat{\B\theta})$ for all $l \geq 0$. Thus, $E(\B\theta^{l}) + \lambda_0 G(\B\theta^{l})  \leq E(\hat{\B\theta}) + \lambda_0 G(\hat{\B\theta})$, which combined with the fact that $ E(\B\theta^{l}) \geq E(\hat{\B\theta})$, implies that $G(\B\theta^{l}) \leq G(\hat{\B\theta})$ for all $l$. Thus, by the definition of $k$ in the assumption, we have $\| \B\theta^{l} \|_0 \leq k$ for all $l$. But since every $k$ columns in $\M W$ are linearly independent, we conclude that $E(\B\theta_S)$ is strongly convex.
    
    After the support stabilizes (by Part 1), Algorithm 1 becomes equivalent to minimizing the strongly convex function $E(\B\theta_S)$ using cyclic CD. By standard results on CD (e.g., see \cite{bertsekas2016nonlinear}), this is guaranteed to converge to a stationary solution $\B\theta^{*}$ of $E(\B\theta_S)$. This establishes \eqref{eq:betastar}.
    
    Finally, we will show that \eqref{eq:insuppineq} and \eqref{eq:outsuppineq} hold. By the definition of the thresholding operator in \eqref{eq:thresholdingop}, we have
    \begin{align}
        \| \B\theta^{l}_g\|_2 > \sqrt{\frac{2 \lambda_0}{\hat{L}_g}}, \quad \forall g \in S.
    \end{align}
    Taking the limit as $l \to \infty$, we arrive to \eqref{eq:insuppineq}. Similarly, we have 
    \begin{align}
        \| \nabla_{\B\theta_g} \ell(\B\theta^{l}) \|_2 \leq \sqrt{2 \lambda_0 \hat{L}_g} + \lambda_1, \quad \forall g \in S^c.
    \end{align}
    Taking the limit $l \to \infty$ leads to \eqref{eq:outsuppineq}.
    
    \item \textbf{Part 3.} 
    After support stabilization, Algorithm 1 is equivalent to performing cyclic CD to minimize the function $E(\B\theta_S)$. Moreover, every iterate of the algorithm after support stabilization, i.e., $\B \theta^{l}_{S}$ for $l \geq K$, belongs to the set $D := \{\B\theta_S \ | \ \| \B\theta_S \|_2 \geq  \sqrt{\frac{2 \lambda_0}{\hat{L}_g}} \}$ (this follows from \eqref{eq:thresholdingop}). Note that $\nabla_{\B\theta_S} E(\B\theta_S)$ is group-wise Lipschitz continuous over $D$, i.e., the following holds for every $g \in [q]$:
    $$\| \nabla_{\B\theta_S} E(\B\theta^{1}_S) - \nabla_{\B\theta_S} E(\B\theta^{2}_S) \|_2 \leq \tilde{L}_g \|\B\theta^{1}_S - \B\theta^{2}_S \|_2, ~~ \forall  \B\theta^{1}_S, \B\theta^{2}_S \in D \text{ s.t. } \B\theta^{1}_i = \B\theta^{2}_i ~  \forall i \neq g $$
    where $\tilde{L}_g = \hat{L}_g + 2 \lambda_1$.
    Similarly, $\nabla_{\B\theta_S} E(\B\theta_S)$ has a (global) Lipschitz constant of $L_S + 2 |S| \lambda_1 $, over $D$. 
    
    Lemma 3.3 of \cite{BeckConvergence} bounds the objective values of cyclic CD after one full cycle. Their result holds for continuously differentiable functions whose gradient is Lipschitz over $\mathbb{R}^n$. Our function's gradient is Lipschitz over $D$, but we note that \cite{BeckConvergence}'s result can be easily extended to $D$, leading to the following bound:
    \begin{align} \label{eq:objimpbd}
        E(\B\theta_S^{lq}) - E(\B\theta_S^{(l+1)q}) \geq \frac{1}{2\eta} \| \nabla_{\B\theta_S} E(\B\theta_S^{lq}) \|_2^2, \quad \forall \ l \geq K,
    \end{align}
    where $\eta$ is defined in the statement of the theorem. In part 2, we have shown that $E(\B\theta_S)$ is strongly convex. Thus, the following holds:
    \begin{align} \label{eq:strongconvexityS}
        E(\B\alpha_S) \geq E(\B\theta_S) + \langle \nabla E(\B\theta_S), \B\alpha_S - \B\theta_S \rangle + \frac{\sigma_{S}}{2} \| \B\alpha_S - \B\theta_S \|_2^2, \quad \forall \B\alpha_S, \B\theta_S.
    \end{align}
    Minimizing both sides in \eqref{eq:strongconvexityS} w.r.t. $\B\alpha_S$ and rearranging terms, we get
    \begin{align} \label{eq:strongconvmin}
        E(\B\theta_S) - E(\B\theta_S^{*}) \leq \frac{1}{2 \sigma_S}  \|\nabla_{\B\theta_S} E(\B\theta_S) \|_2^2, \quad \forall \B\theta_S.
    \end{align}
    Inequalities \eqref{eq:objimpbd} and \eqref{eq:strongconvmin} lead to:
    \begin{align}
        (E(\B\theta_S^{lq}) - E(\B\theta_S^{*})) - (E(\B\theta_S^{(l+1)q}) - E(\B\theta_S^{*})) & \geq \frac{1}{2\eta} \| \nabla_{\B\theta_S} E(\B\theta_S^{lq}) \|_2^2 \\
        & \geq \frac{\sigma_S}{\eta} (E(\B\theta_S^{lq}) - E(\B\theta_S^{*})).
    \end{align}
    Rearranging the terms in the above yields:
    \begin{align}
        E(\B\theta^{(l+1)q}) - E(\B\theta^{*}) \leq \Bigg(1 - \frac{\sigma_S}{\eta} \Bigg)  \Big( E(\B\theta^{lq}) - E(\B\theta^{*}) \Big).
    \end{align}
    Finally, we note that the function $E$ in the above can be replaced by $h$ (because of support stabilization), which establishes part 3.
\end{itemize}

\subsection{Proof of Theorem \ref{theorem:localsearch}}
By Theorem \ref{theorem:CDConvergence}, the support of the iterates in Algorithm 1 stabilizes, say on a support $S$, and converges to a solution of $\min_{\B{\theta}, \text{Supp}(\theta)=S} h(\B{\theta})$. The latter observation along with the fact that Step 2 of Algorithm 2 ensures strict descent, imply that the sequence of solutions $\B\theta^{t}$ in Algorithm 2 must have distinct supports. Therefore, the algorithm terminates in a finite number of iterations. Note that $\B\theta^{\dagger}$ is the output of Algorithm 1 so it must satisfy the characterization given in part 2 of Theorem \ref{theorem:CDConvergence}. Moreover, the search in Step 2 must fail at $\B\theta^{\dagger}$, and thus \eqref{eq:inescapable} holds.

\subsection{Proof of Proposition \ref{prop:relaxation_quality}}
Let $F_1(\B\theta, \B z)$ and $F_2(\B\theta, \B z, \B{s})$ be the objective functions in \eqref{eq:MIP_BnB} and \eqref{eq:MIP_hybrid}, respectively. Note that by definition, $v_2 = F_2(\B\theta^{*}, \B z^{*}, \B{s}^{*})$. Since $(\B\theta^{*}, \B z^{*})$ is feasible for the problem corresponding to $v_1$, we have:
\begin{align} \label{prop1_proof}
v_2 - v_1 \geq F_2(\B\theta^{*}, \B z^{*}, \B{s}^{*}) - F_1(\B\theta^{*}, \B z^{*})
\end{align}
Since $(\B\theta^{*}, \B z^{*}, \B{s}^{*})$ is optimal for the problem of $v_2$, it must satisfy  $s_g^{*} = 0$ if $z_g^{*} = 0$ and $s_g^{*} = \frac{\| \B\theta^{*}_g \|_2^2}{z_g^{*}}$ otherwise (because this is the smallest value of $s_g$, which satisfies \eqref{eq:cone_constraint}). Plugging $s_g^{*}$ into the term $F_2(\B\theta^{*}, \B z^{*}, \B{s}^{*})$ in \eqref{prop1_proof} and simplifying, leads to the result of the proposition.

\subsection{Proof of Proposition \ref{prop:reformulation}}
The root relaxation of \eqref{eq:MIP_hybrid} can be written as:
\begin{align} \label{eq:equivalent_relaxation_proof}
    \min_{\B\theta} ~~~  & \Bigg \{ \tilde{\ell}(\B\theta) + \lambda_1 \sum_{g=1}^{q} \| \B\theta_{g}\|_{2} + \sum_{g=1}^q \min_{z_g, s_g} (\lambda_0 z_g + \lambda_2 s_g) \Bigg\}  \\
    \text{s.t.}~~& \|\B\theta_{g}\|_2 \leq \MU z_{g},~ g \in [q] \label{eq:bigM_constraint}\\
    & s_g z_g \geq \|\B\theta_{g}\|_2^2, ~~ g \in [q] \label{eq:perspective_constraint} \\ 
    & z_g \in \left[0,1\right], s_g \geq 0, ~~ g \in [q] \label{eq:vars_constraint}
\end{align}
Define 
\begin{align} \label{eq:omega_g}
\omega(\B\theta_g; \B\lambda, \MU) = \min_{z_g, s_g} (\lambda_0 z_g + \lambda_2 s_g) ~~~\text{s.t.}~~~\eqref{eq:bigM_constraint}, \eqref{eq:perspective_constraint}, \eqref{eq:vars_constraint}.
\end{align}
Note that the above optimization problem appears inside the second summation of~\eqref{eq:equivalent_relaxation_proof}. Next, we will derive a closed form expression for \eqref{eq:omega_g}.  Let  $(\B\theta_g, z_g, s_g)$ be some  feasible solution. Then, the solution $(\B\theta_g, \hat{z}_g, s_g)$, where $\hat{z}_g = \max \{ \frac{\| \B\theta_g \|_2^2}{s_g}, \frac{\| \B\theta_g \|_2}{\MU}\}$, has an objective value which is less than or equal to that of $(\B\theta_g, z_g, s_g)$ (since $\hat{z}_g$ is the smallest possible choice of $z_g$ which satisfies all the constraints)---if $\B\theta_g = \M 0$ and $s_g = 0$, we assume that $\frac{\| \B\theta_g \|_2^2}{s_g} = 0$, which leads to $\hat{z}_g = 0$. Thus, replacing constraints \eqref{eq:bigM_constraint} and  \eqref{eq:perspective_constraint} with the constraint $z =  \max \{ \frac{\| \B\theta_g \|_2^2}{s_g}, \frac{\| \B\theta_g \|_2}{\MU}\}$ does not change the optimal objective of the problem. This replacement leads to the following equivalent problem:
\begin{align} 
\omega(\B\theta_g; \B\lambda, \MU) = \min_{z_g, s_g} (\lambda_0 z_g + \lambda_2 s_g) ~~~\text{s.t.}~~~ {z}_g = \max \Big \{ \frac{\| \B\theta_g \|_2^2}{s_g}, \frac{\| \B\theta_g \|_2}{\MU} \Big\}, z_g \in [0,1], s_g \geq 0.
\end{align}
In the above, we can eliminate $z_g$ by plugging its expression into  the the objective and the constraint $z_g \in [0,1]$, which leads to the following equivalent formulation:
\begin{align} \label{eq:omega_simplfied}
    \omega(\B\theta_g; \B\lambda, \MU) = \min_{s_g} \max \Bigg\{ \underbrace{\frac{\lambda_0 \| \B\theta_g \|_2^2}{s_g} + \lambda_2 s_g}_{\text{Term 1}}, \underbrace{ \frac{\lambda_0 \| \B\theta_g \|_2}{\MU} + \lambda_2 s_g}_{\text{Term 2}} \Bigg\} ~~ s.t. ~~ s_g \geq \| \B\theta_g \|_2^2,  \| \B\theta_g \|_2 \leq \MU.
\end{align}
Suppose that Term 1 in \eqref{eq:omega_simplfied} attains the maximum. This holds iff Term 1 $\geq$ Term 2, which simplifies to: $s_g \leq \MU \| \B\theta_g \|_2$. Term 1 is convex in $s_g$, so the solution of \eqref{eq:omega_simplfied} (obtained via solving the first order optimality condition, assuming $s_g \leq \MU \| \B\theta_g \|_2$) is given $s_g^{*} = \sqrt{\lambda_0/\lambda_2} \| \B\theta_g \|_2$ if $\| \B\theta_g \|_2 \leq \sqrt{\lambda_0/\lambda_2} \leq M$, and $s_g^{*} = \| \B\theta_g \|_2^2 $ if $ \sqrt{\lambda_0/\lambda_2} \leq \| \B\theta_g \|_2 \leq M$. Plugging $s_g^{*}$ into \eqref{eq:omega_simplfied}, leads to $\omega(\B\theta_g; \B\lambda, \MU) = 2 \lambda_0 \mathcal{H} (\sqrt{\lambda_2/\lambda_0} \| \B\theta_g \|_2)$, for $ \sqrt{\lambda_0/\lambda_2} \leq \| \B\theta_g \|_2 \leq \MU$.
\\ \\
Now suppose Term 2 attains the maximum in \eqref{eq:omega_simplfied}. 
There are two lower bounds on $s_g$ in this case: $s_g \geq \MU \| \B\theta_g \|_2$ (from Term 1 $\leq$ Term 2) and $s_g \geq \| \B\theta_g \|^2_2$ (from the feasible set in \eqref{eq:omega_simplfied}). Since $\| \B\theta_g \|_2 \leq \MU$, we have $\MU \| \B\theta_g \|_2 \geq \| \B\theta_g \|^2_2$, which implies that $s_g \geq \MU \| \B\theta_g \|_2$ is the only lower bound needed. Thus, we can simplify  \eqref{eq:omega_simplfied} to:
$$
    \omega(\B\theta_g; \B\lambda, \MU) = \min_{s_g}  \frac{\lambda_0 \| \B\theta_g \|_2}{\MU} + \lambda_2 s_g ~~ s.t. ~~ s_g \geq \MU \| \B\theta_g \|_2,  \| \B\theta_g \|_2 \leq \MU.
$$
The optimal solution of the above is given by $s_g^{*} = \MU \| \B\theta_g \|_2$, and this holds for $\sqrt{\lambda_0/\lambda_2} \geq \MU$. Plugging $s_g^{*}$ into \eqref{eq:omega_simplfied} leads to $\omega(\B\theta_g; \B\lambda, \MU) = (\lambda_0/\MU + \lambda_2 \MU) \| \B\theta_g \|_2$, for $\sqrt{\lambda_0/\lambda_2} \geq \MU$. Finally, we replace the inner minimization in \eqref{eq:equivalent_relaxation_proof} by the closed form expression of $\omega(\B\theta_g; \B\lambda, \MU)$, which leads to the result of the proposition.



\subsection{Proof of Proposition~\ref{prop.ineqs1}}

Because $\kappa_{k,c}\ge \kappa_{k,1}$ for $c\ge1$, it is sufficient to derive the stated inequality for $c=1$.

We consider an arbitrary $\bbeta$ satisfying $\bbeta\ne\M{0}$ and $G(\bbeta)\le 2k$.  We let~$J_0\subseteq[q]$ index the~$k$ largest values in the set $\{\|\bbeta_{g}\|_{2,1}\}_{g\in [q]}$, noting that $|J_0|=k$ and $\|\bbeta_{J_0^c}\|_{2,1}\le \|\bbeta_{J_0}\|_{2,1}$.  The stated inequality follows from an observation that
\begin{equation*}
\frac{\sqrt{2k}\|\M{X}\bbeta\|_2}{\sqrt{n}\|\bbeta\|_{2,1}}\ge 
\frac{\sqrt{2k}\|\M{X}\bbeta\|_2}{2\sqrt{n}\|\bbeta_{J_0}\|_{2,1}}\ge
\frac{\kappa_{k,1}}{\sqrt{2}}.
\end{equation*}
\qed

\subsection{Proof of Theorem~\ref{gen.thm}}

Optimality of~$\widehat\bbeta$ and feasibility of~$\bbeta^*$ imply  $\|\bY-\M{X}\widehat\bbeta\|_2^2\le\|\bY-\M{X}\bbeta^*\|_2^2$, which leads to
\begin{equation}
\label{basic.ineq}
\|\M{X}(\widehat\bbeta-\bbeta^*)\|^2_2\le 2\bepsilon^\top\M{X}(\widehat\bbeta-\bbeta^*).
\end{equation}
We will derive a bound for the right hand side of inequality~\eqref{basic.ineq}.  

First, we consider a fixed subset $J\subseteq [q]$ such that $|J|=2k$.  We define $I_J=\cup_{g\in J}\mathcal{G}_g$ and $s=\bar{T}_k k$, noting that $|I_J|\le 2s$.
We choose an orthonormal basis $\M{\Phi}=[{\boldsymbol \phi}_1,...,{\boldsymbol \phi}_{2s}]$, such that the corresponding linear space contains the one spanned by features $\{\M{x}_j\}_{j\in I_J}$. Then, $\|\M{\Phi}^\top\bepsilon\|_2^2/\sigma^2$ has chi-square distribution with at most~$2s$ degrees of freedom, and
\begin{equation*}
\bepsilon^\top\M{X}\btheta \le \|\M{\Phi}^\top\bepsilon\|_2 \|\M{X}\btheta\|_2,
\end{equation*}
for all $\btheta\in\mathbb{R}^p$ with $\text{supp}(\btheta)\subseteq I_J$. Applying a chi-square tail bound \citep[for example, the one in Section 8.3.2 of][]{Bbuhl1}, we derive that $|\M{\Phi}^\top\bepsilon|^2 \lesssim  \sigma^2 s(1+a)$ with probability at least $1-\exp(-2s a)$. Consequently, with probability at least $1-\exp(-2s a)$, inequality
\begin{equation}
\label{fixed.S.bnd}
\bepsilon^T\M{X}\btheta \lesssim   \Big[ \sigma^2 s(1+a) \Big]^{1/2} \|\M{X}\btheta\|_2
\end{equation}
holds uniformly for all $\btheta\in\mathbb{R}^{p}$ with $\text{supp}(\btheta)\subseteq I_J$.

We now extend this bound to \textit{all} subsets $J\subseteq [q]$ that have size $2k$.  Note that the number of such subsets is bounded by $(q e/2k)^{2k}$.   Applying the union bound, we deduce that inequality~(\ref{fixed.S.bnd}) holds uniformly over both such~$J$ and~$\btheta$ with probability at least $1-\exp(-2s a+2k\log(qe/2k))$.  We note that $G(\widehat \bbeta - \bbeta^*)\le 2k$ and take $a=\bar{T}_k^{-1}\log(eq/2k)+[2s]^{-1}\log(1/\delta_0)$.  It follows that
\begin{equation}
\label{ineq.Gk}
\bepsilon^T\M{X}(\widehat\bbeta-\bbeta^*)\lesssim  \Big[\sigma^2 k[\bar{T}_k+\log( eq /k)]+ \sigma^2\log(1/\delta_0)\Big]^{1/2}\|\M{X}(\widehat\bbeta-\bbeta^*)\|_2,
\end{equation}
with probability at least $1-\delta_0$.  We complete the proof by combining the above bound with inequality~(\ref{basic.ineq}). \qed

{\color{black}
\subsection{Proof of Corollary~\ref{cor.opt.gap}}

To simpify the presentation, we define $L(\bbeta)=\| \M{y} - \M{X} \B\beta \|_{2}^2$. Because $UB=L(\widetilde{\bbeta})$, $LB\le L(\bbeta^*)$, and $UB=LB/(1-\tau)$, we derive
\begin{equation*}
L(\widetilde{\bbeta}_2)\le L(\bbeta^*)/(1-\tau).
\end{equation*}
As $L(\bbeta^*)= \|\bepsilon\|^2_2$, we can rewrite the above inequality as follows:
\begin{equation*}
\|\bY-\bX\widetilde{\bbeta}_2\|^2_2\le \|\bepsilon\|^2/(1-\tau).
\end{equation*}
Repeating the steps in the proof of Theorem~\ref{gen.thm}, taking ito account the optimality gap, and letting $\delta_0=(k/q)^k/2$ we arrive at inequality
\begin{equation*}
\|\M{X}(\widetilde{\bbeta}-\bbeta^*)\|^2_2\lesssim \sigma^2 k[\bar{T}_k+\log( eq /k)]+\|\bepsilon\|^2\tau/(1-\tau),
\end{equation*}
which holds with probability at least $1-(k/q)^k/2$. Standard chi-square tail bounds \citep[for example, those in Section 8.3.2 of][]{Bbuhl1} imply that, wich an appropriate multiplicative constant, inequality $\|\bepsilon\|^2/n\lesssim \sigma^2[1\vee k\log(eq/k)/n]$ holds with probability at least $1-(k/q)^k/2$. Because $\tau\le1$ and $1/(1-\tau)$ is upper-bounded by a universal constant, we then conclude that inequality
\begin{equation*}
\frac1n\|\M{X}(\widetilde{\bbeta}-\bbeta^*)\|^2_2\lesssim \sigma^2 k\Big[\frac{\bar{T}_k+\log( eq /k)}{n}\Big]+\sigma^2\tau
\end{equation*}
holds with probability at least $1-(k/q)^k$. 
}

\subsection{Proof of Corollary~\ref{lin.cor2}}

We let~$c_0$ be the universal constant from the error bound in Theorem~\ref{gen.thm} and define
\begin{equation*}
W=\|\M{X}\widehat\bbeta-\M{X}\bbeta^*\|^2_2 - c_0\sigma^2k\big[\bar{T}_k+\log( q /k)\big].
\end{equation*}
By Theorem~\ref{gen.thm} we have $W\le c_0 \sigma^2\log(1/\delta_0)$ with probability at least $1-\delta_0$.  Hence,
\begin{equation*}
\mathbb{P}\big(W>w\big)\le e^{-w/[c_0\sigma^2]},
\end{equation*}
for every non-negative~$w$.  Consequently,
\begin{equation*}
\mathbb{E}W \le \int_0^{\infty}\mathbb{P}\big(W>w\big)dw \le \int_0^{\infty}e^{-w/[c_0\sigma^2]}dw \le c_0\sigma^2.
\end{equation*}
Thus, by the definition of~$W$, we have
\begin{equation*}
\mathbb{E}\|\M{X}\widehat\bbeta-\M{X}\bbeta^*\|^2_2 \le c_0\sigma^2k\big[\bar{T}_k+\log( q /k)\big]+c_0\sigma^2,
\end{equation*}
which implies the bound in the statement of Corollary~\ref{lin.cor2}.
\qed

{\color{black}

\subsection{Proof of Theorem~\ref{BIC.thm}}

We let $B(\bbeta)=G(\bbeta)[\check{T}+\log(q/G(\bbeta))]$ to simplify the presentation. Using the definitions of~$\widehat\bbeta_k$ and~$\widehat\bbeta^B$, we derive
\begin{equation*}
\|\bY-\M{X}\widehat\bbeta^B\|^2_2 + aB(\widehat\bbeta^B) \le \|\bY-\M{X}\widehat\bbeta^*\|^2_2+aB(\bbeta^*),
\end{equation*}
which implies
\begin{equation}
\label{basic.ineq.BIC}
\|\M{X}(\widehat\bbeta^B-\bbeta^*)\|^2_2 + aB(\widehat\bbeta^B) \le 2\bepsilon^\top\M{X}(\widehat\bbeta^B-\bbeta^*)+aB(\bbeta^*).
\end{equation}

Revisiting the derivation of inequality~(\ref{ineq.Gk}) in the proof of Theorem~\ref{gen.thm}, we note that~$\widehat\bbeta$ only played a role through its sparsity bound $G(\widehat\bbeta)\le k$. Thus, the corresponding probability lower-bound applies to the event of inequality~(\ref{ineq.Gk}) holding for each~$\bbeta$ with $G(\bbeta)\le k$, rather than just~$\widehat\bbeta$.  We let~$\delta_0=(k/[eq])^k\epsilon_0$, where $\epsilon_0\in(0,1)$ is an arbitrary value, and conclude that for each $k\in[q]$, event
\begin{equation*}
\mathcal{A}_k=\left\{\frac{\bepsilon^T\M{X}(\bbeta-\bbeta^*)}{\|\M{X}(\bbeta-\bbeta^*)\|_2}\lesssim  \sigma\sqrt{ B(\bbeta-\bbeta^*)+ \log(1/\epsilon_0)},\;\forall \bbeta\in\mathbb{R}^p \;\text{s.t.}\; G(\bbeta)\le k\right\}
\end{equation*}
holds with probability at least~$1-\epsilon_0$. We note that neither~$\epsilon_0$ nor the universal multiplicative constant in the definition of~$\mathcal{A}_k$ depends on~$k$. We define $\mathcal{A}=\cap_{k=1}^q\mathcal{A}_k$ and note that
\begin{equation*}
\mathbb{P}(\mathcal{A}^c)\le\sum_{k=1}^q\mathcal{P}(\mathcal{A}_k^c)\le\sum_{k=1}^q(k/[eq])^k\epsilon_0\le\sum_{k=1}^q e^{-k}\epsilon_0\le \epsilon_0.
\end{equation*}
On the event~$\mathcal{A}$, we have
\begin{eqnarray*}
\bepsilon^\top\bX(\widehat\bbeta^B-\bbeta^*) &\lesssim& \sigma\sqrt{B(\widehat\bbeta^B-\bbeta^*)+\log(1/\epsilon_0)}\|\bX(\widehat\bbeta^B-\bbeta^*)\| \\
&\lesssim&  \sigma\sqrt{B(\widehat\bbeta^B)+B(\bbeta^*)+\log(1/\epsilon_0)}\|\bX(\widehat\bbeta^B-\bbeta^*)\|.
\end{eqnarray*}
Consequently, making the universal constant~$a$ in the BIC penalty sufficiently large and using inequality~(\ref{basic.ineq.BIC}), we deduce that
\begin{equation}
\label{or.bnd.bic}
\|\M{X}\widehat\bbeta^B-\M{X}\widehat\bbeta^*\|^2+\sigma^2B(\widehat\bbeta^B)\lesssim \sigma^2B(\bbeta^*)+\sigma^2\log(1/\epsilon_0)
\end{equation}
with probability at least $1-\epsilon_0$.
Repeating the argument in the proof of Corollary~\ref{lin.cor2}, we derive
\begin{equation}
\label{exp.bnd.pen2}
\mathbb{E}\|\M{X}\widehat\bbeta^B-\M{X}\widehat\bbeta^*\|^2\lesssim  \sigma^2B(\bbeta^*)+\sigma^2,
\end{equation}
which establishes the prediction error bound in the statement of Theorem~~\ref{BIC.thm}.

It is only left to derive the group sparsity bound for~$\widehat\bbeta^B$. Treating expressions of the form $0\cdot\infty$ as~$0$, we define function $b(x)=x[\check{T}+\log(q/x)]$ for $x\in[0,q]$ and note that~$b(x)$ is monotone increasing with $b(x)\ge x$ for all~$x\in[0,q]$. We note that inequality~(\ref{or.bnd.bic}) implies  $b(\hat k)\lesssim b(k^*\vee1)[1+\log(1/\epsilon_0)]$. Further exploiting the properties of deterministic function~$b(x)$, we can then deduce that $\hat k\lesssim[1+\log(1/\epsilon_0)]^2(k^*\vee1)$ with probability at least $1-\epsilon_0$. Following the argument in the proof of Corollary~\ref{lin.cor2}, we derive the bound $\mathbb{E}\hat k/(k^*\vee1)\lesssim 1+\int_0^{\infty}e^{-w^{1/2}}dw\lesssim 1$.
\qed

\subsection{Proof of Theorem~\ref{slow.rt.thm}}

To simplify the presentation, we write $P(\bbeta)=\sum_{g \in [q]} \sqrt{T_g}\| \bbeta_{g}\|_{2}$ for the penalty function in optimization problem~(\ref{grp-l0-1-const1-card-l2}). 
By the optimality of~$\widehat\bbeta$, we have 
\begin{equation}
\label{basic.ineq.slow}
\|\M{X}(\widehat\bbeta-\bbeta^*)\|^2_2+ \lambda P(\widehat\bbeta)\le 2\bepsilon^\top\M{X}(\widehat\bbeta-\bbeta^*)+ \lambda P(\bbeta^*).
\end{equation}
We define $\widetilde{\btheta}_g=(\widehat\bbeta_g-\bbeta^*_g)/\|\widehat\bbeta_g-\bbeta^*_g\|_2$, write $\M{X}_g$ for the submatrix of~$\M{X}$ corresponding to the predictors in group~$g$, and observe the following inequalities:
\begin{equation}
\label{emp.proc.bnd.slow}
\bepsilon^\top\M{X}(\widehat\bbeta-\bbeta^*)\le
\sum_{g\in[q]}\|\bepsilon^\top\M{X}_g\widehat\btheta_g\|_2\|\widehat\bbeta_g-\bbeta^*_g\|_2\le \max_{g\in[q]}\left(\frac{\|\bepsilon^\top\M{X}_g\widehat\btheta_g\big\|_2}{\sqrt{T_g}}\right) P(\widehat\bbeta-\bbeta^*).
\end{equation}

Given a $\btheta\in \mathbb{R}^p$, we write $\theta_1^\sharp,...,\theta_p^\sharp$ for a non-increasing rearrangement of $|\theta_1|,...,|\theta_p|$. By Theorem~4.1 in \cite{bellec2018slope}, event
\begin{equation*}
\mathcal{F} = \big\{\bepsilon^\top\bX\btheta\le[4+\sqrt{2}]\sigma\max\big(\sum\nolimits_{j=1}^p \theta_j^\sharp\sqrt{n\log(2p/j)}\,,\,\sqrt{\log(1/\delta_0)}\|\bX\btheta\|_2\big), \, \forall\btheta\in \mathbb{R}^p\big\}
\end{equation*}
holds with probability at least~$1-\delta_0/2$. Using Stirling's formula together with Cauchy-Schwartz inequality, we derive $\sum_j\theta_j^\sharp\sqrt{n\log(2p/j)}\le\sqrt{n\|\btheta\|_0\log(2ep/\|\btheta\|_0)}\|\btheta\|_2$; we also note inequalities $\|\bX\btheta\|_2\le\sqrt{n}\|\btheta\|_1\le\sqrt{n\|\btheta\|_0}\|\btheta\|_2$, which rely on the normalization $\|\bx_j\|_2=\sqrt{n}$. Hence, as $\|\widehat\btheta_g\|_2=1$ for all $g\in[q]$, we deduce that event~$\mathcal{F}$ implies that 
\begin{equation*}
\max_g (T_g^{-1/2}\|\bepsilon^\top\M{X}_g\widehat\btheta_g\|_2)\le[4+\sqrt{2}]\sigma\sqrt{n}\max\big(\sqrt{\log(2e\tilde{q})},\sqrt{\log(1/\delta_0)}\big).    
\end{equation*}
Taking~$\delta_0=1/[2e\tilde{q}]$, we conclude that
$\max_g (T_g^{-1/2}\|\bepsilon^\top\M{X}_g\widehat\btheta_g\|_2)\le[4+\sqrt{2}]\sigma\sqrt{n\log(2e\tilde{q})}$ with probability at least $1-1/[4e\tilde{q}]$. Restricting our attention to the corresponding high-probability event, and taking into account inequalities~(\ref{basic.ineq.slow}) and~(\ref{emp.proc.bnd.slow}), we derive
\begin{equation*}
\|\M{X}(\widehat\bbeta-\bbeta^*)\|^2_2+ \lambda P(\widehat\bbeta)\le [8+2\sqrt{2}]\sigma\sqrt{n\log(2e\tilde{q})}P(\widehat\bbeta-\bbeta^*)+ \lambda P(\bbeta^*).
\end{equation*}
We note that $P(\widehat\bbeta-\bbeta^*)\le P(\widehat\bbeta)+P(\bbeta^*)$, chose the universal constant~$c_0$ in the statement of Theorem~\ref{slow.rt.thm} to satisfy $c_0>8+2\sqrt{2}$, and conclude that $\|\M{X}(\widehat\bbeta-\bbeta^*)\|^2_2\le 2\lambda P(\bbeta^*)$.
\qed

}

\subsection{Proof of Proposition~\ref{prop.ineqs2}}

Consider an arbitrary $f\in A_{2k,\xi}$.  Let $J_0$ be the index set corresponding to the $k$ components $f_j$ with the largest $\|\cdot\|_n$ norm.  Write~$r_n$ for $n^{-m/(2m+1)}$.  Note that
$$r_n \text{Pen}_g(f)\le (\xi/2-1/2)\sum\nolimits_{j=1}^q\|\bff_{j}\|_n\le (\xi-1)\sum_{j\in J_0}\|\bff_{j}\|_n,$$ and hence
\begin{equation*}
\sum_{j\notin J_0}\|\bff_{j}\|_n+r_n \text{Pen}_{\text{gr}}(f)\le \xi\sum_{j\in J_0}\|\bff_{j}\|_n.
\end{equation*}
Consequently, $f\in B(J_0,\xi)$.  To complete the proof, we note that
\begin{equation*}
\frac{\sqrt{2k}\|\bff\|_n}{\sum_{j=1}^q\|\bff_j\|_n}\ge \frac{\sqrt{2k}\|\bff\|_n}{2\sum_{j\in J_0}\|\bff_{j}\|_n}\ge \phi(k,\xi)/\sqrt{2}.
\end{equation*}
\qed

\subsection{Proof of Theorem~\ref{add.thm}}

By analogy with the $\|\cdot\|_n$ notation, we define $(\bepsilon,\bv)_n=(1/n)\sum_{i=1}^n\epsilon_i v_i$, for each $\bv\in\RR^n$.
%
The global optimality of~$\widehat f$, together with the feasibility of~$f^*$, implies the following inequality:
\begin{equation}
\label{basic.ineq.fn}
\|\widehat \bff-\bff^*\|_n^2+\lambda_n  \text{Pen}_{\text{gr}}(\widehat f) \le 2(\bepsilon,\widehat\bff - \bff^*)_n+\lambda_n  \text{Pen}_{\text{gr}}(f^*).
\end{equation}
To control the term $(\bepsilon,\widehat\bff - \bff^*)_n$ we need the following result, which is proved in Section~\ref{prf.lem.max.ineq.unif}.

\medskip

\begin{lem}
\label{lem.max.ineq.unif}
Let $\mathcal{F}_{s}=\{f: \; f\in\mathcal{C}_{\text{\rm gr}}, \; G(f)\le s\}$ and let~$\gamma$ be a fixed constant such that $r_n \le s^{\gamma}$.  Then, with probability at least $1-\epsilon$, inequality
\begin{eqnarray*}
(\bepsilon/\sigma,\bff)_n&\lesssim& \Big[ s^{1/2 + \gamma/(2m)}r_n + \sqrt{\frac{s\log(eq/s)}{n}} + \sqrt{\frac{\log(1/\epsilon)}{n}} \Big] \|\bff\|_n \\
&&+\Big[ s^{1/2-\gamma(2m-1)/(2m)}r_n^2 + s^{-\gamma}r_n\sqrt{\frac{s\log(eq/s)}{n}} + s^{-\gamma}r_n\sqrt{\frac{\log(1/\epsilon)}{n}} \Big] \text{\rm Pen}_{\text{\rm gr}}(f)
\end{eqnarray*}
holds uniformly over ${f\in\mathcal{F}_s}$.
\end{lem}

We now prove inequalities~\eqref{np.or.bnd.gen} and~\eqref{np.or.bnd.cc} in the statement of Theorem~\ref{add.thm}.

\medskip

\textbf{Proof of inequality~\eqref{np.or.bnd.gen}}.
We note that $G(\widehat f-f^*)\le2k$. Applying Lemma~\ref{lem.max.ineq.unif} with $f=\widehat f-f^*$, $s=2k$ and $\epsilon=(k/q)^k$, we conclude that, with probability at least $1-(k/q)^k$,
\begin{eqnarray}
(\bepsilon/\sigma,\widehat\bff - \bff^*)_n&\le& \tilde{c}_1k^{1/2}\Big[ k^{\gamma/(2m)}r_n + \sqrt{\frac{\log(eq/k)}{n}} \Big] \|\widehat\bff - \bff^*\|_n \nonumber\\
\label{max.ineq.fhat} && + (c_1/4)\Big[ k^{1/2-\gamma(2m-1)/(2m)}r_n^2 + k^{1/2-\gamma}r_n\sqrt{\frac{\log(eq/k)}{n}}  \Big] \text{\rm Pen}_{\text{\rm gr}}(\widehat f-f^*)
\end{eqnarray}
for some universal constants~$\tilde{c}_1$ and~$c_1$.

For the remainder of the proof we restrict our attention to the random event on which~(\ref{max.ineq.fhat}) holds.  We will establish a general prediction error bound, from which inequality~\eqref{np.or.bnd.gen} will follow by setting\footnote{Setting $\gamma$ equal to~$m/(2m+1)$, or any other positive value, does not violate the conditions imposed on~$\gamma$ in the statement of Lemma~\ref{lem.max.ineq.unif}.} $\gamma=m/(2m+1)$.  We let
\begin{equation*}
\begin{aligned}
\tau_n:=2\tilde{c}_1\sigma k^{1/2}\Big[ k^{\gamma/(2m)}r_n + \sqrt{\frac{\log(eq/k)}{n}} \Big] \;&~~ \text{and} \; \\
\lambda_n\ge c_1\sigma\Big[ k^{1/2-\gamma(2m-1)/(2m)}r_n^2 + k^{1/2-\gamma}r_n\sqrt{\frac{\log(eq/k)}{n}}  \Big],
\end{aligned}
\end{equation*}
noting that when $\gamma=m/(2m+1)$, the last inequality matches the corresponding lower-bound on~$\lambda_n$ in the statement of Theorem~\ref{add.thm}.
Multiplying inequality~(\ref{basic.ineq.fn}) by two and then applying~(\ref{max.ineq.fhat}) with $f=\widehat f-f^*$, we derive
\begin{eqnarray*}
2\|\widehat \bff-\bff^*\|_n^2+\lambda_n  \text{Pen}_{\text{gr}}(\widehat f) &\le& 2\tau_n\|\widehat\bff - \bff^*\|_n+3\lambda_n \text{Pen}_{\text{gr}}(f^*)\\
&\le&\|\widehat\bff - \bff^*\|_n^2 + \tau_n^2+3\lambda_n \text{Pen}_{\text{gr}}(f^*).
\end{eqnarray*}
Consequently,
\begin{equation*}
\|\widehat \bff-\bff^*\|_n^2 \lesssim  \sigma^2 k \Big[ k^{\gamma/m}r_n + {\frac{\log(eq/k)}{n}} \Big] + \lambda_n \text{Pen}_{\text{gr}}(f^*).
\end{equation*}
Inequality~\eqref{np.or.bnd.gen} then follows from the above bound by letting~$\gamma=m/(2m+1)$.  We note that this choice of~$\gamma$ optimizes the prediction error rate in the setting where $\text{\rm Pen}_{\text{\rm gr}}(f^*)\asymp \sigma k$, however, the rate can be improved when $\text{\rm Pen}_{\text{\rm gr}}(f^*)$ and~$\sigma k$ have different orders of magnitude.

\medskip

\textbf{Proof of inequality~\eqref{np.or.bnd.cc}}. Applying Lemma~\ref{lem.max.ineq.unif} with~$s=1$ and $\epsilon=1/q$, we deduce that with probability at least $1-1/q$, inequality
\begin{equation*}
(\bepsilon/\sigma,\bff_j)_n\lesssim \Big[ r_n + \sqrt{\frac{\log(q)}{n}}\Big] \Big[\|\bff_j\|_n+
r_n \text{Pen}(f_j)\Big]
\end{equation*}
holds uniformly over $f\in\mathcal{C}_{\text{\rm gr}}$ and $j\in [q]$. The above bound implies that there exists a universal constant~$c_0$, such that
\begin{equation*}
(\bepsilon/\sigma,\bff)_n=\sum_{j=1}^q (\bepsilon/\sigma,\bff_j)_n\le c_0\Big[ r_n + \sqrt{\frac{\log(q)}{n}}\Big] \Big[\sum_{j=1}^q\|\bff_j\|_n+
r_n \text{Pen}_{\text{\rm gr}}(f)\Big].
\end{equation*}
Letting $f=\widehat{f}-f^*$, we conclude that
\begin{equation}
\label{max.ineq1}
(\bepsilon/\sigma,\widehat \bff-\bff^*)_n\le c_0\Big[ r_n + \sqrt{\frac{\log(q)}{n}}\Big] \Big[\sum_{j=1}^q\|\widehat \bff_j-\bff^*_j\|_n+
r_n \text{Pen}_{\text{\rm gr}}(\widehat{f}-f^*)\Big]
\end{equation}
with probability at least $1-1/q$.

For the remainder of the proof we restrict our attention to the random event on which~(\ref{max.ineq1}) holds.
We define $\mu_n=4 c_0\sigma\big[ r_n + \sqrt{{\log(q)}/{n}}\big]$ and let $\lambda_n\ge 4\mu_n r_n \xi/(\xi-1)$.  Applying inequality~(\ref{max.ineq1}), we rewrite inequality~(\ref{basic.ineq.fn}) as follows:
\begin{equation}
\label{basic.rewr}
2\|\widehat \bff-\bff^*\|_n^2+\lambda_n  \text{Pen}_{\text{gr}}(\widehat f-f^*) \le \mu_n\sum_{j=1}^q \|\widehat\bff_j - \bff^*_j\|_n+3\lambda_n \text{Pen}_{\text{gr}}(f^*).
\end{equation}
We now consider two possible cases.

{\bf Case i)}: $\mu_n\sum_{j=1}^q \|\widehat\bff_j - \bff^*_j\|_n\ge 3\lambda_n \text{Pen}_{\text{gr}}(f^*)$.
It follows that
\begin{equation}
\label{case1.eq}
2\|\widehat \bff-\bff^*\|_n^2+ \lambda_n\text{Pen}_{\text{gr}}(\widehat f-f^*) \le 2\mu_n\sum_{j=1}^q \|\widehat\bff_j - \bff^*_j\|_n,
\end{equation}
and, consequently, $2r_n \text{Pen}_{\text{gr}}(\widehat f-f^*)\le 4(\mu_n r_n/\lambda_n)\sum_{j=1}^q \|\widehat\bff_j - \bff^*_j\|_n\le (\xi-1)\sum_{j=1}^q \|\widehat\bff_j - \bff^*_j\|_n$.  Taking into account inequality $G(\widehat f - f^*)\le 2k$ and Definition~\ref{nonparam.def}, we then derive
\begin{equation}
\label{comp.bnd}
\sum_{j\le q}\|\widehat\bff_j - \bff^*_j\|_n\le[2k]^{1/2} [\psi(2k,\xi)]^{-1}\|\widehat\bff - \bff^*\|_n.
\end{equation}
Combining this bound with inequality~(\ref{case1.eq}), we colclude
\begin{equation*}
\|\widehat \bff-\bff^*\|_n^2\le \mu_n[2k]^{1/2}[\psi(2k,\xi)]^{-1}\|\widehat\bff - \bff^*\|_n,
\end{equation*}
which implies the stated prediction error bound.


{\bf Case ii)}: $\mu_n \sum_{j=1}^q \|\widehat\bff_j - \bff^*_j\|_n < 3 \lambda_n \text{Pen}_{\text{gr}}(f^*)$.
Going back to inequality~(\ref{basic.rewr}), we derive
\begin{equation*}
2\|\widehat \bff-\bff^*\|_n^2+\lambda_n  \text{Pen}_{\text{gr}}(\widehat f-f^*) \le 6\lambda_n  \text{Pen}_{\text{gr}}(f^*),
\end{equation*}
which implies the stated prediction error bound. \qed

\subsection{Proof of Lemma~\ref{lem.max.ineq.unif}}
\label{prf.lem.max.ineq.unif}

Given $J\subseteq [q]$, we define a functional class $\mathcal{F}(J)=\{f:\; f(\M{x})=\sum_{j\in J} f_j(x_j),\;f_j\in\mathcal{C}\}$. We will need the following result, which is proved in Section~\ref{prf.lem.max.ineq}.

\medskip

\begin{lem}\label{lem.max.ineq}
Let $J\subseteq [q]$ and let~$\gamma$ be a fixed constant such that $r_n \le |J|^{\gamma}$.  Then, with probability at least $1-e^{-t}$, inequality
\begin{equation*}
\label{hp.ineq.add}
(\bepsilon/\sigma,\bff)_n\lesssim \Big[ |J|^{1/2 + \gamma/(2m)}r_n + \sqrt{t/n} \Big] \|\bff\|_n + \Big[ |J|^{1/2-\gamma(2m-1)/(2m)}r_n^2 + |J|^{-\gamma}r_n\sqrt{t/n} \Big] \text{Pen}_{\text{\rm gr}}(f)
\end{equation*}
holds uniformly over ${f\in\mathcal{F}(J)}$.
\end{lem}

Let~$M_s$ denote the number of distinct subsets of $[q]$ that have size~$s$. We note that $\log(M_s)\le s\log(eq/s)$ and, thus, $M_s e^{-t} \le e^{s\log(eq/s)-t}$.  Applying Lemma~\ref{lem.max.ineq} together with the union bound, we derive that, with probability at least $1 - e^{s\log(eq/s)-t}$, inequality
\begin{equation*}
(\bepsilon/\sigma,\bff)_n\lesssim \Big[ s^{1/2 + \gamma/(2m)}r_n + \sqrt{t/n} \Big] \|\bff\|_n + \Big[ s^{1/2-\gamma(2m-1)/(2m)}r_n^2 + s^{-\gamma}r_n\sqrt{t/n} \Big] \text{Pen}_{\text{\rm gr}}(f)
\end{equation*}
holds uniformly over ${f\in\mathcal{F}_{s}}$.   We complete the proof by noting that for $t=s\log(eq/s)+\log(1/\epsilon)$ the above inequality becomes
\begin{eqnarray*}
(\bepsilon/\sigma,\bff)_n&\lesssim& \Big[ s^{1/2 + \gamma/(2m)}r_n + \sqrt{\frac{s\log(eq/s)}{n}}+ \sqrt{\frac{\log(1/\epsilon)}{n}} \Big]\|\bff\|_n \\
&& + \Big[ s^{1/2-\gamma(2m-1)/(2m)}r_n^2 + s^{-\gamma}r_n\sqrt{\frac{s\log(eq/s)}{n}}+ s^{-\gamma}r_n\sqrt{\frac{\log(1/\epsilon)}{n}} \Big] \text{Pen}_{\text{\rm gr}}(f),
\end{eqnarray*}
and the corresponding lower-bound on the probability simplifies to $1-\epsilon$. \qed

\subsection{Proof of Lemma~\ref{lem.max.ineq}}
\label{prf.lem.max.ineq}

Given a positive constant~$\delta$ and a metric space~$\mathcal{H}$ endowed with the norm $\|\cdot\|$, we use the standard notation and write $H(\delta,\mathcal{H},\|\cdot\|)$ for the $\delta$-entropy of $\mathcal{H}$ with respect to~$\|\cdot\|$.  More specifically, $H(\delta,\mathcal{H},\|\cdot\|)$ is the natural logarithm of the smallest number of balls with radius~$\delta$ needed to cover~$\mathcal{H}$.

With a slight abuse of notation, we extend the domain of~$\|\cdot\|_n$ from vectors in~$\mathbb{R}^n$ to real-valued functions on~$[0,1]^q$ by letting $\|\cdot\|_n$ be the empirical $L_2$-norm.  Thus, given a function~$h$, we let $\|h\|_n=[\sum_{i=1}^n h(\M{x}_i)^2/n]^{1/2}$.  This extension is consistent in the sense that $\|f\|_n=\|\bff\|_n$ and $\|f_j\|_n=\|\bff_j\|_n$ for $f\in\mathcal{C}_{\text{\rm gr}}$, $j\in [q]$.

We let $\mathcal{H}({J})=\{h: \; h\in\mathcal{F}(J),\; \|h\|_n/(r_n|J|^{-\gamma})+\text{\rm Pen}_{\text{\rm gr}}(h)\le 1\}$, noting that $\|h\|_n\le r_n|J|^{-\gamma}$ and $\text{\rm Pen}_{\text{\rm gr}}(h)\le 1$ for every $h\in \mathcal{H}({J})$.  By Corollary 8.3 in \cite{vdg2000applications} (cf. Lemma~12 in the supplementary material for \cite{tan2019doubly}),
\begin{equation}
\label{tail.bnd.entr}
\sup_{h\in\mathcal{H}({J})}(\bepsilon/\sigma,\bh)_n\lesssim  n^{-1/2}\int_0^{r_n|J|^{-\gamma}}\sqrt{H(u,\mathcal{H}({J}),\|\cdot\|_n)}du + r_n|J|^{-\gamma}\sqrt{t/n}
\end{equation}
with probability at least $1-e^{-t}$.  To bound the entropy, we will use the following result, proved in Section~\ref{prf.lem.entr.bnd}.

\medskip

\begin{lem}\label{lem.entr.bnd}
$H(u,\mathcal{H}({J}),\|\cdot\|_n)\lesssim |J|(1/u)^{1/m}$ for~$u\in(0,1)$.
\end{lem}

Noting that $r_n=n^{-m/(2m+1)}$ and, thus, $n^{-1/2}=r_n^{(2m+1)/(2m)}$, we derive
\begin{eqnarray*}
n^{-1/2}\int_0^{r_n|J|^{-\gamma}}\sqrt{H(u,\mathcal{H}({J}),\|\cdot\|_n)}du&\lesssim& n^{-1/2}\int_0^{r_n|J|^{-\gamma}}|J|^{1/2}u^{-1/(2m)}du\\
\\
&\lesssim& |J|^{1/2}n^{-1/2}\Big[r_n|J|^{-\gamma}\Big]^{(2m-1)/(2m)}\\
\\
&=& r_n^{(2m+1)/(2m)+(2m-1)/(2m)}   |J|^{1/2-\gamma(2m-1)/(2m)}\\
\\
&=& r_n^2 |J|^{1/2-\gamma(2m-1)/(2m)}.
\end{eqnarray*}
Applying bound~(\ref{tail.bnd.entr}), we conclude that
\begin{equation*}
\sup_{h\in\mathcal{H}({J})}(\bepsilon/\sigma,\bh)_n\lesssim  r_n^2|J|^{1/2-\gamma(2m-1)/(2m)} + r_n|J|^{-\gamma}\sqrt{t/n}
\end{equation*}
with probability at least $1-e^{-t}$.  The statement of the lemma is then a consequence of the fact that for every $f\in\mathcal{F}({J})$, function
$f/\big[\|f\|_n/(r_n|J|^{-\gamma})+\text{\rm Pen}_{\text{\rm gr}}(f)\big]$ falls in the class~$\mathcal{H}({J})$.
\qed

\subsection{Proof of Lemma~\ref{lem.entr.bnd}}
\label{prf.lem.entr.bnd}
We will establish the stated entropy bound for the functional space ${\mathcal{H}}_{J}'=\{h: \; h\in\mathcal{F}(J),\; \|h\|_{n}+\text{\rm Pen}_{\text{\rm gr}}(h)\le 1\}$.  The same bound will then automatically hold for~$\mathcal{H}({J})$, because $r_n|J|^{-\gamma}\le1$ and, hence, ${\mathcal{H}}_{J}'\subseteq\mathcal{H}({J})$. We treat~$m$ as fixed, so that universal constants in inequalities below are allowed to depend on~$m$.

Consider an arbitrary~$g\in\mathcal{C}$. By the Sobolev embedding theorem \citep[for example,][Theorem 3.13]{oden1976reddy}, we can write $g$ as a sum of a polynomial of degree~$m-1$ and a function~$\tilde{g}$ that satisfies $\|\tilde{g}\|_{L_2}\lesssim \text{\rm Pen}(g)$, where we note that $\text{\rm Pen}(g)=\text{\rm Pen}(\tilde{g})$. Applying Lemma 10.9 in \cite{vdg2000applications}, which builds on the interpolation inequality of \cite{agmon1965lectures}, we derive $\|\tilde{g}\|_{\infty}\lesssim \text{\rm Pen}(\tilde{g})$. Thus, ${\mathcal{H}}_{J}'\subseteq \{p+\tilde{h}: \; p\in\mathcal{P}_{J},\; \tilde{h}\in\tilde{\mathcal{H}}_{J}\}$, where
\begin{eqnarray*}
\mathcal{P}_{J}&=&\{p: \; p(\M{x})=\alpha_0+\sum_{j\in J}\sum_{l=1}^{m-1} \alpha_{jl} x_{j}^l, \; \alpha_0\in\mathbb{R},\;\alpha_{jl}\in\mathbb{R}\;\forall j,k, \; \|p\|_{n}\le 2 \}\\
\tilde{\mathcal{H}}_{J}&=&\{\tilde{h}:\; \tilde{h}\in \mathcal{F}(J), \; \text{\rm Pen}_{\text{\rm gr}}(\tilde{h})\le 1, \;  \|\tilde{h}_j\|_{\infty}\lesssim \text{\rm Pen}(\tilde{h}_j)\; \forall j\in J \}.
\end{eqnarray*}
We are able to impose the bound $\|p\|_{n}\le 2$ in the definition of $\mathcal{P}_{J}$, because if $h=p+\tilde{h}$ for $h\in {\mathcal{H}}_{J}'$ and $\tilde{h}\in\tilde{\mathcal{H}}_{J}$, then $\|p+\tilde{h}\|_n\le 1$ and $\|\tilde{h}\|_n\le \text{\rm Pen}_{\text{\rm gr}}(\tilde{h})\le 1$. Consequently,
\begin{equation}
\label{entr.sum.bnd} H(u,\mathcal{H}_{J}',\|\cdot\|_{n})\le H(u/2,\mathcal{P}_{J},\|\cdot\|_{n}) + H(u/2,\tilde{\mathcal{H}}_{J},\|\cdot\|_{\infty}),
\end{equation}
where we used the fact that the unit ball with respect to the $\|\cdot\|_{\infty}$-norm is contained within the corresponding ball with respect to the $\|\cdot\|_{n}$-norm. We note that $\mathcal{P}_{J}$ is a ball of radis~$2$, with respect to the $\|\cdot\|_{n}$-norm, in a linear functional space of dimension $|J|(m-1)+1$. Hence, $H(u/2,\mathcal{P}_{J},\|\cdot\|_{n})\lesssim |J|+|J|\log(1/u)$ by, for example, Corollary 2.6 in \cite{vdg2000applications}.  Thus, the result of Lemma~\ref{lem.entr.bnd} follows from~\ref{entr.sum.bnd} if we also establish that $H(\delta,\tilde{\mathcal{H}}_{J},\|\cdot\|_{\infty})\lesssim |J|(1/\delta)^{1/m}$ for $\delta\in(0,1)$.

It is only left to derive the stated bound on $H(\delta,\tilde{\mathcal{H}}_{J},\|\cdot\|_{\infty})$.  Note that we can represent functional class~$\tilde{\mathcal{H}}_{J}$ as follows:
\begin{equation*}
\tilde{\mathcal{H}}_{J} = \left\{\tilde{h}:\; \tilde{h}(\M{x})=\sum_{j\in J} \lambda_j g_j(x_j), \; \sum_{j\in J} |\lambda_j| \le 1, \; g_j\in \mathcal{C}, \; \text{\rm Pen}(g_j)\le 1, \; \|g_j\|_{\infty}\le 1 \; \forall j\in J \right\}.
\end{equation*}
Given functions $\tilde{h}(\M{x})=\sum_{j\in J} \lambda_j g_j(x_j)$ and $\tilde{h}'(\M{x})=\sum_{j\in J} \lambda_j' g_j'(x_j)$ in~$\tilde{\mathcal{H}}_{J}$, we have
\begin{eqnarray*}
\|\tilde{h} - \tilde{h}'\|_{\infty} &\le&  \|\sum_{j\in J} \lambda_j g_j - \sum_{j\in J} \lambda_j g_j'\|_{\infty} + \|\sum_{j\in J} \lambda_j g_j' - \sum_{j\in J} \lambda_j' g_j'\|_{\infty} \\
&\le&  \max_{j\in J}\|g_j - g_j'\|_{\infty} \sum_{j\in J} |\lambda_j|  + \max_{j\in J}\|g_j'\|_{\infty} \sum_{j\in J} |\lambda_j - \lambda_j'| \\
&\le&   \max_{j\in J}\|g_j - g_j'\|_{\infty} + \sum_{j\in J} |\lambda_j - \lambda_j'|.
\end{eqnarray*}
Consequently, if we let $\mathcal{G}=\{g: \; g\in \mathcal{C}, \; \text{\rm Pen}(g)\le 1, \; \|g\|_{\infty}\le 1\}$, let $\|\cdot\|_1$ denote the~$\ell_1$-norm and let $B^{d}_1$ denote a unit $\ell_1$-ball in~$\mathbb{R}^d$, then
\begin{equation*}
H(\delta,\tilde{\mathcal{H}}_{J},\|\cdot\|_{\infty})\le |J|H(\delta/2,\mathcal{G},\|\cdot\|_{\infty}) + H(\delta/2,B^{|J|}_1,\|\cdot\|_1).
\end{equation*}
By the results in \cite{birman1}, $H(\delta/2,\mathcal{G}_j,\|\cdot\|_{\infty})\lesssim (1/\delta)^{1/m}$.  By the standard bounds on the covering numbers of a norm ball, $H(\delta/2,B^{|J|}_1,\|\cdot\|_1)\lesssim|J|+|J|\log(1/\delta)$.  Thus, $H(\delta,\tilde{\mathcal{H}}_{J},\|\cdot\|_{\infty})\lesssim |J|(1/\delta)^{1/m}$ for $\delta\in(0,1)$. \qed

\section{Additional Experimental Results}\label{sec:additional_experiments}

\subsection{Performance for Varying Number of Observations}
\begin{figure}[h!]
    \centering
    High Correlation Setting ($\rho = 0.9$). Tuning: Validation MSE for $\ell_0$ and Lasso; HBIC for SCAD and MCP.
    \includegraphics[width=\textwidth]{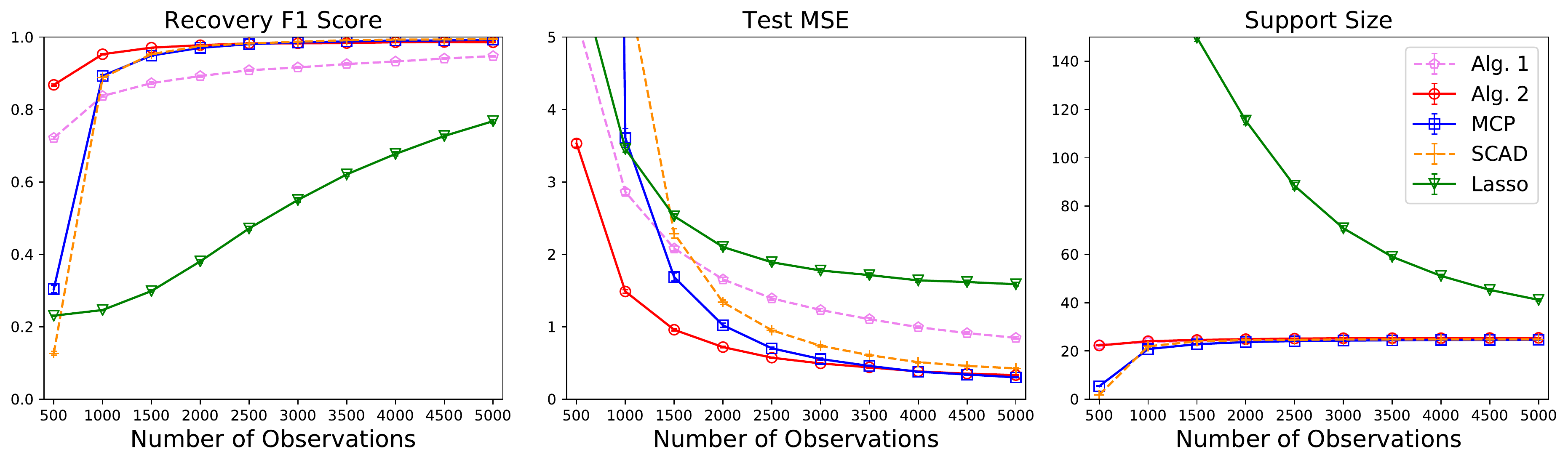}
    \caption{\textcolor{black}{Performance measures for varying number of observations on a synthetic dataset with a correlation coefficient $\rho = 0.9$. We use HBIC-based tuning for Group SCAD and MCP, and validation MSE-based tuning for other estimators. 
    The standard error of the mean is represented using error bars. Alg. 1 and Alg. 2 are our proposed algorithms. Here, ``Lasso" is a shorthand for Group Lasso, we use the same convention for SCAD, MCP.}}
    \label{fig:appendix_high_corr_hbic}
\end{figure}

\begin{figure}[h!]
    \centering
    Mild Correlation Setting ($\rho = 0.5$). Tuning: Validation MSE for all methods.
    \includegraphics[width=\textwidth]{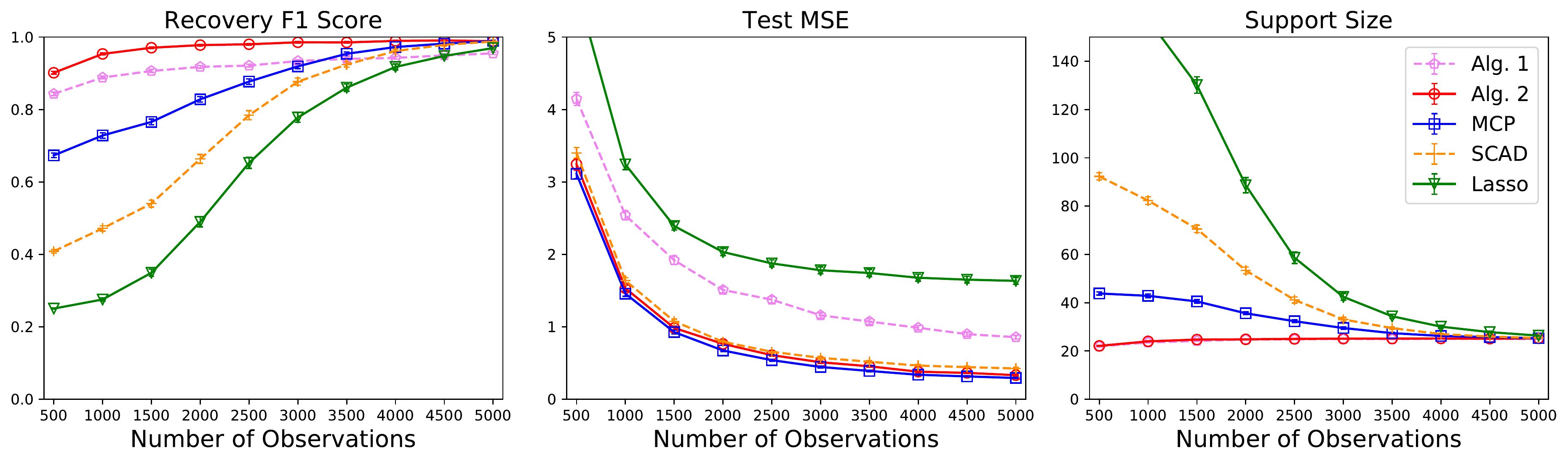}

\medskip    

    Mild Correlation Setting ($\rho = 0.5$). Tuning: Validation MSE for $\ell_0$ and Lasso; HBIC for SCAD and MCP.
    \includegraphics[width=\textwidth]{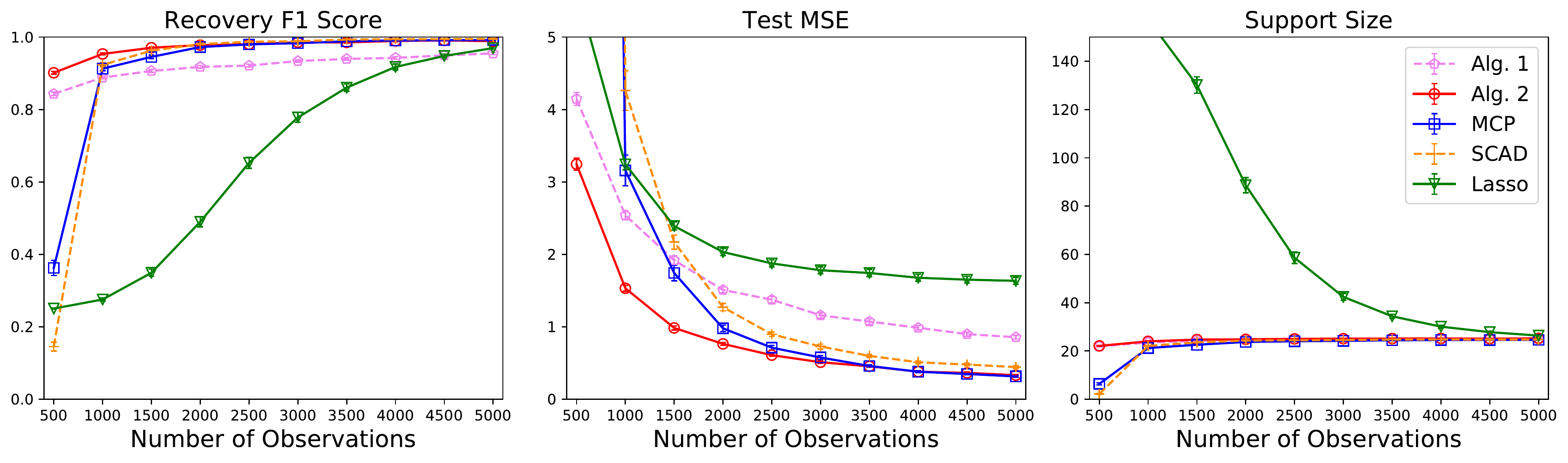}
    \caption{\textcolor{black}{Performance measures for varying number of observations on a synthetic dataset with a correlation coefficient $\rho = 0.5$. Top panel shows validation MSE-based tuning for all methods, bottom panel shows HBIC-based tuning for SCAD and MCP.  The standard error of the mean is represented using error bars. Alg. 1 and Alg. 2 are our proposed algorithms. Here, ``Lasso" is a shorthand for Group Lasso, we use the same convention for SCAD, MCP.}}
    \label{fig:appendix_0.5_corr}
\end{figure}

\begin{figure}[htbp]
    \centering
    Uncorrelated Setting ($\rho = 0$). Tuning: Validation MSE for all methods.
    \includegraphics[width=\textwidth]{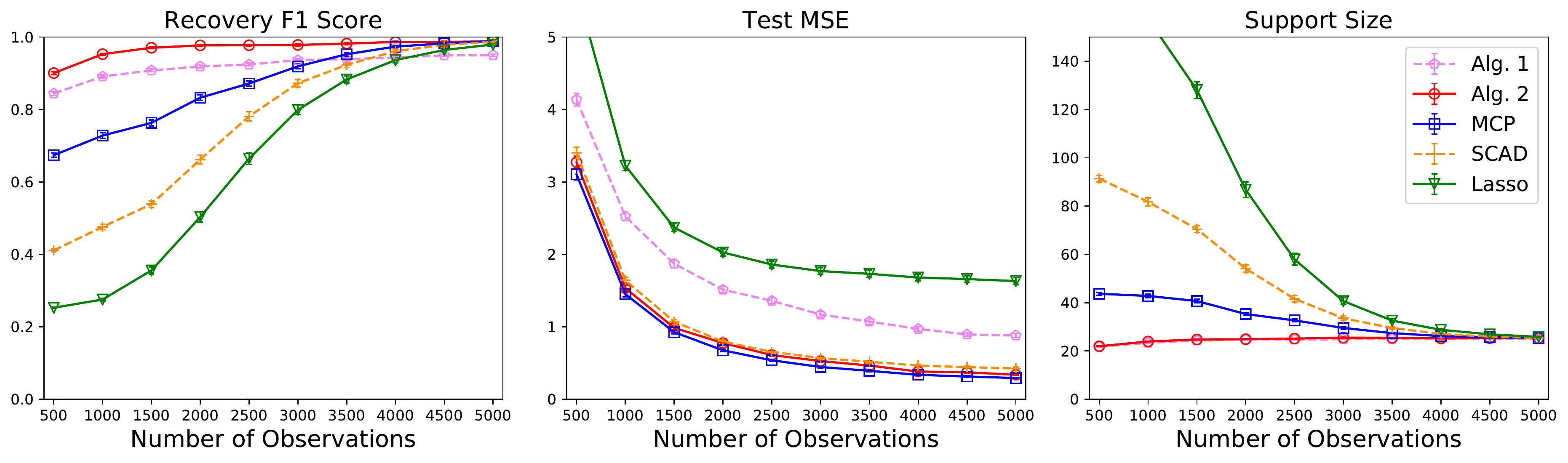}

\medskip 

    Uncorrelated Setting ($\rho = 0$). Tuning: Validation MSE for $\ell_0$ and Lasso; HBIC for SCAD and MCP.
    \includegraphics[width=\textwidth]{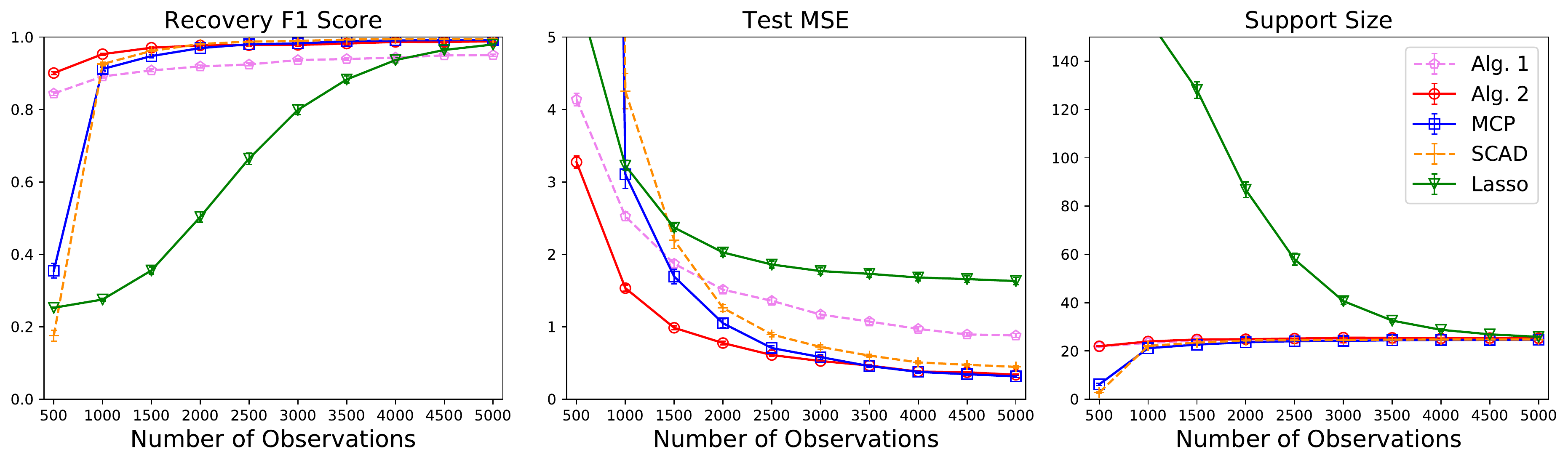}
    \caption{\textcolor{black}{Performance measures for varying number of observations on a synthetic dataset with a correlation coefficient $\rho = 0$.
    Top panel shows validation MSE-based tuning for all methods, bottom panel shows HBIC-based tuning for SCAD and MCP. 
    The standard error of the mean is represented using error bars. Alg. 1 and Alg. 2 are our proposed algorithms. Here, ``Lasso" is a shorthand for Group Lasso, we use the same convention for SCAD, MCP.}}
    \label{fig:appendix_0_corr}
\end{figure}

\textcolor{black}{In Figure \ref{fig:appendix_high_corr_hbic}, we report the results of the experiment of Section \ref{sec:vary_n} with HBIC tuning~\cite{wang2013calibrating} for Group MCP and SCAD. With HBIC tuning, solutions from Group MCP and SCAD are more sparse with worse prediction accuracy, compared to validation MSE tuning. 
In Figures  \ref{fig:appendix_0.5_corr} and \ref{fig:appendix_0_corr}, we report the results of the same experiment in Section \ref{sec:vary_n} but with lower correlation coefficients $\rho=0.5$ and $\rho=0.0$, respectively. Each of the figures  \ref{fig:appendix_0.5_corr} and \ref{fig:appendix_0_corr} presents results based on MSE validation tuning and HBIC tuning (for Group SCAD and MCP).}

\subsection{Performance for Varying SNR}
\begin{figure}[htbp]
    \centering
    Varying SNR: $n=5000, p=5000$. Tuning: Validation MSE for all methods.
    \includegraphics[width=\textwidth]{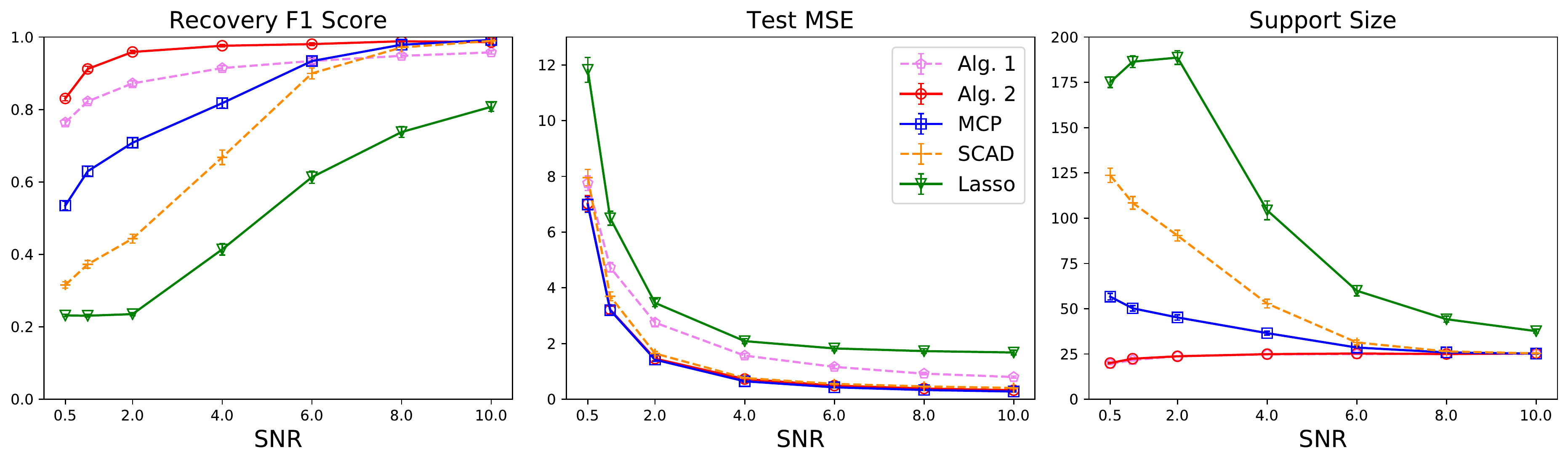}
    
    \medskip
    
    Varying SNR: $n=5000, p=5000$. Tuning: Validation MSE for $\ell_0$ and Lasso; HBIC for SCAD and MCP.
    \includegraphics[width=\textwidth]{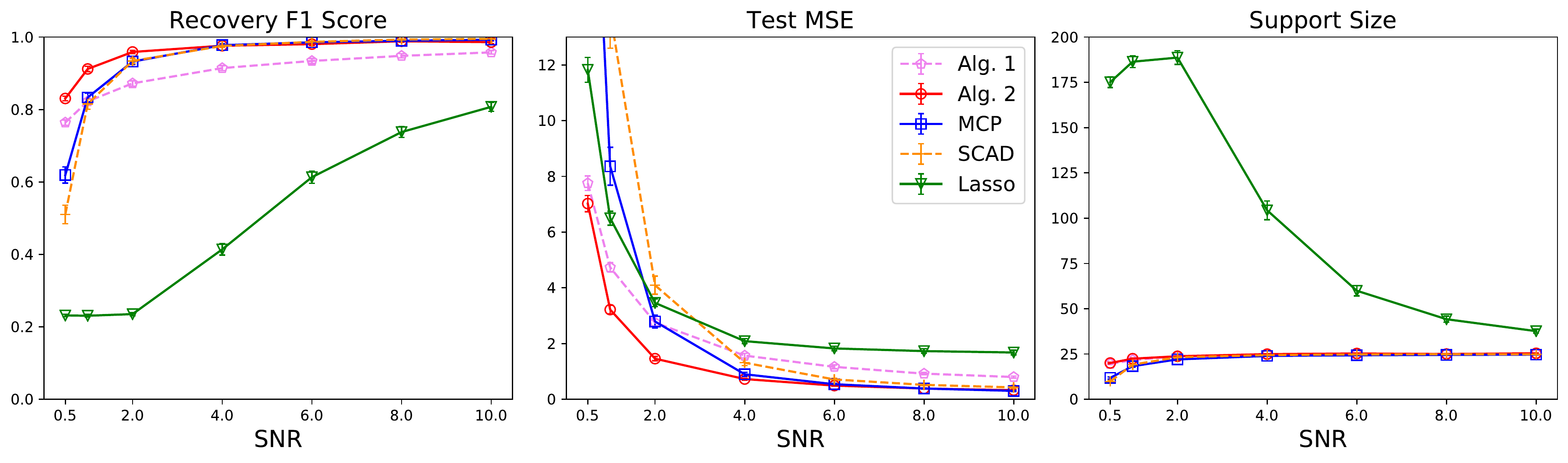}
    \caption{\textcolor{black}{Performance measures for varying SNR on a synthetic dataset with $n=5000$, $p=5000$, $\rho=0.9$. Top panel shows validation MSE-based tuning for all methods, bottom panel shows HBIC-based tuning for SCAD and MCP. The standard error of the mean is represented using error bars. Alg. 1 and Alg. 2 are our proposed algorithms.}}
    \label{fig:appendix_vary_snr_n_5000}
\end{figure}

\begin{figure}[htbp]
    \centering
    Varying SNR: $n=1000, p=5000$. Tuning: Validation MSE for all methods.
    \includegraphics[width=\textwidth]{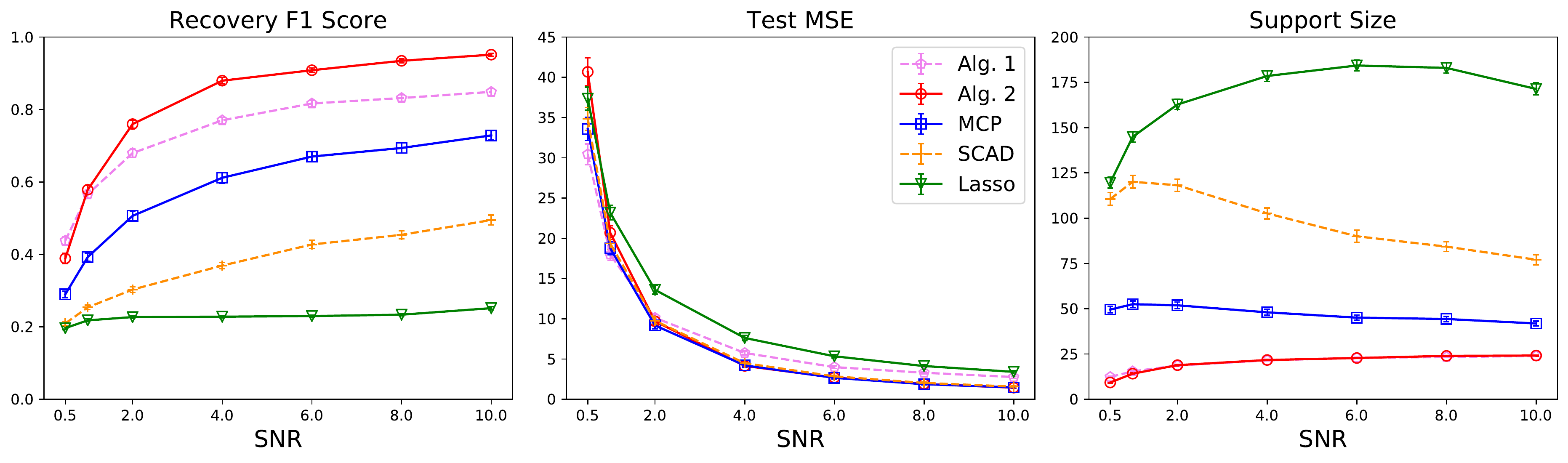}
    
    \medskip
    
    Varying SNR: $n=1000, p=5000$. Tuning: Validation MSE for $\ell_0$ and Lasso; HBIC for SCAD and MCP.
    \includegraphics[width=\textwidth]{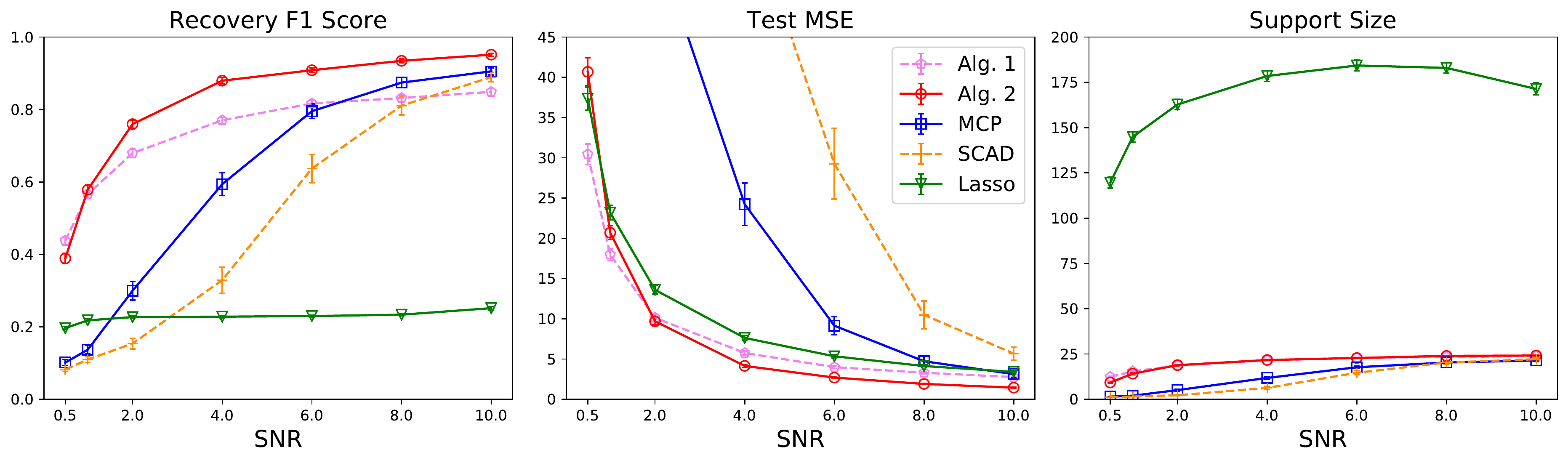}
    \caption{\textcolor{black}{Performance measures for varying SNR on a synthetic dataset with $n=1000$, $p=5000$, $\rho=0.9$. Top panel shows results based on validation MSE-tuning for all methods; and bottom panel shows results for HBIC-based tuning for Group MCP and SCAD.  The standard error of the mean is represented using error bars. Alg. 1 and Alg. 2 are our proposed algorithms. Here Lasso is a shorthand for Group Lasso (similar convention applies to MCP, SCAD). }}
    \label{fig:appendix_vary_snr_n_1000}
\end{figure}

\textcolor{black}{Here we study the performance of the different algorithms for varying SNR. Similar to the experiment of Section \ref{sec:vary_n}, we fix a correlation parameter $\rho = 0.9$, $p = 5000$, a group size of $4$, number of nonzero groups $k_{*} = 25$. We vary the SNR in $\{0.5, 1, 2, 4, 6, 8, 10 \}$ and the sample size $n \in \{1000, 5000\}$. The results for $n=5000$ and $n=1000$ are shown in Figures \ref{fig:appendix_vary_snr_n_5000} and \ref{fig:appendix_vary_snr_n_1000}, respectively. Each figure presents results based on MSE validation tuning and HBIC tuning (for Group SCAD and MCP).}

\subsection{Statistical Performance on High-dimensional Instances}

\textcolor{black}{In Table \ref{table:appendix_hbic}, we report the results of the experiment of Section \ref{sec:large_instances} using HBIC tuning for Group MCP and SCAD estimators and validation MSE tuning for other estimators.}

\begin{table}[h!]
\centering
\caption{\textcolor{black}{Performance measures for Setting 1 (top panel) and Setting 2 (bottom panel) where, tuning parameters for Group MCP and SCAD are selected by HBIC tuning, and other estimators are chosen by validation MSE tuning. Means are reported along with their standard errors (we consider 500 replications).}}
    \vspace{0.5cm}
\label{table:appendix_hbic}
\begin{tabular}{cc}
\rotatebox[origin=c]{90}{Setting 1}&
\begin{tabular}{lccccc} 
\toprule
Algorithm &  $\| \hat{\B\beta} \|_0$ &  TP &  FP &  MSE &  $\| \hat{\B\beta} - \B\beta^{*} \|_{\infty}$ \\
\midrule
Group $\ell_0$ &         $ 100.3 ~ (1.2) $ &           $ 8.9 ~ (0.09) $ &              $ 1.1 ~ (0.1) $ &          $ 19.2  ~ (1.1) $ &      $ 1.17 ~ (0.03) $ \\
Group Lasso      &        $ 2086.1 ~ (27.8) $ &  $ 9.6 ~ (0.05)$ &            $ 199.0 ~ (2.8) $ &          $ 26.9  ~ (1.1) $ &      $ 1.44 ~ (0.02) $ \\
Group MCP        &         $ 162.0 ~ ( 5.5 ) $ &           $ 6.0 ~ ( 0.13 ) $ &            $ 10.1 ~ ( 0.5 ) $ &          $ 48.8 ~ ( 1.4 ) $ &      $ 1.7 ~ ( 0.03 ) $ \\
Group SCAD    &   $ 368.0 ~ ( 13.4 ) $  &         $ 6.9 ~ ( 0.12 ) $ &            $ 29.9 ~ ( 1.3 ) $ &          $ 147.8 ~ ( 20.2 ) $ &      $ 1.57 ~ ( 0.03 ) $ \\
\bottomrule
\end{tabular} \\[2.3em]
\rotatebox[origin=c]{90}{Setting 2}&
\begin{tabular}{lccccc}
Group $\ell_0$ &           $ 79.4 ~ ( 0.1 ) $ &            $ 19.7 ~ ( 0.03 ) $ &              $ 0.2 ~ ( 0.02 ) $ &           $ 1.12 ~ ( 0.03 ) $ &      $ 0.356 ~ ( 0.007 ) $ \\
Group Lasso      &        $ 1126.5 ~ ( 11.7 ) $ &           $ 19.8 ~ ( 0.02 ) $ &           $ 261.8 ~ ( 2.9 ) $ &           $ 5.06 ~ ( 0.10 ) $ &      $ 0.703 ~ ( 0.007 ) $ \\
Group MCP        &         $ 109.4 ~ ( 1.5 ) $ &           $ 19.4 ~ ( 0.07 ) $ &            $ 8.0 ~ ( 0.3 ) $ &          $ 2.62 ~ ( 0.26 ) $ &      $ 0.438 ~ ( 0.010 ) $ \\
Group SCAD       &        $ 196.2 ~ ( 4.4 ) $&            $ 19.5 ~ ( 0.05 ) $ &          $ 29.5 ~ ( 1.1 ) $ &           $ 3.27 ~ ( 0.22 ) $ &      $ 0.534 ~ ( 0.009 ) $ \\
\bottomrule
\end{tabular}
\end{tabular}
\end{table}

\subsection{Performance on the Birthweight Dataset} \label{sec:birthweight}
\begin{figure}[htbp]
    \centering
    \includegraphics[scale=0.45]{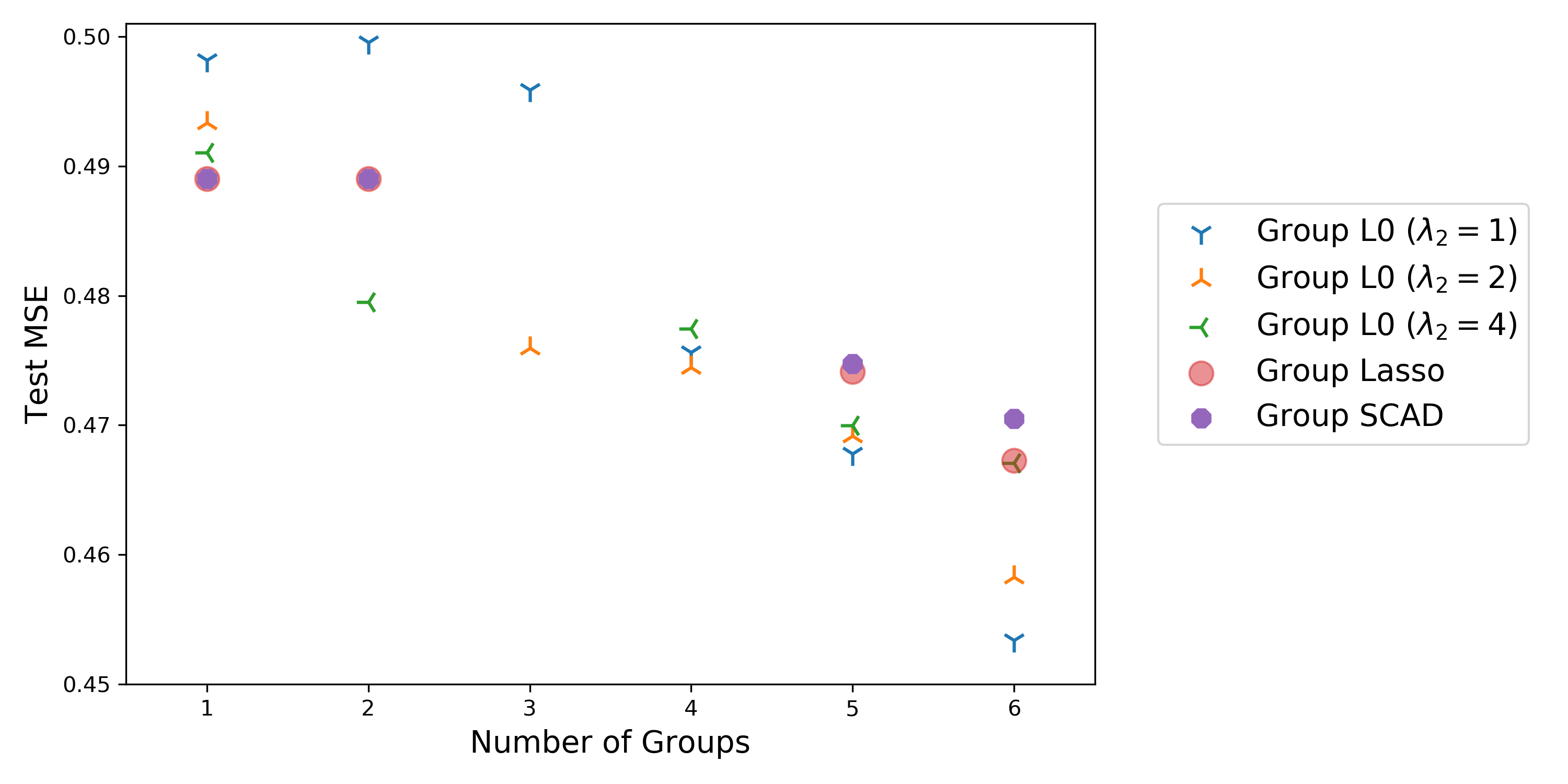}
    \caption{Test MSE on the Birthweight dataset. For \grp, we consider additional ridge regularization and vary the corresponding regularization parameter $\lambda_2 \in \{1, 2, 4\}$. Group sizes $3$ and $4$ could not be attained using Group Lasso and SCAD.}
    \label{fig:birth_weight}
\end{figure}

We study the Birthweight dataset, taken from the \texttt{R} package {\texttt{grpreg}}.
Here, we predict birth weight using 7 grouped covariates.
The dataset has~$189$ observations, which we randomly split into $75\%$ for training and $25\%$ for testing. On this dataset, we fit regularization paths for  \grp, Lasso, and SCAD. For \grp, we use an additional $\ell_2$ regularization and consider $\lambda_2 \in \{1, 2, 4\}$. In Figure~\ref{fig:birth_weight}, we plot the test MSE versus the sparsity level for the different methods. The results show that the \grp-based methods outperform Group Lasso and SCAD when the group size is $2$ or more.

\subsection{Additional Timing Comparisons} \label{sec:additional_timings}
Here we consider the same setup as in the experiment of Section \ref{sec:experiment_timings}, and we report the running times for additional values of $\MU$ to demonstrate the sensitivity of the runtime to $\MU$. Let $M^{*}$ be the value of $\MU$ used in Section \ref{sec:experiment_timings} -- note that this is the smallest value of $\MU$. We express our choices of $\MU$ in terms of $M^{*}$. We report the results for cases (i) and (ii) in Tables \ref{table:timings} and \ref{table:timings_pureL0}, respectively.

\subsubsection{Timings with different $n$}\label{sec:app-timings-n}
\textcolor{black}{To understand the sensitivity of runtimes of our BnB procedure for different values of $n$, we 
ran the same experiment of Section~\ref{sec:experiment_timings} with different values of $n \in \{1000, 5000, 7000, 10000\}$ with $p=10^4$ held fixed. This experiment was carried out on a machine with a 6-core Intel Core i7-8750H processor and 16GB of RAM---due to memory limits, we did not consider larger values of $n$.
         The running time is reported in 
         Table~\ref{table:reply} below.  The table shows that the runtime increases with increasing $n$ --- we believe this is mainly due to the increased runtimes in solving the node relaxations of the BnB tree.}


\begin{table}[h!]
\caption{Running time in seconds for solving case (i), i.e., the MIP in \eqref{eq:MIP_hybrid} with $\lambda_2 = \lambda_2^{*}$, to optimality. A dash (-) indicates that Gurobi cannot solve the problem in 24 hours and has an optimality gap of~$100\%$ upon termination.}
    \vspace{0.5cm}
 \label{table:timings}
\centering
\setlength{\tabcolsep}{8pt}
\begin{tabular}{|c|cc|cc|cc|}
\hline
\multirow{2}{*}{$p$} & \multicolumn{2}{c|}{$\MU = M^{*}$} & \multicolumn{2}{c|}{$\MU = 1.5 M^{*}$} & \multicolumn{2}{c|}{$\MU = \infty$} \\
                     & Ours            & Gurobi           & Ours              & Gurobi             & Ours            & Gurobi            \\ \hline
$10^3$               & 96              & 24223            & 186               & 12320              & 192             & 2399              \\
$10^4$               & 199             & -                & 245               & -                  & 333             & -                 \\
$10^5$               & 231             & -                & 404               & -                  & 421             & -                 \\
$10^6$               & 386             & -                & 1014              & -                  & 1250            & -                 \\
$5 \times 10^6$      & 1922            & -                & 3686              & -                  & 4036            & -                 \\ \hline
\end{tabular}
\end{table}

\begin{table}[h!]
\caption{Running time in seconds for solving case (ii), i.e., the MIP in \eqref{eq:MIP_hybrid} with $\lambda_2 = 0$, to optimality. A star or dash (-) indicates that the solver cannot solve the problem in 24 hours. For star, the optimality gap (in percent) is shown in parenthesis, whereas the gap is $100\%$ for dash.}
    \vspace{0.5cm}
 \label{table:timings_pureL0}
\centering
\setlength{\tabcolsep}{8pt}
\begin{tabular}{|c|cc|cc|cc|}
\hline
\multirow{2}{*}{$p$} & \multicolumn{2}{c|}{$\MU = M^{*}$} & \multicolumn{2}{c|}{$\MU = 1.5 M^{*}$} & \multicolumn{2}{c|}{$\MU = 2 M^{*}$}  \\
                     & Ours            & Gurobi           & Ours              & Gurobi             & Ours            & Gurobi            \\ \hline
$10^3$               & 373              & 8737            & 913               & 10675              & 1010             & 13901            \\
$10^4$               & 466             & -                & 2813               & -                  & *(3.9)             & -         \\
$10^5$               & 1136             & -                & *(4.7)               & -                  & *(20.7)            & -            \\
$10^6$               & 1628            & -                & *(5.1)               & -                  & *(21.6)            & -            \\ \hline
\end{tabular}
\end{table}

    \begin{table}[h]
    \centering
    \caption{Running time (in seconds) of our BnB method for different values of $n$ and $p=10^4$. Additional details can be found in Section~\ref{sec:app-timings-n}.}
    \vspace{0.5cm}
    \label{table:reply}
    \begin{tabular}{@{}ccc@{}}
    \toprule
    n      & p      & Time (seconds) \\ \midrule
    $1000$ & $10^4$ & 199            \\
    $5000$ & $10^4$ & 340            \\
    $7000$ & $10^4$ & 556            \\
    $10000$ & $10^4$ & 5796           \\ \bottomrule
    \end{tabular}
    \end{table}

\section{Additional Details on the Datasets}\label{data-examples}

\subsection{Description of the Amazon Reviews Dataset} This dataset is a subset of the Amazon Grocery and Gourmet Food dataset \cite{he2016}. To obtain $\M X$ and $\M y$, we follow the same steps described in \cite{fastsubset}, and we restrict $\M X$ to the top $5500$ words in the corpus. Here $\M X$ is a TF/IDF representation of the text reviews and $\M y$ is a continuous variable which measures  review helpfulness. \textcolor{black}{To obtain the groups, we employ an unsupervised method that only makes use of the covariates. We draw inspiration from the work of~\cite{buhlmann2013correlated}, who use a clustering on the features followed by a group Lasso procedure on the selected groups.} 
We run Latent Dirichlet Allocation (LDA) \cite{blei2003latent} on the corpus using \texttt{scikit-learn} \cite{pedregosa2011scikit}, where we set the number of groups to $100$. We then use the LDA solution to construct a collection of probability vectors $\{\B \pi^{(i)}\}_{i=1}^{100}$, each corresponding to a topic. Here $\pi^{(i)}_j$ refers to the probability of encountering word $j$ in topic $i$. We assign word $j$ to the group with index $\argmax_{i} \{\pi^{(i)}_j\}_{i=1}^{100}$ (i.e., to the group that allocates $j$ the highest probability). For example, the top $5$ words in group 1 are ``coffee roast cup keurig cups'' so the topic is on coffee. Group 2 has ``bpa worse cans dented claim'', which refers to problems with the packaging of the product. To obtain the training set, we sub-sample uniformly at random from the corpus and remove any covariates with zero variance (after sub-sampling), which reduces the number of covariates from $5500$ to $3482$. Note that the $100$ groups have different sizes, ranging between $9$ and $85$.

\textcolor{black}{We note that the above grouping procedure is one of many possible ways to obtain a grouping of the features. Our goal here is to obtain a partition of the features, to be used as an input for all the group sparse estimators, so that we are able to compare the performances of the different estimators. The downstream results depend upon the input groups. It may be interesting to see if one can simultaneously learn the grouping structure and build a sparse prediction model so as to optimize a suitable joint estimation criterion. This, of course, goes beyond the scope of the group-selection problem that we are studying in this paper, and is left as future work.}

\end{appendix}

\bibliographystyle{plainnat}
\bibliography{HusseinBiB,Peterbib}

\end{document}